\documentclass[12pt]{article}
\usepackage{psfig,epsfig}

\renewcommand{\theequation}{\thesection.\arabic{equation}}
 \newcommand{\EQ}{\begin{equation}}
 \newcommand{\EEQ}{\end{equation}}
 \newcommand{\EA}{\begin{eqnarray}}
 \newcommand{\EEA}{\end{eqnarray}}
 \newcommand{\D}{{\rm d}}
 \renewcommand{\r}{{\bf r}}

\newcommand{\Z}{{\cal Z}}
\newcommand{\tc}{\tilde t}
\newcommand{\rc}{\tilde r}
\newcommand{\hs}{\hspace{1mm}}
\newcommand{\HS}{\hspace*{1cm}}
\newcommand{\BS}{\hspace*{-1cm}}        
\newcommand{\ep}{\varepsilon}

\newcommand{\Ref}[1]{\ref{#1}}
\newcommand{\Label}[1]{\label{#1}}

\begin{document}

\begin{titlepage}
\vspace*{1cm}
 
\begin{center}

{\Large ON GROWTH, DISORDER, AND FIELD THEORY}
\\
\vspace{3cm}
{\large Michael L\"assig}
\\
\vspace{1cm}
Max-Planck-Institut f\"ur Kolloid-und Grenzfl\"achenforschung
\\
Kantstr.55, 14513 Teltow, Germany \\
lassig@mpikg-teltow.mpg.de 
\vspace{3cm}
\abstract{
This article reviews recent developments in statistical field theory far from  
equilibrium. It focuses on the Kardar-Parisi-Zhang equation of
stochastic surface growth and its mathematical relatives, namely
the stochastic Burgers equation in fluid mechanics and directed polymers
in a medium with quenched disorder. At strong stochastic driving --
or at strong disorder, respectively -- these systems develop {\em
nonperturbative} scale-invariance. Presumably exact values of the 
scaling exponents follow from a self-consistent asymptotic theory. 
This theory is based on the concept of an {\em operator product expansion}
formed by the local scaling fields. The key difference to standard
Lagrangian field theory is the appearance of a {\em dangerous irrelevant}
coupling constant generating {\em dynamical anomalies} in the continuum
limit.} 

\end{center}
\end{titlepage}

\newpage
\thispagestyle{empty}
\tableofcontents

\newpage

\section{Introduction}
\setcounter{page}{1}

The last 25 years have seen a very fruitful application of continuum
field theory to statistical physics~\cite{Cardy.book}. A system in
thermodynamic equilibrium tuned to a critical point ``looks the same
on all scales'', that is, it shows an enhanced symmetry called scale 
invariance. Since there is no characteristic scale, the correlation
functions in a critical system can only be power laws. 
Microscopically quite different systems share certain aspects of their
critical behavior which 
are therefore independent of the small-scale details. This 
phenomenon is called universality. Among the universal features 
are the exponents characterizing the power laws of correlation
functions. 

The renormalization group has provided the first satisfactory 
explanation of universality as well as calculational tools to 
compute, for example, critical exponents. 
The central concept is that of a flow on the ``space of theories''.
In equilibrium systems, this flow is parametrized by the coupling
constants of an effective Hamiltonian determining the partition
function. It can be understood as a link
between the microscopic coupling constants defining a model and the
effective parameters governing its large-distance limit. (For an Ising
model, e.g., the former are the spin couplings on the lattice and
the latter are the reduced temperature and the magnetic field.)
The fixed points 
of this flow represent the universality classes. At a fixed point,
a system can be described by a so-called renormalized continuum field
theory, that is, a theory where all microscopic quantities such as
the underlying lattice constant no longer play any role. The universal
properties can often be calculated in a systematic perturbation expansion.

About a decade after Wilson's seminal work on renormalization, the
advent of conformal field theory triggered a new interest in
two-dimensional critical phenomena~\cite{ItzyksonDrouffe.book}. 
It was realized that these systems have an even larger symmetry
called conformal invariance. (A conformal transformation is 
any coordinate transformation that {\em locally} reduces to a 
combination of scale
transformation, rotation and translation. The rescaling 
factor is now allowed to vary from place to place.) Conformal
invariance severely constrains the structure of continuum field theories
in two dimensions, yielding a (partial) classification of the
universality classes and their exact scaling properties. Under some
additional conditions, there is only a discrete set of solutions: the 
possible values of the critical exponents  are {\em quantized}.

The subsequently developed $S$-matrix theory has extended the exact 
solvability to theories close to a critical point, i.e., with
a large but finite correlation length.  
For the first time, it has been possible to verify the
ideas of scaling beyond perturbation theory for a whole class of
strongly interacting field theories. On the other hand, the conformal 
formalism entails a shift of focus from {\em global} quantities (such
as the Hamiltonian of a theory and its coupling constants) to
{\em local} observables, i.e., the correlation functions. Their structure 
is encoded by the so-called operator product expansion, a rather 
formidable name for a simple concept (explained in Section~2).
The operator product expansion is at the heart of a 
field theory. Conformal symmetry imposes a set of constraints 
on the operator product expansion which lead to expressions 
for the correlation functions. This theory makes no 
reference to a Hamiltonian. Without a renormalization
group flow, however, the link to microscopic model parameters is lost.
Hence, it has to be determined a posteriori which lattice models belong to
the universality class of a given conformal field theory.

Another important shift of focus has taken place in recent years.
Traditionally scale invariance has been associated with  second order
phase transitions, which requires fine-tuning of the model parameters to
a critical manifold. However, it became clear that there are many 
systems that have generic scale invariance in a region of their parameter
space. Perhaps the simplest such systems are interfaces. For example,
the surface of a crystal in thermal equilibrium fluctuates around an
ideal symmetry plane of the crystal. The displacement can be described
by the ``height'' field $h(\r)$, the two-dimensional vector~$\r$
denoting  the position in the reference plane. Above a certain
temperature, the surface is  rough, i.e., it develops mountains 
and valleys whose size is typically some power of the system
size. The correlation functions of the height field become 
power laws as well. The
roughness turns out to be even stronger if the crystal is growing,
which drives the surface out of equilibrium. There is another important
difference between equilibrium surfaces and driven surfaces. In the 
former case, the height pattern shows an up-down symmetry between 
mountains and valleys. Out of equilibrium, that symmetry is
lost.  

Obviously, such open systems
are quite common: any growth, pattern formation or reaction process
propagating through some localized boundary or front generates a driven
interface, often with long-ranged correlations~\cite{KrugSpohn.review, 
HalpinHealeyZhang.review}. An example is the kinetic roughening
of thin metal films by vapor deposition. Fig.~1(a) shows the ST microscope
analysis of a gold film at room temperature from which the
authors were able to extract power law behavior of the height
correlations, indicative of a scale-invariant surface 
state~\cite{HerrastiAl.gold}.  
\begin{figure}
\vspace*{-1.5cm}
\epsfig{file=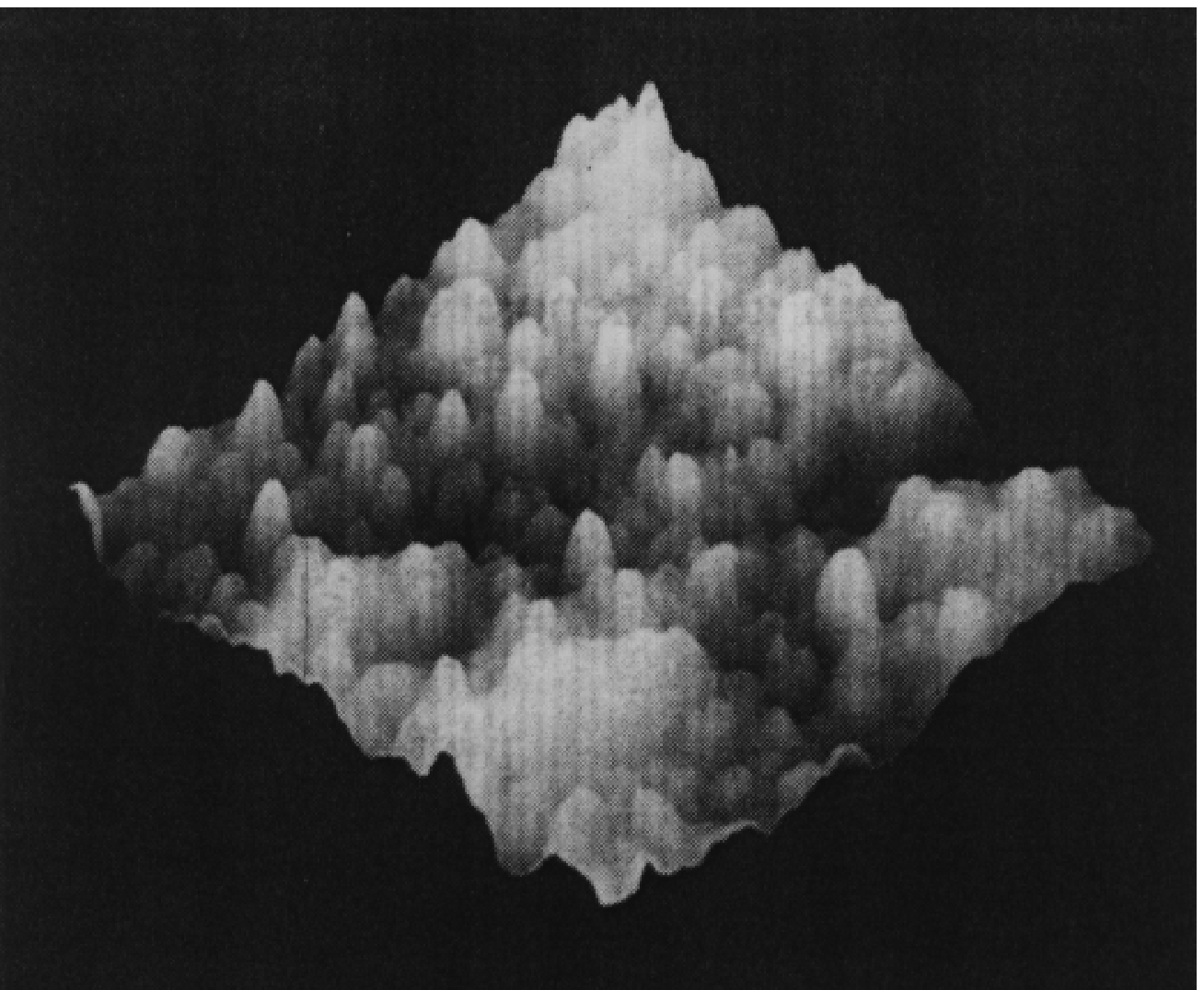,height=5cm}
\hspace{2cm} 
\raisebox{-2cm}{\epsfig{file=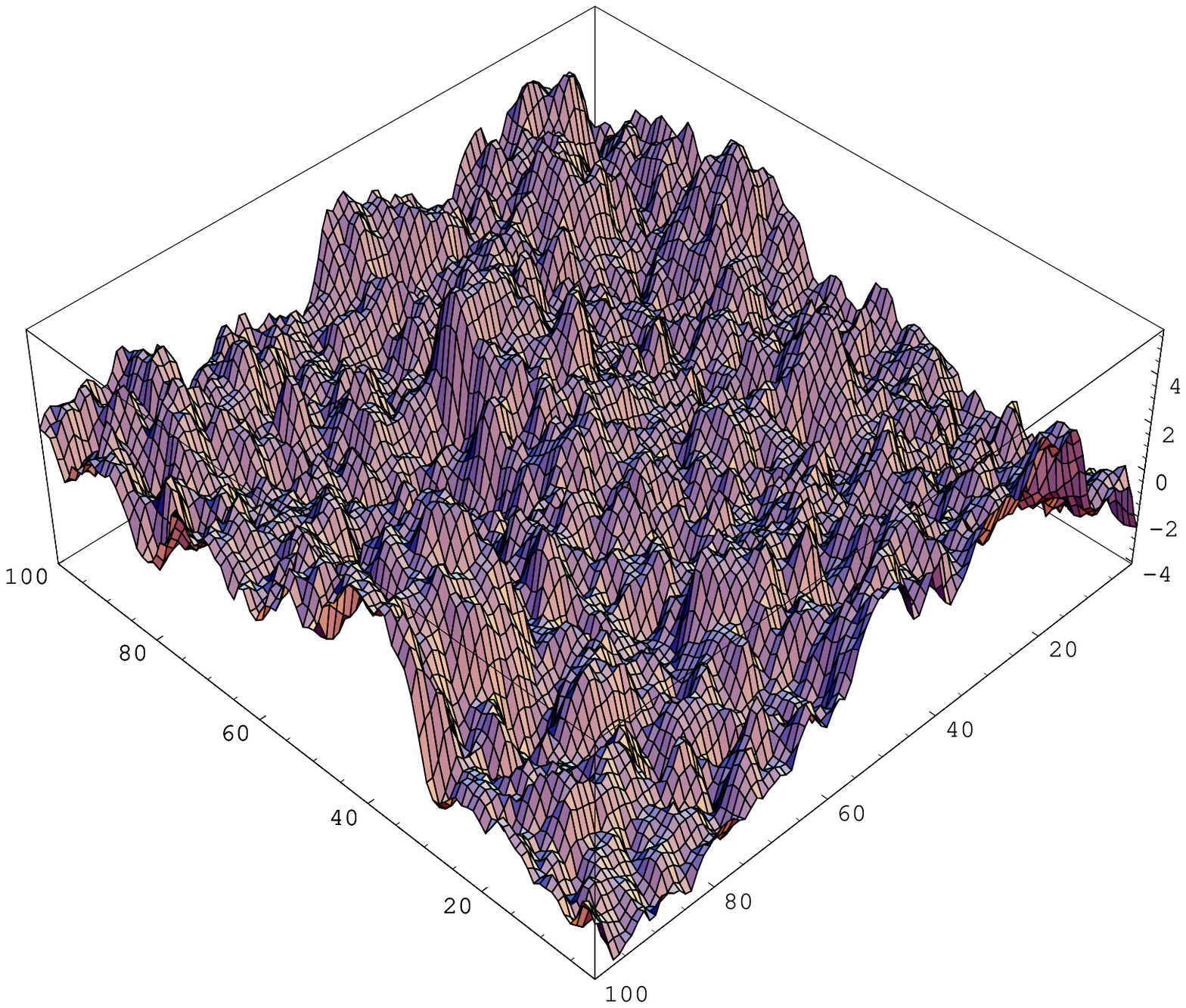,height=9cm}}
\vspace*{-1cm}
\baselineskip=12pt

{\small Fig. 1: 
(a) STM snapshot of a kinetically 
roughened gold film. The projected area of the sample is 
$ 510 \times 510 \mbox{nm}^2$. Courtesy of Herrasti 
et al.~\cite{HerrastiAl.gold}.
(b) Part of a discretized KPZ surface. The entire surface
has $400 \times 400$ lattice points and is shown after 1000 time steps.}
\end{figure}
This state is seen to be {\em directed} (i.e., it has no up-down symmetry) and
{\em stochastic}. Surface inhomogeneities increase with time, signaling
a {\em nonlinear} evolution. However, since there are no significant 
overhangs, the growth mechanism should be essentially 
{\em local}; that is, the growth
rate at a given point depends only on the surface pattern in the
neighborhood of that point. (Kinetic roughening can also produce 
quite different surface patterns with branched tree-like structures.
Their growth is strongly nonlocal since the large trees shield the smaller 
ones from further deposition of material~\cite{KahandraAl.electrochem}.)
 
Of course, generic
scale invariance far from equilibrium is not limited to interfaces.
Other important examples are hydrodynamic turbulence or slowly driven
systems with so-called self-organized criticality:  dynamical
processes  such as the stick-slip motion of an earthquake fault
generate a power law distribution of ``avalanches'' with long-ranged
correlations in space and time. A simple lattice model
with a self-organized critical state is the so-called forrest fire
model~\cite{BakAl.ff,DrosselSchwabl.ff}.
On a given lattice site, a tree grows with a small probability
per unit time. With an even smaller probability, the tree is 
hit by a lightning which destroys it along with all the trees 
in the same contiguous forrest cluster~\cite{DrosselSchwabl.ff}.
These events are the avalanches. The dynamics leads to a self-similar 
stationary pattern of forrests and voids.

A satisfactory theory of non-equilibrium scale invariance should
classify the different universality classes and provide calculational
methods to obtain the scaling exponents exactly or in a controlled
approximation. To establish such a theory is obviously a complex task
that will challenge statistical physicists probably over the next 25 years.  
The power law
correlations on large scales of space and time should again be
described by continuum field theories, although the proper continuum
formulation is far from clear for many dynamical systems defined
originally on a lattice. It is also an open issue which of the
presently known field-theoretic concepts will continue to play an
important role.  For example, perturbative renormalization may fail to
produce a fixed point describing the large-distance regime, as will be
shown below for the example of driven surface growth. The seemingly
easier task of describing the time-independent scaling in a stationary
state is still involved since there is no simple Hamiltonian generating
these correlations.

This article collects a few results that may become part of an eventual
field theory of nonequilibrium systems. 
We limit ourselves to models related to the Kardar-Parisi-Zhang (KPZ)
equation~\cite{KPZ}
\EQ
\partial_t h(\r,t) = 
 \nu \nabla^2 h(\r,t) + \frac{\lambda}{2} (\nabla h(\r,t))^2 +
 \eta(\r,t)
\Label{KPZ}
\EEQ
for a $d$-dimensional height field $h(\r,t)$ driven by a force 
$\eta(\r,t)$ random in space and time. This 
equation has come to fame as the ``Ising model'' of nonequilibrium
physics. It is indeed the  simplest equation capturing nevertheless
the main determinants of the growth dynamics in Fig.~1(a):
{\em directedness, nonlinearity, stochasticity}, and {\em locality}.
The  KPZ surface shown in Fig.~1(b) has been produced  by a discretized version of
the growth rule (\ref{KPZ}) and looks indeed qualitatively 
similar to these experimental data. 

The  theoretical richness of the KPZ model is partly due to  
close relationships with other areas of statistical physics --
notably hydrodynamic turbulence and disordered systems -- which are
briefly reviewed below. Many more details can be found in
refs.~\cite{KrugSpohn.review, HalpinHealeyZhang.review}.
Due to the famous problem of quenched averages,
disordered systems share some of the conceptual problems mentioned above.
The observables are correlation functions averaged over the distribution 
of the random  variables, which is not given by a Boltzmann weight.
These correlations may be regarded as an
abstract field theory but there is again no simple effective Hamiltonian. 
Such systems can have scale-invariant states at zero temperature for
which the very existence of a continuum limit needs to be re-established.

Despite considerable efforts, the KPZ equation
has so far defied attempts at an exact solution or a systematic
approximation in dimensions $d > 1$.  The main reason is the failure of
renormalized perturbation theory. Renormalization aspects of
Eq.~(\Ref{KPZ}) and its theoretical relatives will be discussed below.
The main emphasis lies, however, on structures beyond perturbation theory.
We analyze the internal consistency of the strong-coupling field theory
expressed by the operator product expansion of its local fields. This 
approach turns out to be quite powerful. Using phenomenological 
constraints and the symmetries of the equation, it produces  
a quantization condition on the scaling indices from
which their exact values in $d = 2$ and $d = 3$ can be deduced. This 
quantization is somewhat reminiscent of what happens in conformal field
theory and suggests the KPZ equation 
possesses an infinite-dimensional symmetry as well.  

It is not clear to what extent the results carry over to other
nonequilibrium systems. Yet, the approach used here is fairly general and
should be applicable in a wider context. It transpires that
incorporating nonequilibrium phenomena into the 
framework of field theory will require yet another shift of focus 
to nonperturbative concepts and methods. This is likely to change
our view of field theory as well. Negative scaling dimensions, dangerous
variables, anomalies etc.~are oddities today but may become an essential
part of its future shape.

\subsection{Directed growth, Burgers equation, and polymers}

The time evolution of a KPZ surface depends only 
on the local configuration of the surface itself (and not, for example, on the bulk
system beneath the surface). Hence, the r.h.s.~of Eq.~(\Ref{KPZ})
contains only terms that are invariant under translations $h \to h +
\mbox{const.}$: 
\newline (a) The dissipation term $\nabla^2 h$ is the divergence
of a downhill current and acts to smoothen out the inhomogeneities 
of the height field. 
\newline (b) The nonlinear term  $(\nabla h)^2$ arises
from expanding the tilt dependence of the local growth rate and acts
to increase the inhomogeneities of the surface. A linear term 
$ {\bf b} \cdot \nabla h$ would be redundant since it could be absorbed 
into a tilt  $h \to h + {\bf b} \cdot \r$. The higher  powers 
$(\nabla h)^3, \dots$ turn out to be irrelevant in the presence of the quadratic
term, as well as terms containing higher gradients such as $(\nabla^2
h)^2$ or $\nabla^4 h$. 
\newline (c) The stochastic driving term $\eta(\r,t)$
describes the random adsorption of molecules onto the surface. It is
taken to have a spatially uniform Gauss distribution with correlations 
only over microscopic distances,
\EQ
\overline{\eta(\r,t)} = 0 \;, \HS
\overline{\eta(\r,t) \eta(\r',t')} = 
         \sigma^2 \delta(\r - \r') \delta(t - t') \;.
\Label{etaeta}
\EEQ
A uniform average $\overline{\eta(\r,t)} = \overline \eta$ would again
be redundant since it could be absorbed into the transformation 
$h(\r,t) \to h(\r,t) - \overline \eta t$.  

Eq.~(\Ref{KPZ}) is by no means the only model for a driven surface, and
many experimental realizations of crystal growth are probably governed
by related equations with additional symmetries and conservation 
laws~\cite{WolfVillain.cons}
or with different correlations of the driving force 
(see, for example,~\cite{Krug.review}). As the simplest
nonlinear model, however, the KPZ equation remains a
cornerstone for the theoretical understanding of stochastic growth.

The morphology  of a rough surface is characterized by 
the asymptotic scaling of the spatio-temporal height correlations. 
In a stationary state, the mean square height difference is expected
to take the form
\begin{equation}
\langle ( h({\bf r}_1, t_1) - h({\bf r}_2, t_2) )^2 \rangle
    \sim |\r_1 - \r_2|^{2 \chi} \, 
    {\cal G} \left ( \frac{t_1 - t_2}{|\r_1 - \r_2|^z} \right ) \;;
\Label{hh}
\end{equation}
the higher moments  
$\langle ( h({\bf r}_1, t_1) - h({\bf r}_2, t_2) )^k \rangle$
are of similar form. (In a  system of finite size $R$, Eq.~(\ref{hh}) is
valid for $|\r_1 - \r_2| \ll R$ and $|t_1 - t_2| \ll R^z$.)
The scaling function $\cal G$ parametrizes 
the crossover between the power laws 
$\langle ( h({\bf r}_1, t) - h({\bf r}_2, t) )^2 \rangle 
 \sim |\r_1 - \r_2|^{2 \chi}$ and
$\langle ( h({\bf r}, t_1) - h({\bf r}, t_2) )^2 \rangle 
 \sim |t_1 - t_2|^{2 \chi/z}$
of purely spatial and purely temporal correlations, respectively. 
These relations define the roughness exponent $\chi > 0$ and the dynamic
exponent $z$. In the marginal case $\chi = 0$, the surface may still be
logarithmically rough. 

A surface governed by the linear 
dynamics (\ref{KPZ}) with $\lambda = 0$  
has $\chi = (2 - d)/2$ and $z = 2$: it is 
rough for $d = 1$, marginally rough for $d = 2$ and smooth for $d > 2$.
The phase diagram  is well known also for $\lambda \neq 0$. In
dimensions $d \leq 2$, any small nonlinearity $(\lambda/2) (\nabla
h)^2$ is a relevant perturbation of the linear theory and induces
a crossover to a different rough state called the strong-coupling regime. 
For $d > 2$,
a small nonlinearity does not alter the
smooth state of a linear surface. There is now a roughening transition 
to the strong coupling regime at finite values $\pm
\lambda_c$~\cite{ImbrieSpencer,CookDerrida,EvansDerrida}. 

This phase diagram corresponds to the following renormalization group
flow. For $d \leq 2$, the Gaussian fixed point ($\lambda = 0$) is 
{(infrared-)}un\-stable, and there is a crossover to the stable
strong-coupling fixed point. For $d >2$, a third fixed point exists,
which represents the roughening transition. It is unstable and 
appears between
the Gaussian fixed point and the strong-coupling fixed point, which are
now both stable~\cite{TangNattermannForrest,
NattermannTang,FreyTaeuber,Lassig.kpz}.

In one dimension, the critical indices of the strong-coupling 
regime take the exact values $\chi = 1/2$ and 
$z = 3/2$~\cite{FNS,KPZ,GwaSpohn.Bethe}. Their
values in higher dimensions as well as the properties of the
roughening transition have been known only 
numerically~\cite{KimKosterlitz.num,ForrestTang.num,KimAl.num,
KimMooreBray.num,TangAl.num,AlaNissilaAl.num,AlaNissila.ucd,Kim.ucd}
and in various approximation 
schemes~\cite{HalpinHealey.fren,NattermannLeschhorn.fren,MooreAl.modecoupling}. 
In particular, it has been controversial 
whether there is a finite upper critical dimension $d_>$ at and above 
which KPZ surfaces are only marginally rough ($\chi = 0$
and $z = 2$). These issues are discussed in detail
in Sections~3 and~4. 

Experiments on growing surfaces require delicacy since crossover and saturation
effects can mask the asymptotic scaling. However, several experiments have
produced scaling consistent with KPZ growth. The fire fronts in slow
combustion of paper have $\chi = 0.50$ and $z =
1.53$~\cite{MaunukselaAl.paper}, in very good agreement with the KPZ values
in $d = 1$.
A recent study of kinetically roughened Fe/Au multilayers~\cite{PaniagoAl.exp}
obtains $\chi = 0.43 \pm 0.05$, 
which should be compared to the current numerical
estimate $\chi \approx 0.39$ for $d = 2$ and to the presumably exact value
(\ref{chi2}).

Eq.~(\Ref{KPZ}) is formally equivalent to Burgers' equation
\begin{equation}
\partial_t {\bf v} + ({\bf v} \cdot {\bf \nabla}) {\bf v}
          = \nu \, {\bf \nabla}^2 {\bf v}  + {\bf \nabla} \eta 
\Label{Burgers}
\end{equation}
for the driven dynamics of the vortex-free velocity field ${\bf v}({\bf r},t) =
{\bf \nabla} h({\bf r},t)$ describing a randomly stirred fluid 
(with $\lambda = -1$)~\cite{FNS}. In this formulation, Galilei invariance
becomes obvious: the substitutions 
\EQ
h(\r,t) \to h(\r - {\bf u} t,t) + {\bf u} \cdot \r - \frac{1}{2} {\bf u}^2 t
\;, \HS 
{\bf v}(\r,t) \to {\bf v}(\r - {\bf u} t,t) + {\bf u}
\EEQ
leave Eqs.~(\Ref{KPZ}) and (\Ref{Burgers}) invariant. Due
to this invariance, the roughness exponent and the dynamical exponent 
obey the scaling relation~\cite{MeakinAl.scal}
\EQ
\chi + z = 2 \;,
\Label{2}
\EEQ
which guarantees that the total derivative 
$ \D_t \equiv \partial_t + {\bf u} \cdot \nabla$ 
behaves consistently under scale transformations. 

It should be emphasized, however, that the velocity field of a stirred 
Burgers fluid looks quite different from the gradient field of a
KPZ surface since the driving force in a fluid 
is correlated over {\em macroscopic} spatial distances $R$. This generates
turbulence~\cite{BouchaudAl.dp} (somewhat different, of course, from
Navier-Stokes turbulence). The velocity correlations show
multiscaling. For example, the stationary moments
$\langle (v_\|(\r_1) - v_\|(\r_2))^k \rangle$ of the longitudinal
velocity difference 
have a $k$-dependent singular dependence on $R$ for $|\r_1 - \r_2| \ll R$.
Multiscaling is not expected for driving forces with short-ranged 
correlations (\ref{etaeta}). This point will become important below.  

Via the well-known Hopf-Cole transformation, 
\EQ
Z(\r,t) \equiv \exp \left [ \frac{\lambda}{2 \nu} h ({\bf r} ,t) \right ] \;,
\Label{HopfCole}
\EEQ
Eq.~(\Ref{KPZ}) can be mapped onto the imaginary-time Schr\"odinger
equation
\EQ
\beta^{-1} \partial_t Z = \frac{\beta^{-2}}{2} \,\nabla^2 Z + \lambda \eta Z 
\Label{Zt}
\end{equation}
with $\beta = 1/2\nu$~\cite{KPZ}.
The solution can be represented as a path integral 
\EQ
Z(\r',t') = \int {\cal D} {\bf r} \, \delta ( {\bf r}(t') - {\bf r'}) 
\exp ( - \beta S ) 
\Label{Zrt}
\end{equation}
with the action
\begin{equation}
S = \int^{t'} {\rm d} t
           \left ( \frac{1}{2} \left(  \frac{{\rm d}{\bf r}}{{\rm d} t}
                                \right )^2 -
                    \lambda \eta({\bf r} (t), t) \right ) \;,
\Label{Seta}
\end{equation}
describing a {\em string} or directed polymer ${\bf r}(t)$ (i.e., the 
world line of
a random walk)  in the quenched random potential 
$\lambda \eta({\bf r}, t)$ at temperature $\beta^{-1} = 2 \nu$.
A configuration of the string is shown in Fig.~2.  
The stochastic driving term now appears as quenched disorder in a
$(1 + d)$-dimensional equilibrium system. This system is of
conceptual importance as one 
of the simplest problems with quenched disorder.  
\begin{figure}
\epsfig{file=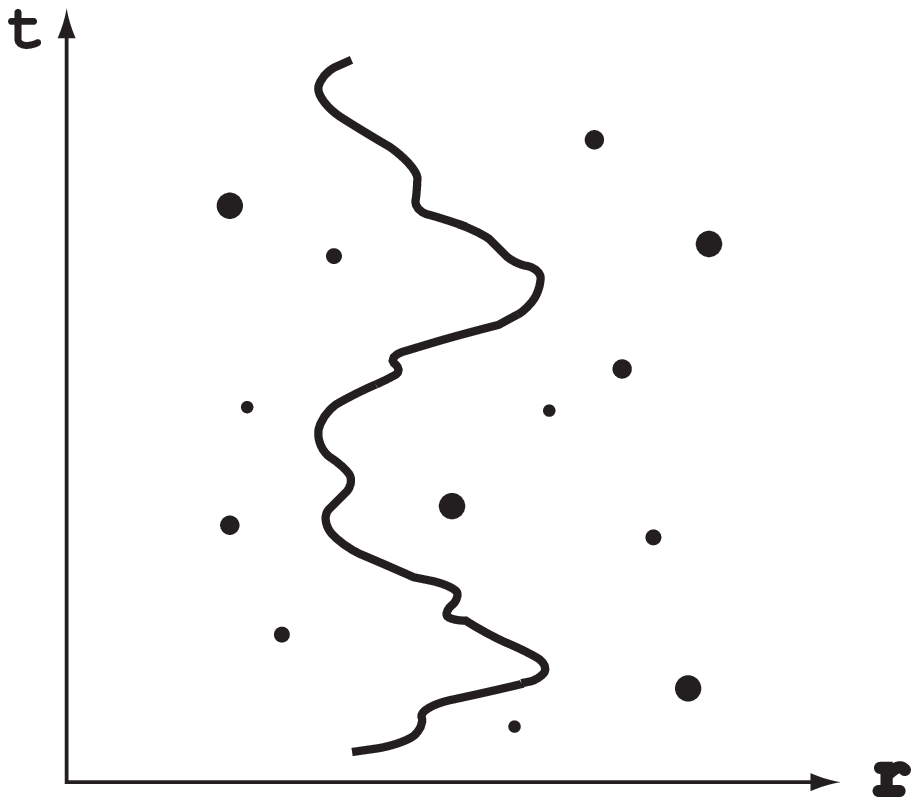,height=5cm}

\vspace{10pt}
\baselineskip=10pt
{\small Fig. 2: A configuration $\r(t)$ of a string (or directed polymer) in 
a medium with quenched point disorder. Due to the inhomogeneities of the
medium, a typical path takes larger excursions to the left and to the right 
than an ordinary random walk.}
\end{figure}

The transversal displacement of the string,
\begin{equation}
\overline{ \langle ( {\bf r}(t_1) - {\bf r}(t_2))^2 \rangle } 
\sim |t_1 - t_2|^{2 \zeta} \;,
\Label{r2} 
\end{equation}
defines its roughness exponent $\zeta$. (Averages over the disorder are denoted
by overbars, thermal averages by brackets $\langle \dots \rangle$.) 
The rough strong-coupling 
regime (\Ref{hh}) of the growing surface corresponds
to a superdiffusive state of the string 
($\zeta > 1/2$)~\cite{HuseHenley.paths,Kardar.dp,FisherHuse.paths}. In this
state, the universal part of its free energy in a system of longitudinal size $L$
and transversal size $R$ has the scaling form
\EQ
\overline F(L,R) \sim L^\omega {\cal F} (L R^{-1/\zeta}) \;.
\Label{FLR}
\EEQ
In particular, the ``Casimir'' term 
\begin{equation}
\overline f (R) \equiv 
\lim_{L \to \infty} \partial_L  \overline F(L,R)
\sim R^{(\omega - 1) / \zeta} 
\Label{fbar}
\end{equation}
measures the free energy cost per unit of $t$ of confining a long string 
to a tube of width $R$. 
The exponents $\zeta$ and $\omega$ are related to the growth exponents by 
\EQ
\zeta = 1/z \;, \HS  \omega = \chi/z \;.
\EEQ
The scaling relation (\Ref{2}) now reads 
\EQ
\omega = 2 \zeta - 1 \;. 
\label{scaling2}
\EEQ
Replica methods yield the exact exponents $\zeta = 2/3$ and
$\omega = 1/3$ for $d = 1$ but fail in higher dimensions. 

In the superdiffusive state, the free 
energy acquires an anomalous dimension $-\omega < 0$. (At an ordinary
critical point, the free energy is scale-invariant ($\omega = 0$), which 
implies a set of hyperscaling relations. Such relations are no longer
valid for quenched averages.) 

The disorder-induced fluctuations (\Ref{r2}) 
persist in the limit $\beta^{-1} \to 0$, that
is, in the ensemble of minimum energy paths $\r_0(t)$. In the
weak-coupling (high-temperature) regime for $d > 2$, thermal
fluctuations dominate ($\zeta = 1/2$) and hyperscaling is preserved
($\omega = 0$).  The roughening transition between these two phases
takes place at a finite temperature $\beta_c^{-1}$. For $d \geq d_>$,
the Gaussian exponents govern the low-temperature phase as well,
albeit with possible logarithmic corrections.

\subsection{Overview of this article}

As emphasized already KPZ growth defines a field theory that is
{\em non-Lagrangian} and {\em non-perturbative}. This article focuses
on exact properties of its local correlation functions. We mention only
briefly the results of various approximation schemes. In particular,
functional renormalization (\cite{HalpinHealeyZhang.review} and
references therein) and mode-coupling theory 
(\cite{MooreAl.modecoupling, FreyAl.modecoupling} and references therein)
are important theoretical tools in a number of strong-coupling problems
but their status in field theory is not yet fully understood.

The first part of this article describes directed strings. 
These systems have a fascinating spectrum of physical applications 
and the language of directed strings proves to be an ideal framework 
to address some of the theoretical issues of directed growth.

In Section 2, we discuss directed strings in thermal equilibrium
without quenched disorder. A single such string describes a free random
walk and is thus generically rough.  Interactions of a single string
with an external defect or mutual interactions between strings,
however, can induce a localization transition destroying the
long-ranged correlations of the rough state. (De-)localization 
phenomena are an essential feature of strings and 
surfaces~\cite{ForgacsAl.DG}; they will appear in several
contexts throughout this article. We use renormalized perturbation
theory to derive the phase diagram of directed strings with
short-ranged interactions and the critical behavior at the
(de-)localization 
transition~\cite{LassigLipowsky.depinning,Lassig.fermions,LassigLipowsky.Altenberg}. 
The results are, of course, well known
and can be derived in various other ways.  The approach used here
stresses that the response of the system to perturbations probes the
correlations in the unperturbed, rough state. It is based on the
{\em operator product expansion} of the local interaction fields,
a familiar concept in Lagrangian field theory (see, e.g.,
ref.~\cite{Cardy.book}).

In Section 3, this approach is extended to the field theory of
directed strings in a random medium. In the replica formalism, a single
such string is represented by a system of many strings 
without quenched disorder but with mutual interactions. The 
localized many-string
state corresponds to the strong-coupling regime
of the random system. Perturbative renormalization of these
interactions turns out to produce the exact scaling at the roughening
transition for $2 < d < 4$ but fails to describe the strong-coupling
regime~\cite{Lassig.kpz,BundschuhLassig.kpz}. Insight  can be 
gained by studying several strings in 
a random medium with additional direct interactions. These probe the
disorder-induced correlations in the scale-invariant strong-coupling
regime. The temperature becomes a {\em dangerous irrelevant}
variable at the strong-coupling fixed point; this is the field-theoretic
fingerprint of quenched randomness. The direct interactions 
can be treated in renormalized perturbation theory about that
fixed point~\cite{KinzelbachLassig.depinning,KinzelbachLassig.fermions},
assuming the existence of an operator product expansion.
We find again (de-)localization transitions, which are relevant
for various applications. Their critical properties 
are given in terms of the single-string exponents.
Comparing the effect of pair interactions in the strong-coupling phase
and at the roughening transition of a single string then shows that the
single-string system -- corresponding to the standard KPZ dynamics -- 
has an upper critical dimension 
$d_> \leq 4$~\cite{LassigKinzelbach.ucd,LassigKinzelbach.reply}.

Section 4 returns to growing surfaces. The dynamical field theory of
KPZ systems and its renormalization are discussed. 
Perturbative renormalization of the dynamic functional is compared to the 
string renormalization of Section 3~\cite{Lassig.kpz}, and it is
shown why perturbation theory fails for the strong-coupling regime in 
$d > 1$. However, the scaling in this regime can be studied directly
using the operator product expansion of the height field. We find that the 
KPZ equation can have only a discrete set of solutions 
distinguished by field-theoretic {\em anomalies}~\cite{Lassig.anomaly}. 
Comparing this set with numerical estimates of the exponents $\chi$ 
and $z$ then gives their exact values in $d = 2$ and $d = 3$.

\bigskip 

\section{The field theory of directed strings}

Ensembles of interacting directed strings describe a surprising variety
of statistical systems in a unifying way.  Examples are 
interfaces between different bulk phases in a 2D system~\cite{ForgacsAl.DG},  
steps on crystal surfaces~\cite{Jayapakrash.TSK}, 
flux lines in a type-II superconductor~\cite{NelsonAl.fluxlines}, 
or 1D  elastic media~\cite{CuleHwa.tribology}.  
Directed strings are also related to
mathematical algorithms detecting similarities between DNA 
sequences~\cite{HwaLassig.dna,DrasdoAl.global,HwaLassig.local1,DrasdoAl.local2,%
Olsen}.

At finite temperatures, a
single string (or a collection of independent ones) would simply perform
Gaussian fluctuations.   It is the interactions of the
strings with each other and with external objects that generate the
thermodynamic complexity of these systems. Attractive forces lead to
 wetting transitions of interfaces, bunching transitions of
steps, and depinning transitions of flux lines. All of these are transitions 
between a  delocalized high-temperature state with unconstrained
fluctuations and a  localized low-temperature state whose
displacement fluctuations are constrained to a finite width
$\xi$; for a review, see ref.~\cite{ForgacsAl.DG}. 
In this and the next Section, we discuss a few such systems, 
emphasizing their common field-theoretic aspects.

A single thermally fluctuating string  
is given by the partition function 
\EQ
Z = \int {\cal D} r \exp ( - \beta S[\r] )
\EEQ
with the Gaussian action 
\EQ
S[r] = \int \D t \, \frac{1}{2} \left ( \frac{\D \r}{\D t} \right )^2 
\EEQ
for the $d$-component displacement field $\r(t)$.
In a finite system ($0 \leq t \leq L$, $0 \leq r_1, \dots, r_d \leq R$),
the universal part of the free energy has the scaling form
\EQ
F(L,R) = {\cal F}(L/\beta R^2) \;.
\EEQ
This defines in particular the Casimir amplitude
\EQ
{\cal C}(R) \equiv \beta ^2 R^2  \lim_{L \to \infty} \partial_L F(L,R) \;,
\Label{C0}
\EEQ
measuring the scaled free energy cost per unit of $t$ of confining 
a long string to a tube of width $R$. 
It depends only on the boundary conditions in
transversal direction. For periodic boundary conditions, ${\cal C} = 0$.

The displacement field $\r(t)$  has the negative scaling dimension 
$-\zeta_0 = -1/2$. Its two-point function
\EQ
\langle \r(t_1) \r(t_2) \rangle = 
 \int \D \omega \, \frac{ e^{i \omega (t_1 - t_2)}}{\omega^2}
\Label{rr}
\EEQ 
requires an infrared regularization by appropriate boundary conditions.
It is the difference correlation function 
\EQ 
\langle (\r(t_1) - \r(t_2))^2 \rangle =
-2 \langle \r(t_1) \r(t_2) \rangle 
+ \langle \r^2(t_1) \rangle 
+ \langle \r^2(t_2) \rangle \sim 
|t_1 - t_2|^{2 \zeta_0}
\EEQ
that remains well-defined in the thermodynamic limit $L,R \to \infty$ and 
becomes scale-invariant. 
The exponent $\zeta_0$ is called  the thermal roughness exponent. 

Consider now a directed string interacting with a rigid linear 
defect at $\r = 0$ as shown in Fig.~3.
 \begin{figure}
\epsfig{file=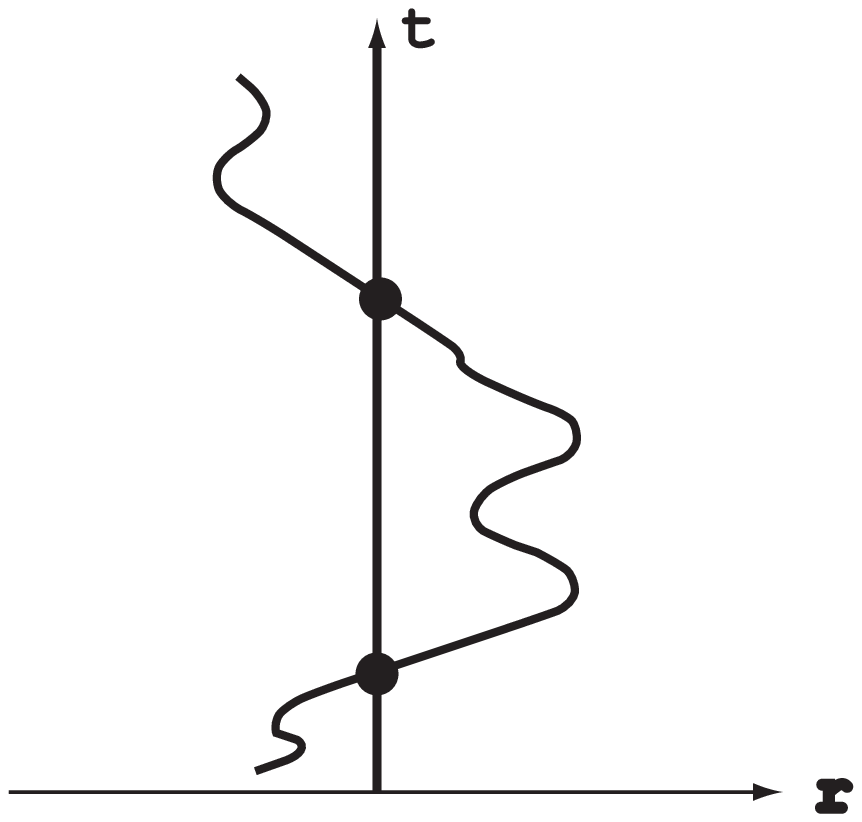,height=4cm}

\vspace{10pt}
\baselineskip=12pt
{\small Fig. 3: 
A thermally fluctuating directed string $\r(t)$ and a rigid linear defect
at $\r = 0$. Contact interactions between these objects are described
by the local scaling field $\Phi(t)$.}
 \end{figure}
If the interaction decays on the microscopic scale $|\r| \sim a$, the
system has the action
\EQ
S[r] = \int \D t \left ( \frac{1}{2} \left ( \frac{\D \r}{\D t} \right )^2 +
                g_0 \, \Phi (t) \right ) \;,
\Label{S1}
\EEQ
in the continuum limit $a \to 0$. The local interaction of the string
with the defect is proportional to the contact field 
$\Phi (t) \equiv \delta (\r(t))$ of canonical scaling
dimension $x_0 = d \zeta_0$. The conjugate coupling constant $g_0$ has the dimension
$y_0 = 1 - x_0$ with $t$ as the basic scale.

Of course, this system can be treated exactly, for example by solving
the imaginary-time Schr\"odinger equation
\EQ
\beta^{-1} \partial_t Z = \frac{\beta^{-2}}{2} \, \nabla^2 Z + g_0 \, \delta(\r) Z
\Label{Schr}
\EEQ
for the wave function (\Ref{Zrt}); see~\cite{Lipowsky.lines} in the
context of directed strings. Here we discuss a different way of
solution~\cite{LassigLipowsky.depinning,Lassig.fermions,LassigLipowsky.Altenberg} 
that can be generalized to problems with quenched disorder.
For the purposes of this Section, it is convenient to set $\beta = 1$,
which amounts to the substitution $t \to \beta^{-1} t$ in the action 
(\Ref{S1}).

\bigskip

\subsection{Correlation functions and the operator product expansion}

The perturbative analysis of the interaction in (\Ref{S1}) 
is based on the correlation functions $\langle \Phi (t_1) \dots
\Phi(t_N) \rangle$ in the unperturbed state, which can be calculated explicitly. 
We take each component of the
displacement vector to be compactified on a circle of circumference $R$.
The  scale $R$ also serves to generate the renormalization group flow
defined below. With this regularization, longitudinal translation invariance emerges
for ``bulk'' values $0 \ll t_1, \dots, t_N \ll L$ in the limit $L \to \infty$
independently of the boundary conditions at $t = 0$ and $t = L$.

The translation invariant one-point function
\EQ
\langle \Phi (t) \rangle \equiv \langle \Phi \rangle = R^{-x_0 / \zeta_0}
\Label{Phi0}
\EEQ
is simply the probability (density) of finding the fluctuating string $\r(t)$ at the
origin $\r = 0$ for a given $t$. Similarly, the $N$-point function 
$\langle \Phi(t_1) \dots \Phi(t_N) \rangle$ is the joint probability of 
the configurations with $N$ intersections of the origin at given values
$t_1, \dots, t_N$. 

These correlation functions develop singularities
as some of the points approach each other.
For example, the joint probability of intersecting
the origin $\r = 0$ both at $t$ and at $t'$ equals the single-event
probability (\Ref{Phi0}) times the probability
of return to the origin after a ``time'' $|t-t'|$, which becomes
singular as $t' \to t$:
\EQ
\langle \Phi (t) \Phi(t') \rangle = 
  C_0  | t - t'|^{-x_0}  \, \langle \Phi(t) \rangle + \dots 
\Label{Phi02}
\EEQ
with $C_0 = (2 \pi)^{-x_0}$. 

The structure of this singularity and the
coefficient $C_0$ are local properties: they appear 
in any (connected) $N$-point function as two of the arguments $t_i, t_j$
approach each other, independently of the other points remaining
at a finite distance and of the infrared regularization.
This can be expressed by the field 
relation~\cite{LassigLipowsky.depinning} 
\EQ
\Phi (t) \Phi(t') = C_0  | t - t'|^{-x_0}  \, \Phi(t) + \dots 
\Label{ope0}
\EEQ
illustrated in Fig.~4. It is 
called an operator product expansion (a familiar concept in
field theory; see, e.g., ref.~\cite{Cardy.book}).  
The dots denote 
less singular terms involving gradient fields. 
Such terms are generated, for example, if the r.h.s.~of (\Ref{ope0})
is expressed in terms of 
$\Phi(t') = \Phi(t) + (t' - t) \Phi'(t) + \dots $,
which leaves the leading singularity invariant. 
\begin{figure}
 \vspace*{0.5cm}
 \epsfig{file=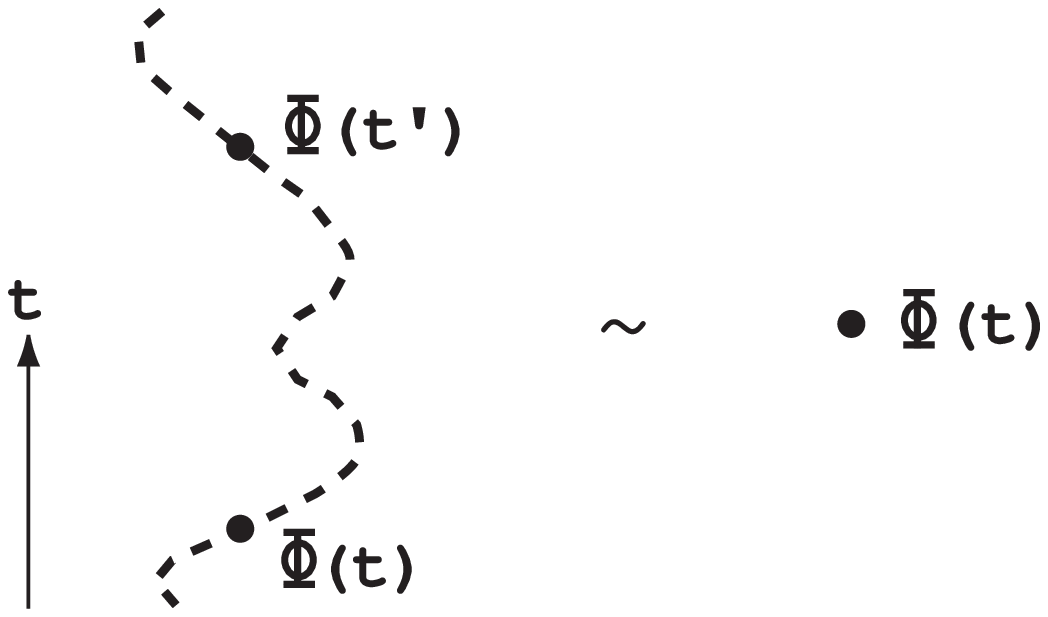,height=4cm}
 \vspace{10pt}
 
 \baselineskip=12pt
 {\small Fig. 4: 
 Operator product expansion of contact fields for a thermally fluctuating
 string. The short-distance asymptotics of the pair of fields $\Phi(t)$ 
 and $\Phi(t')$ is given by the single field $\Phi(t)$ times a singular
 prefactor. The dashed line indicates the string configurations generating
 the singularity $|t - t'|^{-x_0}$.  
 }
 \end{figure}

\bigskip

\subsection{Renormalization}

It is convenient to set up the perturbation theory for the Casimir
amplitude (\Ref{C0}). Since ${\cal C}$ is a 
dimensionless number, the contribution of the interaction  
\EQ
\Delta C (u_0) \equiv {\cal C}(g_0,R) - {\cal C}(0,R)
\label{C0g}
\EEQ
depends only on the dimensionless coupling constant
\EQ
u_0 \equiv g_0  R^{y_0 / \zeta_0} \;.
\EEQ  

The perturbation expansion 
\EQ
\Delta {\cal C}(u_0) = 
   -  R^2 \sum_{N = 1}^{\infty} \frac{ (- g_0)^N }{N!}
   \int \D t_2 \dots \D t_N \langle \Phi (t_1) \dots \Phi (t_N) \rangle^c \;.
\Label{C0series}
\EEQ
contains integrals over connected correlation 
functions of the contact field in the Gaussian theory ($g_0 = 0$).
Hence, the singularities of the operator product expansion 
(\ref{ope0}) lead to poles in (\Ref{C0series}). 
Inserting (\Ref{Phi0}) and (\Ref{ope0}) into (\Ref{C0series}), we obtain 
\EQ
\Delta {\cal C}(u_0) = 
    R^{x_0 / \zeta_0} \langle \Phi \rangle 
    \left (u_0 - \frac{C_0}{y_0} u_0^2 \right ) + O(y_0^0 u_0^2, u_0^3) \;.
\Label{C0series2}    
\EEQ
The same type of singularity (with different combinatoric factors)
occurs in the expansion of correlation functions 
$\langle \Phi(t_1) \dots \Phi(t_N) \rangle (u_0)$. For example,
\EA
\langle \Phi \rangle (u_0,R)  & = &
\sum_{N = 0}^{\infty} \frac{ (- g_0)^N }{N!}
   \int \D t_1 \dots \D t_N 
        \langle \Phi (t) \Phi (t_1) \dots \Phi (t_N) \rangle^c
\nonumber 
\\ & = &
\langle \Phi \rangle 
    \left (1 - \frac{2C_0}{y_0} u_0 \right ) + O(y_0^0 u_0, u_0^2) \;.
\Label{Phi0series}
\EEA

Perturbative renormalization consists in absorbing the singularities of the 
``bare''  series (\ref{C0series}) and (\ref{Phi0series}) 
into new variables $u_P$ and $\Phi_P$ defined order by order. (We use
the subscript $P$ to distinguish perturbatively defined couplings
and fields from their nonperturbatively renormalized counterparts.
This distinction is not necessary in the present context but
will become crucial below.) To leading (one-loop) order, the 
coupling constant 
renormalization can be read off directly from Eq.~(\ref{C0series2}).
Defining 
\EQ
u_P \equiv \Z_P u_0 
\Label{uP}
\EEQ
with 
\EQ
\Z_P (u_P) = 1 - \frac{C_0}{y_0} u_P  + O(y_0^0 u_P, u_P^2)
\Label{Z01}
\EEQ
the Casimir 
amplitude as function of $u_P$ becomes finite to order $u_P^2$, 
\EQ
\Delta {\cal C}(u_P) = 
    R^{x_0 / \zeta_0} \langle \Phi \rangle \, u_P 
    + O(y_0^0 u_P^2, u_P^3) \;.
\Label{C0series3}
\EEQ
The coupling constant renormalization (\ref{Z01}) implies
a renormalization of the conjugate field, 
\EQ
\Phi_P (t) \equiv \tilde \Z_P \Phi (t) 
\EEQ
with
\EQ
\tilde \Z_P (u_P) = \frac{\D u_0}{\D u_P} = 
1 + \frac{2 C_0}{y_0} u_P  + O(y_0^0 u_P, u_P^2) \;.
\Label{Z01tilde}
\EEQ
This renders also the correlation functions
$\langle \Phi_P(t_1) \dots \Phi_P(t_N) \rangle (u_P)$ finite.
The scale dependence of $u_P$ is governed by the flow equation
\EQ
\dot{u}_P \equiv \zeta_0 R \partial_R \, u_P
          = \frac{y_0 u_P}{1 - u_P (\D/ \D u_P) \log \Z_P} \;.
\EEQ
Using (\ref{Z01}), we obtain
\EQ
\dot{u}_P = y_0 u_P - C_0 u_P^2 +O(y_0 u_P^2, u_P^3) \;. 
\Label{beta01} 
\EEQ
Hence, the one-loop $\Z$-factors and the resulting flow equation
are entirely determined by the constants $y_0$ and
$C_0$ encoding local properties of the unperturbed theory. 

For this particular system, the one-loop equations turn out
to be very powerful because the pole at order $u_0^2$
is the only primitive singularity for $y_0 \to 0$ in the bare perturbation 
series. Hence, the theory is 
{\em one-loop renormalizable}~\cite{Duplantier.oneloop,RajBhatt.oneloop,
Lassig.kpz}: it can be described by the flow equation
\begin{equation}
\dot{u}_P = y_0 u_P - C_0 u_P^2  
\Label{beta0}
\end{equation}  
terminating at second order. This property leads to exact
results for local observables of the perturbed theory. In the Appendix, 
it is derived for the more general  many-string system of Section 3.2.

Of course, the form (\ref{beta0}) of the flow equation is not
unique. It depends on the infrared regularization of the bare
perturbation series and on the renormalization conditions defining
the coupling constant $u_P$. Changing either amounts to finite 
reparametrizations of $u_P$. Linear reparametrizations change the 
coefficient of
$u_P^2$ in (\ref{beta0}), while nonlinear reparametrizations introduce
higher order terms (see the discussion in the Appendix). 
However, local observables of the perturbed theory are ``gauge
invariant''; i.e., independent of these choices. They can
be computed exactly from (\ref{beta0}) and the associated 
$\Z$-factors (\ref{Z0}). 
The simplest such observable is the anomalous dimension 
\EQ
x^\star = 
 x_0 - \dot{u}_P \left. \frac{\D}{\D u_P} \log \tilde \Z_P (u_P) 
                 \right |_{u_P^\star} 
    = 1 + y_0 \;.
\Label{x0star}    
\EEQ
It governs the correlation functions of the contact field at the
nontrivial fixed point $u_P^\star$ of the flow equation,
for example,
\EQ
\langle \Phi(t) \rangle \sim R^{- x^\star/\zeta_0} 
\Label{Phi0star}
\EEQ
and 
\EQ
\langle \Phi (t) \Phi(t') \rangle 
\sim |t - t'|^{-x^\star} \langle \Phi(t) \rangle
\Label{Phi02star}
\EEQ
for $ |t - t'| \ll  R^2$.
By simple scaling arguments, it follows from (\Ref{Phi0star}) 
that the normalized string density 
\EQ
P(|\r'|) \equiv 
          \frac{\langle \delta(\r(t) - \r') \rangle}{
      \int \D \r \langle \delta(\r(t) - \r') \rangle} 
\Label{P}
\EEQ
has the singularity 
\EQ
P(r) \sim r^\theta R^{-d - \theta}
 \hspace{1cm} \mbox{for}\;\; r \ll R
\Label{Psing0}
\EEQ
with 
\EQ
\theta =  - \frac{x^\star}{\zeta} - d = 4 y_0 = 2 (2 - d)\;.
\Label{theta}
\EEQ
Of course, the exponent $\theta$ can be obtained in a simpler way.
The Schr\"odinger equation (\ref{Schr}) has the singular ground state 
wave function $Z \sim r^{2 - d}$, and $P \sim Z^2$.

\bigskip

\subsection{Results and discussion}

For $d < 2$, the Gaussian fixed point $u_P = 0$ is unstable
under the flow (\Ref{beta0}) and governs the (de-)localization
transition. Close to the transition, the localization width has the
singularity
\EQ
\xi \sim (-g_0)^{-\zeta_0 / y_0} \;\;\;\; (g_0 < 0)\;.
\label{xi0.1}
\EEQ
The fixed point $u^\star$ is stable and determines the asymptotic scaling 
with repulsive contact interactions in the limit 
$R \to \infty$ or $g_0 \to \infty$. 
The string density $P(r)$ then has a long-ranged depletion 
given by (\Ref{Psing0}) with $\theta >0$. 

At the borderline dimension $d = 2$, where the two fixed points coalesce,
the theory is asymptotically free with
\EQ
\xi \sim \exp (- C_0 \zeta_0 / g_0) \;\;\;(g_0 < 0)\;.
\label{xi0.2}
\EEQ

For $2 < d < 4$, the transition is governed by the nontrivial fixed point 
with
\EQ
\xi \sim (g_c - g_0)^{-\zeta_0 / y^\star} \;\;\;(g_0 < g_c) 
\label{xi0.3}
\EEQ
and $y^\star = 1 - x^\star = - y_0$. This fixed point now has a negative 
value of $\theta$, describing a divergence of
the string density $P(r)$ for $r \to 0$ 
due to the attractive interaction. 

It is obvious that the same scaling occurs in a number of related 
systems. For a directed string $(r_1 >0, r_2, \dots, r_d) (t)$
confined to  half space, the term $g_0 \delta (r_1)$ describes a 
short-ranged interaction with the boundary of the system.  The 
scaling of the string close to the boundary is described by the
above two fixed points for $d = 1$ with the transition point shifted 
to a value $g_c < 0$ and the fluctuations parallel to the boundary 
decoupled. Similarly, a single directed string interacting with
a linear defect is equivalent to the two-string system
$Z = \int \D \r_1 \D \r_2 \exp (- \beta S[\r_1,\r_2])$ with the action
\EQ
S [\r_1, \r_2] = 
   \int \D t \left ( \frac{1}{2} \left ( \frac{\D \r_1}{\D t} \right )^2 +
                     \frac{1}{2} \left ( \frac{\D \r_2}{\D t} \right )^2 +
                      g_0 \, \Psi (t) \right )
\Label{S2}
\end{equation}
containing pair interactions 
$\Psi (t) \equiv \delta (\r_1 (t) - \r_2 (t))$. The center-of-mass
fluctuations are decoupled, and (\Ref{P}) is the pair density
with $\r(t) = \r_1(t) - \r_2 (t)$. In $d = 1$, Gaussian strings 
with contact repulsions are known to behave asymptotically like free 
fermions~\cite{fermions}, 
and (\Ref{theta}) then gives the correct scaling 
$P(r) \sim r^2$ of the pair density 
imposed by the antisymmetry of the fermionic 
wave function. 

Letting aside these specifics of Gaussian strings, the qualitative
features of the phase diagram are seen to rest on two properties
of the local interaction field $\Phi(t)$: its scaling dimension is
depends on $d$ in a continuous way and it obeys an operator product
expansion with a self-coupling term (\Ref{ope0}). These properties
are found to be preserved for directed strings in a random medium
despite the lack of a local action.

It turns out that the renormalization discussed in this Section
is also applicable to 
tem\-pera\-ture-driven transitions in systems of many directed strings. An
example of current experimental interest is {\em vicinal surfaces}, i.e.,
crystal surfaces miscut at a small angle with respect to one of the symmetry
planes. A vicinal surface can be regarded as an ensemble of terraces and 
steps~\cite{Jayapakrash.TSK}. The steps are noncrossing (fermionic) directed
strings. They are coupled by mutual forces that turn out to have a short-ranged
attractive and a long-ranged repulsive  part~\cite{RedfieldZangwill}. Typical
step configurations are shown in Fig.~5.  At high
temperatures, the ensemble of steps is homogeneous.  
\begin{figure}
 \psfig{file=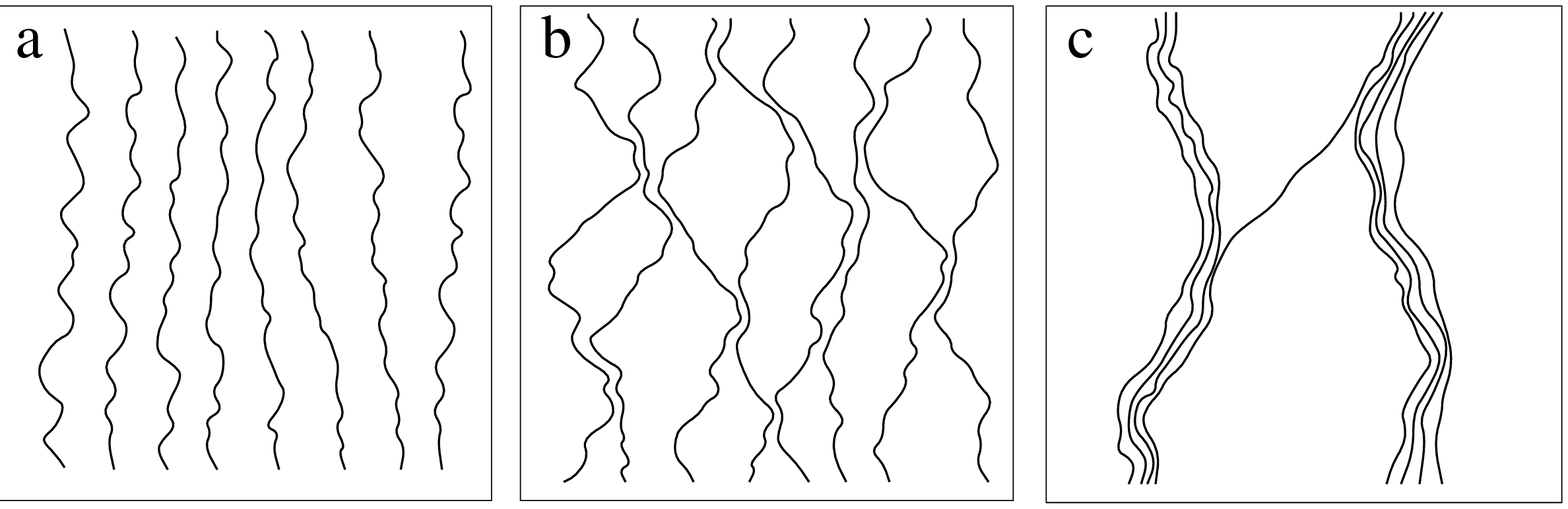,width=0.2\textwidth,angle=270}
 
 \vspace*{10pt}
 \baselineskip=12pt
 {\small Fig. 5: 
(De-)localization in a system of many strings. In this example, the strings
are steps on a crystal surface coupled by inverse-square and short-ranged 
forces. Typical step configurations at different temperatures:
 (a)~Well above the critical temperature $T_c$, the steps are dominated
by the no-crossing constraint and the long-ranged repulsion. Hence, they are
well separated from each other with relatively small fluctuations. 
 (b)~In the critical regime near $T_c$, the probability of a step
being close to one of its neighbors is substancially enhanced. This
is accompanied by increased step fluctuations and a broader distribution
of terrace widths.
 (c)~Below the faceting temperature, the steps
form local bundles. 
On average, the distance between two neighboring bundles is larger than
the width of an individual bundle. The fluctuations of these
``composite'' steps are smaller than those of individual steps.}
 \end{figure}
Below a faceting
temperature (that depends on the step density), the steps are found to form
local bundles.  Hence, the surface splits up into domains of an increased and
temperature-dependent step density alternating with step-free
facets~\cite{SongMochrie}.  A critical regime containing the {\em faceting
transition} (the analogue of the (de-)localization transition of two strings)
separates the  high- and low-temperature regimes. In the renormalization group,
one still finds a pair of fixed points linked by an exact one-loop
renormalization group for the strength of the contact interaction. These fixed
points determine the faceting transition and the high-temperature regime,
respectively~\cite{Lassig.vicinal}. The long-ranged forces influence the
universal features (e.g., the scaling of the contact field) of both fixed
points in a characteristic way. This determines, for example, the distribution
of terrace widths and the density of steps in a bundle. The theoretical results
compare favorably with recent experiments on Si
surfaces~\cite{SongMochrie,vanDijkenAl.SudohAl}.

The thermodynamic complexity of this many-string system is due
to an interplay of attractive interactions, Fermi statistics, and
repulsive forces. In the next Section, we shall discuss the much simpler 
case of bosonic strings (i.e., strings allowed to intersect) with 
contact attractions only. In $d = 1$, the latter will collapse to a bound state 
at any temperature. However,  qualitatively different behavior 
emerges in the formal limit of vanishing number of strings. This 
limit turns out to describe a single string in a random medium.

\bigskip

\section{Directed strings in a random medium}

In this Section, interactions between strings play a double role. On the one
hand, a single string in a quenched random medium can formally be
represented as a system of $p$ strings without disorder but with mutual
interactions, in the limit $p \to 0$~\cite{Kardar.dp}. This well-known
replica formalism turns out to be a convenient basis for the
perturbative renormalization of the random system~\cite{Lassig.kpz}. 

On the other hand, additional interactions are important in many 
applications of directed strings in random media. An example is the 
physics of 
superconductors~\cite{NelsonAl.fluxlines,FeigelmanAl.fluxlines,
NelsonLedoussal.fluxlines,Hwa.fluxlines}.
A type-II superconductor in a magnetic
field $h$ above a critical strength $h_{c1}$ contains magnetic
flux lines at a density that depends on $h$. The lines are directed
parallel to the magnetic field. Their thermally activated transversal
fluctuations dissipate energy at the expense of the
supercurrent. The superconductor may be doted with point impurities,
linear or planar defects designed to suppress these fluctuations
by localizing the flux lines. (Linear defects are generated, for example,
by irradiating the material with heavy 
ions~\cite{CivaleAl.defects,BudhaniAl.defects}.) 

As the external field approaches the critical value $h_{c1}$, the
ensemble of flux lines becomes dilute. It is then useful to study the
approximation of a single flux line interacting with a single columnar
defect in the presence of point impurities
~\cite{TangLyuksyutov.depinning,
BalentsKardar.depinning,
KolomeiskyStraley.depinning,
HwaNattermann.depinning,
KinzelbachLassig.depinning}; 
see Fig.~6(a). The next step is to
consider pair interactions in a dilute ensemble of lines as shown in 
Fig.~6(b)~\cite{Mezard.pair,NattermannAl.pair,Tang.pair,Mukherji.pair,
KinzelbachLassig.fermions}.  
\begin{figure}
 \vspace*{-.4cm}
 \epsfig{file=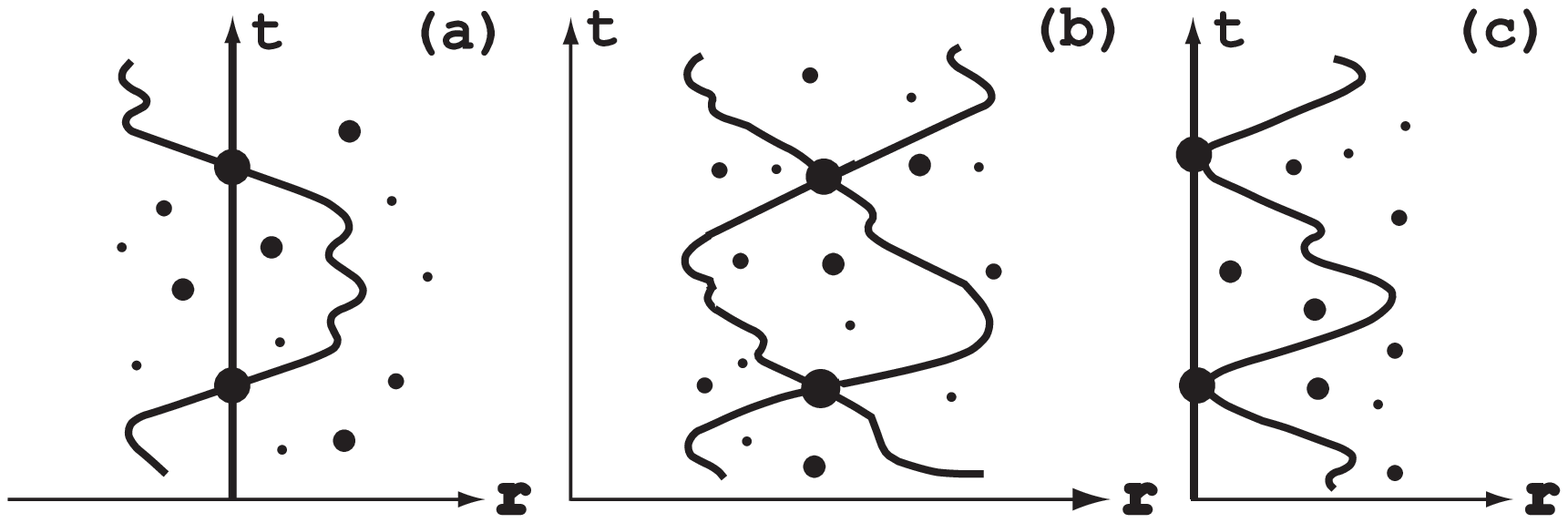,height=4cm}
 
 \vspace*{10pt}
 \baselineskip=12pt
 {\small Fig. 6: 
 Directed strings in a disordered medium with additional contact interactions.
 (a) A string and a rigid linear defect.
 (b) Two strings with mutual  interactions.
 (c) A string and a wall. }
 \end{figure} 

The interaction of strings with the boundaries of the system (Fig.~6(c)) can be
discussed on a similar theoretical footing. The particular case 
of one transversal dimension, where the string becomes an interface
of the system, has been of interest as a simple model for wetting in
a random medium~\cite{Kardar.dp}. The interaction of a string with a linear defect
is also relevant in the context of DNA pattern 
recognition~\cite{HwaLassig.dna,DrasdoAl.global,HwaLassig.local1,%
DrasdoAl.local2,Olsen}. Two 
related DNA sequences in different organisms have mutual correlations
inherited from their common ancestor in the evolution process.
The algorithmic detection of these correlations turns out to
correspond to a localized state of a string.

It is not surprising that a random medium, by changing the displacement
statistics of a single string, modifies also its direct interactions with other
strings and with external objects. In the renormalization group for these
transitions, quenched impurities enter in a characteristic way:  the
strong-coupling fixed point has a  dangerous irrelevant coupling constant that
alters the scaling properties of the direct
interactions~\cite{KinzelbachLassig.depinning}. Consequently, the 
(de-)localization transitions differ from those in pure systems.   In turn, the
response to such interactions becomes a theoretical tool to study the
correlations at the strong-coupling fixed point. This is used at the end of
this Section to show that the single-string  system has an upper critical
dimension less or equal to  four~\cite{LassigKinzelbach.ucd}.

\bigskip

\subsection{Replica perturbation theory}

A medium with quenched pointlike impurities exerts a local random potential 
$\eta (\r,t)$ on a directed string. For a given configuration of the impurities,
the  partition function of the string is
\EQ
Z[\eta] = \int {\cal D} r \exp (- \beta S[\r,\eta])
\label{Zeta}
\EEQ
with the action (\Ref{Seta}),
\EQ
S[\r,\eta] = \int \D t \left (  \frac{1}{2} \left ( \frac{\D \r}{\D t} \right )^2
                       + \eta (\r(t), t)   \right ) \;,
\EEQ
where $\lambda$ has been set to 1. We take the local potential variables 
to have the Gaussian distribution given by (\Ref{etaeta}) and 
compute average quantities like the free energy 
\EQ
\overline{F} = - \beta^{-1}
   \int {\cal D} \eta 
   \exp \left ( -\int \D t \D \r \, \frac{1}{2 \sigma^2} \eta^2 (t,\r)  \right )
   \log Z[\eta] \;.
\Label{Frandom}
\EEQ

At first sight, this system looks quite different from those of the 
previous Section. However, we can write  $\log Z[\eta]$ as a partition 
function of $p$ independent strings labeled by the index $\alpha$, 
\EQ
\log Z [\eta] = 
  \lim_{p \to 0} \frac{1}{p} \left ( \prod_{\alpha = 1}^{p} Z^{(\alpha)} [\eta] - 1
                             \right ) \;.
\EEQ 
In the replicated system, the integration over the $\eta$ variables can
be performed~\cite{Kardar.dp}. This gives 
\EQ
\overline F = \lim_{p \to 0} \frac{1}{p} \, F_p
\EEQ
with 
\EQ
F_p = - \beta^{-1} \log \int {\cal D} \r_1 \dots {\cal D} \r_p
                        \exp( - \beta S[\r_1, \dots, \r_p] ) 
\Label{Fp}
\EEQ
and the action 
\EQ
S[\r_1, \dots, \r_p] =  
 \int \D t \left ( \sum_{\alpha = 1}^p \frac{1}{2} 
                      \left ( \frac{\D \r_\alpha}{\D t} \right )^2  
                   - \beta \sigma^2 \sum_{\alpha < \beta} \Phi_{\alpha \beta} (t)
           \right ) \;.
\Label{Sp}
\EEQ
The coupling of the original string $\r(t)$ to the random medium now
appears as a contact attraction 
$\Phi_{\alpha \beta} (t) \equiv \delta (\r_\alpha (t) - \r_\beta (t))$
between the replicas (i.e., phantom copies) 
$\r_1(t), \dots, \r_p(t)$. 
The free energy of the replicated system is related to the cumulant
expansion of the random free energy~\cite{Kardar.dp}, 
\EQ
F_p = \sum_{k = 1}^\infty \frac{p^k}{k!} \, \overline {F^j}^c \;.
\label{Fbarcum}
\EEQ
The physical properties of this system strongly depend on $p$.
For $p > 1$, the attractive interaction {\em reduces} the 
fluctuations of the strings. 
A single string ($p = 1$) undergoes no interaction.
Randomness {\em enhances} the string fluctuations, and this is naturally
associated with values $p < 1$. The existence of the random limit 
$p \to 0$ is not clear a priori. It can be established for $d = 1$,
where the replica system is exactly solvable (see  Section 3.2). In 
higher dimensions, the replica interaction can still be treated 
in perturbation theory. 

For arbitrary values of $p$, the path integral in Eq.~(\ref{Fp}) can be 
rewritten in second 
quantization~\cite{Lassig.kpz},
\EQ
Z =  \int {\cal D} \phi {\cal D} \bar \phi 
                 \exp \left [ - \beta \int \D t \D \r 
                   \left ( 
                   \bar \phi \left ( \partial_{t} 
                                     - \frac{1}{2 \beta}\nabla^2 
                              \right ) \phi -
                   \beta \sigma^2 \bar \phi^2 \phi^2
                   \right ) 
           \right ] \hs.
\label{F2quant}
\EEQ
The contact attraction is described by the
(normal-ordered) vertex $\bar \phi^2 \phi^2 $. 
Since this interaction conserves the number 
of strings, the dependence of (\ref{F2quant}) on $p$ is contained entirely
in the boundary conditions at early and late values of $t$. Hence, the boundary
conditions are important for obtaining the replica limit $p \to 0$.

The representation (\ref{F2quant}) is a convenient framework 
for perturbation theory~\cite{Lassig.kpz}. We now mark 
the parameters $\beta_0^{-1}, \sigma^2_0$, the free energy 
and all longitudinal lengths with the subscript $0$ to
distinguish them from their renormalized counterparts introduced below.
In the Appendix, the renormalization is carried out for  
the disorder-averaged Casimir amplitude 
\EQ
\overline {\cal C} = 
 \beta_0^2 R^2 \, \overline f_0 (R)
\Label{Cbar0}
\EEQ
defined by Eq.~(\Ref{fbar}). The disorder-induced part
\EQ 
\Delta \overline {\cal C}(\sigma_0^2, \beta_0^{-1}, R) \equiv
     \overline {\cal C}(\sigma_0^2,\beta_0^{-1},R) 
     - \overline {\cal C}(0,\beta_0^{-1},R) 
\EEQ
can be expanded in powers of the dimensionless coupling
constant
$u_0 = g_0 R^{y_0/\zeta_0}$, where 
\EQ
g_0 = - \beta_0^3 \sigma_0^2 
\label{g0}
\EEQ
and $y_0 = (2 - d)/2$.
Due to proliferation of replica indices, the perturbation series is more 
complicated than its analogue in the previous
Section~\cite{BundschuhLassig.kpz}. However, as
shown in the Appendix, it is still one-loop renormalizable: the 
Casimir amplitude (\Ref{Cbar0}) is a regular function
of the coupling constant $u_P = \Z_P u_0$ defined by (\Ref{Z0}),
\EQ
\beta^2 \Delta \overline {\cal C} (u_P) = 
     - \frac{1}{4} u_P + O(y_0 u_P, u_P^2) \;.
\Label{Cbarreg}
\EEQ

An immediate consequence of the one-loop renormalizability is that the
strong-coupling regime is beyond the reach of the loopwise perturbation
expansion (\Ref{Cpseries}) since the flow equation (\Ref{beta0}) does not have
a stable fixed point at negative values of $u_P$~\cite{Lassig.kpz}.
(The fixed point $u_P^\star = y_0/C_0 >0$ for $d < 2$ is unphysical in this
context since a repulsive interaction between replicas translates 
into a purely imaginary random potential.) We come back to this failure
of perturbation theory in Section~4.2.

The fixed point $u_P^\star < 0$ for $d > 2$ is to be identified with
the roughening transition. At this fixed point, the Casimir 
amplitude (\Ref{Cbar0}) takes a finite positive value without any explicit
dependence on $R$,
\EQ
\overline {\cal C}^\star = - \frac{1}{4} \frac{y_0}{C_0} + O(y_0^2) \;.
\Label{Cbarstar}
\EEQ

The scaling properties at the
transition can hence be obtained exactly from the one-loop
renormalization group~\cite{Lassig.kpz,BundschuhLassig.kpz}.  
For example, the displacement
fluctuations  are still only diffusive, 
\EQ
\zeta^\star = 1/2 \;, \hspace{1cm} \omega^\star = 0 \;.
\Label{expstar}
\EEQ
This follows by comparing (\Ref{Cbarstar}) with the scaling 
$\overline {\cal C} \sim  R^{2 \omega / \zeta}$ 
at a generic fixed point given by (\Ref{fbar}).
Other exponents do depend on $d$.  Conjugate to $\sigma_0^2$ is the local field
\EQ
\Phi_\eta (t_0) \equiv \int \D \r' \eta (\r',t_0) \delta (\r(t_0) - \r') \;,
\EEQ
which encodes the random potential evaluated along the
string~\cite{LassigKinzelbach.ucd}.  Small variations of $\sigma_0^2$ (i.e., of
$u_P$) are a relevant perturbation at the roughening transition.
The dimension
\EQ
y^\star = \frac{d - 2}{2} \;,
\Label{ystar}
\EEQ
is given by the eigenvalue of the beta function at the fixed point $u_P^\star$.
The dimension of $\Phi_\eta$ is therefore $x^\star
= (4 - d)/2$. In the action (\Ref{Sp}), this field becomes the replica
pair field $\Phi_{\alpha \beta} (t)$. The same dimension $x^\star$ then
follows from (\ref{x0star}) with the field renormalization  (\Ref{Z0}).
One may also define the pair contact field 
$\Psi (t) \equiv \delta (\r_1 (t) - \r_2 (t))$ of two real copies, 
i.e., two independent strings $\r_1(t)$ and $\r_2(t)$ in the same random 
potential. It can be shown that this field and its conjugate coupling constant 
also have dimensions $x^\star$ and $y^\star$, respectively.

The perturbative calculation can only be trusted for $d < 4$. The fact
that $x^\star$ would turn negative for $d > 4$ is clearly unphysical. 
Moreover,  physical quantities become singular as
$d$ approaches $4$; for example, 
$\overline {\cal C}^\star \sim \sqrt{4 - d}$~\cite{BundschuhLassig.kpz}. This
shows that $d = 4$ is a singular dimension for the roughening transition
and ties in with the existence of an upper critical dimension $d_> \leq 4$
of the strong-coupling phase.

\bigskip

\subsection{The strong coupling regime}

In the strong-coupling regime (i.e. for low temperatures or large 
disorder amplitudes), the string has the disorder-induced
fluctuations  
\EQ
\overline{ \langle (\r(t) - \r(t'))^2 \rangle } \sim |t - t'|^{2 \zeta}
\Label{r2.2}
\EEQ
leading to anomalous scaling of the confinement free energy
(\Ref{fbar}),
\EQ
\overline f (R) \sim R^{(\omega - 1)/\zeta} \;,
\Label{fbar.2}
\EEQ
see refs.~\cite{HuseHenley.paths,FisherHuse.paths}.
The exponents satisfy the scaling relation (\ref{scaling2}). 
Superdiffusive scaling ($\omega = 2\zeta - 1 >0$)
is believed to persist up to an upper critical dimension $d_>$ (see 
Section~3.6).

In $d = 1$, the exponents can be obtained exactly from the replica
approach~\cite{Kardar.dp}. The system of $p$ strings is always in a 
bound state for integer $p>1$. The binding energy in a system of 
infinite width $R$,
\EQ
E_p = \lim_{L \to \infty} \partial_L 
                          \left (F_p(\beta_0, \sigma_0^2,L) -
                                 F_p(\beta_0,          0,L)  \right ) \;,
\EEQ
can be computed by Bethe ansatz methods;
one finds $E_p \sim p + O(p^3)$. Analytically continued to $p = 0$
and inserted in (\ref{Fbarcum}), this
gives $\overline {F^3}^c \sim L$, hence $\omega = 1/3$, and
$\chi = 2/3$ by (\ref{scaling2}).

In higher dimensions, the strings form a bound state only
for $\sigma_0^2 > \sigma^2_{0c}$. If we assume $E_p$ is still 
analytic in $p$ and has the form
$E_p \sim p + O(p^{k_0 + 1})$ ($k_0 = 2,3, \dots$), the same argument
yields 
\EQ
\omega = \frac{1}{k_0 + 1} \;;
\label{omegaquant}
\EEQ
see the discussion in~\cite{McKaneMoore.rep,Zhang.rep}.
The exponents of the random system are indeed quantized, as will 
be discussed in the context of the Kardar-Parisi-Zhang equation
in Section~4. The quantization condition (\ref{quant}) is consistent
with (\ref{omegaquant}).

In the continuum theory (\Ref{Zeta}), the large-distance scaling (\Ref{r2.2})
and (\Ref{fbar.2}) is reached in a crossover from diffusive behavior on smaller
scales.  The crossover has characteristic longitudinal and transversal scales
$\tc_0$ and $\rc$ beyond which the disorder-induced fluctuations dominate over
the thermal fluctuations. 
The dependence of these scales on the effective coupling (\Ref{g0}),
\EQ
\rc^2 = \beta_0^{-1} \tc_0 = \left \{ 
 \begin{array}{ll}
 (-g_0)^{-1/y_0}                    & \mbox{ ($d < 2, g_0< 0$)}    \\
 \exp(- C_0 / g_0)                  & \mbox{ ($d = 2, g_0< 0$)}    \\
 (g_c - g_0)^{-1/y^\star}               & \mbox{ ($2 < d < 4, g_0 < g_c$)}
 \end{array} \right. 
\label{rctc}
\EEQ
with $y_0 = 2/(2-d) = - y^\star$, can be obtained from the replica action
(\Ref{Sp}), see (\Ref{xi0.1}) -- (\Ref{xi0.3}).  
The string displacement and the confinement free energy have the form
\EQ
\overline{ \langle (\r(t_0) - \r(t'_0))^2 \rangle } 
  \sim \beta_0^{-1} t_0 \, \tilde{\cal R} (t_0 / \tc_0)
\Label{r2cr}
\EEQ
and 
\EQ
\overline f_0  
  \sim \beta_0^{-2} R^{-2}  \tilde{\cal F} (R / \rc) \;,
\Label{fbarcr}
\EEQ
with scaling functions that are finite in the short-distance
limits 
$t_0  \ll \tc_0 $ and $R  \ll \xi$, respectively.
In the opposite limit, 
comparison with (\Ref{r2.2}) and (\Ref{fbar.2}) exhibits the singular dependence
on the bare parameters $\beta_0^{-1}$ and 
$\sigma_0^2$ contained in the scaling functions,
\EQ
\overline{ \langle (\r(t_0) - \r(t'_0))^2 \rangle }
  \sim \beta_0^{-2 \zeta}  \tc_0^{-2 \omega} |t_0 - t_0'|^{2 \zeta}  
\Label{r2sing}
\EEQ
and 
\EQ
\overline f_0 \sim \beta_0^{-2} \rc^{-1} R^{-1} \;.
\Label{fbarsing}
\EEQ

These singularities can be absorbed into the definition of the 
renormalized quantities
\EQ
t = (\beta / \beta_0) t_0 \;,\HS  
\overline f = (\beta_0 / \beta) \overline f_0
\Label{r2freg}
\EEQ
written in terms of the renormalized temperature
\EQ
\beta^{-1} = \rc^{\omega/\zeta} = \tc^\omega \;;
\Label{betareg}
\EEQ
see the discussion in~\cite{KinzelbachLassig.depinning}. The renormalized  
displacement function and confinement free energy remain finite in the
continuum limit  $ \rc \to 0$  (i.e. $\beta_0^{-1}
\to 0$ or $\sigma_0^2 \to \infty$), as follows by inserting
(\Ref{r2freg}) and (\Ref{betareg}) into (\Ref{r2sing}) and
(\Ref{fbarsing}).

The existence of a zero-temperature continuum limit is crucial if the ensemble
of ground states generated by the quenched disorder is to have universal
features. 
According to Eq.~(\Ref{betareg}), the renormalized temperature
$\beta^{-1}$ is an irrelevant coupling constant of dimension $-\omega$. This
is why the renormalized theory may be called a zero-temperature fixed point. 
It will be shown below that $\beta^{-1}$ is a {\em dangerous}
irrelevant variable which modifies the correlations of other fields in a
characteristic way~\cite{KinzelbachLassig.depinning}.

\bigskip

\subsection{A string and a linear defect}

A single string coupled to a random medium and a linear defect has
the partition function (\Ref{Zeta}) with the action
\EQ
S[\r,\eta] = \int \D t \left (  \frac{1}{2} \left ( \frac{\D \r}{\D t} \right )^2
                       + \eta (\r(t), t)  + g \Phi(t) \right ) 
\Label{Sdefect}
\EEQ
containing the contact field $\Phi (t) \equiv \delta (\r(t))$.
The string configurations are determined by a competition between 
two kinds of interactions. Point defects roughen the string
and make its displacement fluctuations superdiffusive.  An
attractive extended defect, on the other hand, suppresses these
excursions  and, if it is sufficiently strong, localizes the string
to within a  finite transversal distance $\xi$. As in Section~2,
the two regimes are separated by a  second order phase transition where the
localization length $\xi$ diverges. In contrast to
temperature-driven transitions, it involves the competition of two
different configuration energies rather than energy and entropy. Hence,
the transition persists in the zero-temperature limit. 

The properties of this zero-temperature phase transition have been  
controversial~\cite{TangLyuksyutov.depinning,
BalentsKardar.depinning,
KolomeiskyStraley.depinning,
HwaNattermann.depinning,
KinzelbachLassig.depinning}. 
Following the treatment of 
ref.~\cite{KinzelbachLassig.depinning}, we expand 
the defect contribution
\EQ
\Delta \overline {\cal C}(g,R) 
\equiv \overline {\cal C}(g,R) - \overline {\cal C}(0,R)
\Label{Cbarg}
\EEQ
to the zero-temperature Casimir amplitude 
\EQ
\overline {\cal C} \equiv 
 R^{(1 - \omega)/\zeta} \, \overline f(R)
\Label{Cbar}
\EEQ
about the point $g = 0$ given by the strong-coupling continuum theory. 
This leads to a perturbation series formally analogous 
to~(\ref{C0series}), 
\EQ
\Delta \overline {\cal C} (g,R)   
   = - \beta^{-1} R^{(1 - \omega) / \zeta} \sum_{N = 1}^{\infty} 
   \frac{ (- \beta g)^N}{N!} \int \D t_2 \dots \D t_N 
     \overline{\langle \Phi (t_1) \dots \Phi (t_N) \rangle^c} \;.
\Label{Cbarseries}
\EEQ
Of course, the perturbation series cannot be evaluated explicitly since
the multipoint correlation functions of the contact field at the
zero-temperature fixed point are not known exactly. However, one can still
write down the one-point function
\EQ
\overline{\langle \Phi(t) \rangle} \equiv 
\overline{\langle \Phi \rangle} =
 R^{-x/\zeta}
\EEQ
(with $x = d \zeta$) and establish the short-distance structure of
the higher connected correlation functions. 

Consider first the displacement function (\Ref{r2}). It has the low-temperature 
expansion 
\EQ
\overline{ \langle (\r(t) - \r(t'))^2 \rangle }  =
             |t - t'|^{2 \zeta} 
             + \beta^{-1} \, |t - t'|^{2 \zeta - \omega} + \dots \,,
\Label{r2expansion}
\EEQ
assuming analyticity of the crossover scaling form (\Ref{r2cr}) in the 
scaling variable $\beta^{-1}$. 
Hence at zero temperature, (\Ref{r2}) equals its thermally disconnected part 
$ \overline{ \langle \r (t) - \r (t') \rangle^2 } $. The connected part
can be shown to equal that of the Gaussian 
theory~\cite{FisherHuse.paths},
\EQ
\overline{ \langle ( \r(t) - \r(t') )^2  \rangle^c }
\sim \beta^{-1} | t - t' | \, ;
\Label{r2c}
\EEQ
it appears as the leading correction to scaling in (\Ref{r2expansion}).

In analogy to (\Ref{Phi02}), the two-point
function $\overline{ \langle \Phi (t) \Phi (t') \rangle}$ is assumed to 
factorize for \linebreak 
$|t - t'| \ll R^{1/\zeta}$ into the one-point function  
$ \overline{ \langle \Phi (t)} \rangle$ times the $R$-independent return 
probability to the origin,
which is proportional to the inverse r.m.s. displacement given by 
(\Ref{r2expansion}): 
\begin{equation}
\overline{ \langle \Phi (t) \Phi (t) \rangle } 
\sim {\left\vert t - t' \right\vert}^{-x}
( 1 - \beta^{-1} d^{-1} {\left\vert t - t' \right\vert}^{-\omega}
  + \dots )
\overline{ \left\langle \Phi (t) \right\rangle } \, .
\Label{PhiPhi}
\end{equation}
Again the leading singularity
is due to sample-to-sample fluctuations of the minimal energy paths,
while the correction term is due to thermal fluctuations around  these
paths. At zero temperature, the field $\Phi(t)$ can be replaced by its thermal
expectation value $\langle \Phi(t) \rangle$; hence 
$ \overline{ \langle \Phi (t) \Phi (t') \rangle } $
equals its thermally disconnected part 
$ \overline{ \langle \Phi (t) \rangle \langle \Phi (t') \rangle } $
and the connected part 
$ \overline{ \left\langle \Phi (t) \, \Phi (t') \right\rangle^c } $
vanishes, just as the connected displacement function (\Ref{r2c}) does.
The subleading term in (\Ref{PhiPhi}) is the sum of 
$ \overline{ \left\langle \Phi (t) \, \Phi (t') \right\rangle^c } $
and a temperature-dependent correction to 
$ \overline{ \langle \Phi (t) \rangle \langle \Phi (t') \rangle } $.
An analogous argument applies to the singularities in any correlation function
$ \overline{ \langle \dots \Phi(t) \Phi (t') \dots
             \rangle } $ as $ | t - t' | \to 0 $. This leads to the 
operator product expansion
\EQ
\Phi (t) \Phi (t') \sim  C \,  \beta^{ -1 }   {
\left\vert t - t' \right\vert }^ {- x -\omega} \, \Phi (t)
\Label{ope1}
\EEQ
with a constant $C > 0$. Defining the  contact field
$\Phi(\r',t) \equiv \delta (\r(t) - \r')$, (\ref{ope1}) can be generalized
to the spatio-temporal operator product expansion
\EQ
\Phi (\r,t) \Phi (\r',t') \sim  
C \,  \beta^{ \, -1 }   
{\left | t - t' \right | }^ {- x -\omega} \,
{\cal H} \! \left ( \frac{\nu |t - t'|}{|\r - \r'|^z} \right )
\, \Phi (\r,t)
\Label{ope1rt}
\EEQ
producing a spatial singularity of the form 
\EQ
\Phi (\r,t) \Phi (\r',t) \sim   \beta^{ -1 }   
|\r - \r'|^{- d - \omega / \zeta} \, \Phi(\r,t) \;.
\label{ope1r}
\EEQ
The operator product expansion encodes the statistics of rare
fluctuations around the path of minimal 
energy~\cite{HwaFisher.paths,HwaNattermann.depinning} as shown in Fig.~7.
Notice that the {\em leading} singular term is proportional to the 
irrelevant variable $\beta^{-1}$ and hence governed by a 
correction-to-scaling exponent. This is why the temperature
is called a dangerous irrelevant variable.  
\begin{figure}
 \epsfig{file=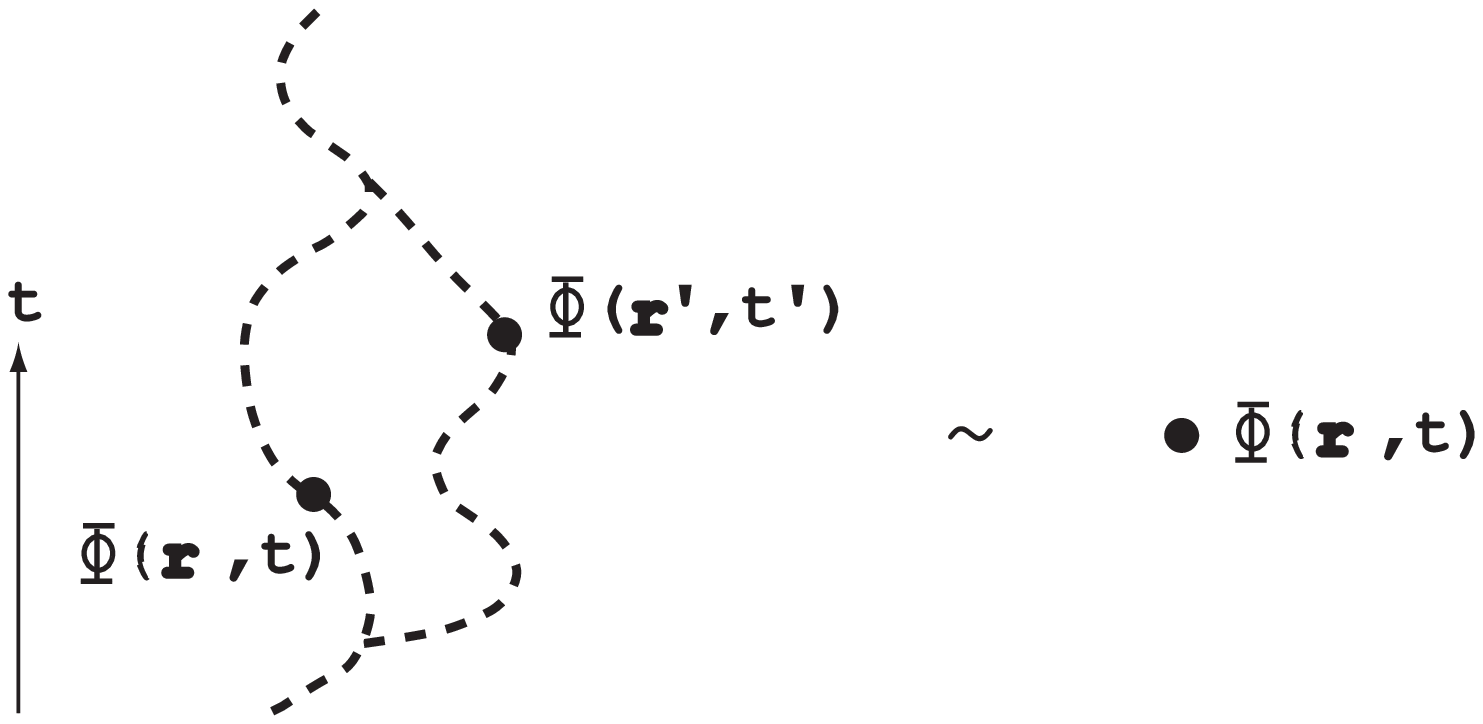,height=7cm}
 
 \vspace*{-2.3cm}
 \baselineskip=12pt
 {\small Fig. 7: 
 Operator product expansion of contact fields for a string in a disordered 
 medium. The short-distance asymptotics of the pair of fields $\Phi(t)$ 
 and $\Phi(t')$ is given by the single field $\Phi(t)$ times a singular
 prefactor. The dashed lines indicate the string configurations generating
 the singularity $|t - t'|^{-x - \omega}$.  }
 \end{figure} 

As before, the operator product expansion (\Ref{ope1}) determines the
leading ultraviolet singularities of the integrals in (\Ref{Cbarseries}),
which appear as poles in $y \equiv 1 - x - \omega$. The defect contribution
to the Casimir amplitude can again be written in terms of a 
dimensionless coupling constant:
\EQ
\Delta \overline{C} (g, R) = 
    R^{x/\zeta } \, \overline{\langle \Phi \rangle } 
    \left ( u - \frac{C}{ y  }  {u}^2 \right )
 +  O (y^0 u^2, u^3)   
\Label{Cbarsing} 
\EEQ 
with
\EQ
u \equiv g R^{y/\zeta}.
\EEQ
Indeed, Eq.~(\Ref{Cbarsing}) remains finite in the limit $\beta^{-1} \to
0$ for fixed $R$ and~$g$. The pole can be absorbed into   
the ``minimally subtracted'' coupling constant 
$u_M = {\cal Z}_M u$ (which should not be confused
with the coupling (\Ref{uP}) defined in perturbation theory about the
Gaussian fixed point). With
\EQ
{\cal Z}_M (u_M) =  1 - \frac{C}{y} u_M + O(u_M^2) \;,
\EEQ
the flow equation 
$\dot{u}_M \equiv \zeta R \partial_R \, u_M$ reads 
\EQ 
\dot{u}_M = y u_M - C u_M^2 + O(u_M^3) \;.
\Label{beta} 
\EEQ 

We conclude that an attractive linear defect in a random system is less
effective in localizing a directed string than in a pure 
system, in agreement with some previous approximate 
renormalization group 
studies~\cite{TangLyuksyutov.depinning,
BalentsKardar.depinning,
HwaNattermann.depinning}.
A weak defect defect is a relevant perturbation of the zero-temperature fixed
point $u_M = 0$ only in dimensions $d < 1$. It localizes the string 
with 
\EQ
\xi \sim (-g)^{-\zeta/y} \HS (g < 0) \;.
\label{xi.1}
\EEQ

At the borderline dimension $d = 1$, the theory is again asymptotically
free, with
\EQ
\xi \sim \exp (-C \zeta / g) \HS (g < 0) \;.
\EEQ

For $d > 1$, a finite defect strength $|g| > |g_c|$ is
necessary to localize the string:
\EQ
\xi \sim (g_c - g)^{-\zeta/y^\star} \HS (g < g_c) \;,
\EEQ
where 
$ y^\star \equiv \lim_{u_M \to u_M^\star} \dot{u}_M / (u_M^\star - u_M) 
= -y + O(y^2) $
is the eigenvalue of the flow equation at the nontrivial fixed point. 
The disorder-averaged string density (\Ref{P}) has the short-distance singularity 
$\overline P(r) \sim r^\theta$ with $\theta = 2 y / \zeta + O(y^2) < 0$.

A weak repulsive defect is an irrelevant perturbation of the strong
coupling fixed point for $d \geq 1$. In particular, it does
not generate a ``fermionic'' zero of the string density $P(r)$
in the limit $R \to \infty$. This prediction of renormalized
perturbation theory is consistent with the scaling at a strong
repulsive defect or {\em barrier}, which can be obtained exactly in 
$d = 1$~\cite{Lassig.barrier}. An 
impenetrable barrier ($g \to \infty$) is equivalent to a system 
boundary, leading to a long-ranged depletion 
$\overline P (r) \sim r$  of the string density (see Section 3.4). A hardly
penetrable barrier has rare crossings that
happen whenever the difference in the random energies on the left
and right sides of the barrier in a given longitudinal interval $\Delta t$
exceeds the barrier penalty. Since the string configurations one side consist 
of essentially uncorrelated pieces of length $\sim R^{3/2}$, this difference scales as 
\EQ
\Delta \overline F (\Delta t, R) 
      = (\Delta t)^{1/3} {\cal F}_d (\Delta t R^{-3/2}) 
      \sim (\Delta t)^{1/2} R^{-1/4}
\EEQ
for $\Delta t \gg R^{3/2}$. Hence the path remains on one side of the
barrier for a typical longitudinal distance
$\Delta t$ given by 
$ (\Delta t)^{1/2} R^{-1/4} \sim g$. This determines the expectation
value of the contact field,
\EQ
\overline{\langle \Phi \rangle}
   \sim \frac{1}{\Delta t} \sim  g^{-2} R^{-1/2} \;.
\EEQ
We conclude that the penetrability $g^{-2}$ conjugate to $\Phi$ is 
a relevant perturbation with eigenvalue $y = 1/3$ at the barrier
fixed point $g^{-2} = 0$. It induces a 
crossover to the delocalization fixed point $g = 0$: the barrier
becomes irrelevant on large scales.

\bigskip

\subsection{A string and a wall}

As mentioned above, a half-space Gaussian string 
$(r_1 > 0, r_2, \dots, r_d)(t)$ in contact interaction with a 
system boundary at $r_1 = 0$ is equivalent to the string $r_1(t)$
in full space coupled to a defect at $r_1 = 0$. In a random background,
this is no longer the case. The disorder potential couples the string 
coordinates $r_1, \dots, r_d$ and the boundary has a nonlocal influence
on the string by cutting off all disorder configurations in the 
half space $r_1 < 0$. Alternatively, the half-space system can be 
understood as an unrestricted system with a defect plane at $r_1 = 0$ and
``mirror'' constraints $\eta(r_1, \dots,r_d,t) = \eta(-r_1, \dots,
r_d,t)$ on the random potential~\cite{HwaLassig.unpubl}.

We restrict ourselves here to the case $d = 1$, where a half-space string
with the action (\Ref{Sdefect}) can be treated exactly by Bethe ansatz
methods~\cite{Kardar.dp}
or by mapping on a lattice gas~\cite{KrugTang.dp}. 
One finds a (de-)localization
transition with the localization length singularity
\EQ
\xi \sim (g_c - g)^{-2} \HS (g < g_c) \;.
\label{xising}
\EEQ
At the transition, the disorder-averaged string density (\Ref{P}) has 
the singularity~(\cite{ForgacsAl.DG}, p.~317) 
\EQ
\overline P(r) \sim r^{-1/2} R^{-1/2} \HS 
 \mbox{ for $\rc \ll r \ll R$.}
\EEQ
Hence, the boundary contact field $\Phi_b (t) \equiv \delta (r(t))$
has the expectation value
\EQ
\overline{\langle \Phi_b (t) \rangle} \sim R^{-x_b /\zeta}
\EEQ
with $x_b = 1/3$. By arguments as in Section 3.3, we then obtain
the operator product expansion 
\EQ
\Phi_b (t) \Phi_b (t') = 
  C_b \beta^{-1} |t - t'|^{-x_b - \omega} \Phi_b (t) + \dots 
\EEQ 
with $C_b > 0$. Hence, the wall changes the statistics of rare fluctuations.
This operator product expansion leads again
to a beta function of the form (\Ref{beta}) with 
$y_b = 1 - x_b - \omega$.

In $d = 1$, the interaction with the wall is a truly relevant perturbation 
with eigenvalue $y_b = 1/3$ at 
the {(de-)}\-localization point. The leads to a singularity 
(\Ref{xi.1}) of the localization length, which agrees with (\Ref{xising}) and
with the result of~\cite{LassigLipowsky.Altenberg} obtained from
a variational scaling argument for the bound-state
free energy.
For $g > g_c$, there is a crossover to the fixed point 
$u_M^\star = y_b / C_b + O(y_b^2)$ with exponents
$x^\star = 1 - \omega + y_b + O(y_b^2)$ and
$\theta = x^\star/\zeta - d > 0$ describing a long-ranged depletion of the 
string density $\overline P(r)$. This is again in agreement with exact 
results~\cite{KrugTang.dp}
and numerical transfer matrix 
studies~\cite{Willbrand.diplom} but the one-loop calculation
underestimates the true value $\theta = 1$.

\bigskip

\subsection{Strings with mutual interactions}

The effective action
\EQ
S [\r_1, \r_2, \eta] = 
   \int \D t \left ( 
       \frac{1}{2} \sum_{i=1}^2 \left ( \frac{\D \r_i}{\D t} \right )^2 +
                   \sum_{i=1}^2 \eta (\r_i (t), t) +
                      g \, \Psi (t) 
              \right )
\end{equation}
with the contact field $\Psi (t) \equiv \delta (\r_1 (t) - \r_2(t))$
describes a pair of directed strings that live in the same random medium
and are coupled by short-ranged mutual 
forces~\cite{Mezard.pair,NattermannAl.pair,Tang.pair,Mukherji.pair,
KinzelbachLassig.fermions}. In the case of flux
lines, for example, this interaction is repulsive. Due to the random
potential, the center-of-mass displacement of the strings does not decouple from
their relative displacement. The two-string system is therefore not
equivalent to a single string and a linear defect. 

The effects of pair interactions turn out to be much stronger than
in a pure system. Qualitatively, this is easy to understand. At zero
temperature, two noninteracting strings in the same random potential 
share a common minimal path $\r_0(t)$. At small but finite
temperatures, it turns out that the strings still follow a ``tube'' of
width $\rc$ around $\r_0 (t)$ with finite probability.
Hence, their overlap probability $\overline{ \langle \Psi (t) \rangle }$
remains finite in the limit $R \to \infty$, in contrast to that of 
noninteracting thermal strings, 
$\langle \Psi (t) \rangle \sim R^{-d \zeta_0}$. This  explains the strong 
sensitivity of the system to repulsive forces. However, the strings 
 make large individual excursions from the tube. The disorder-averaged pair 
density (\Ref{P}) (with $\r = \r_1 - \r_2$) has the 
form~\cite{HwaFisher.paths,KinzelbachLassig.fermions}
\EQ
\overline P(r) \sim \beta^{-1} r^{-d - \omega/\zeta} 
\hspace{1cm} 
\mbox{ for $r 
            \,\raisebox{-.6ex}{$\stackrel{\displaystyle >}{\sim}\,$}
            \rc$
       and $R \to \infty$}
\Label{P2}
\EEQ
dictated by the operator product expansion (\ref{ope1r}).
A strong repulsion ($g \to \infty$)
forces one of the strings onto the lowest excited path $\r_1(t)$ that has
no overlap with $\r_0(t)$ (with fluctuations of the form (\Ref{P2})
around $\r_1 (t)$ at finite temperatures). The paths $\r_1(t)$ and $\r_0(t)$
have an average distance of order $R$, and the pair density 
should have a long-ranged depletion $\overline P(r) \sim r^\theta$ with
$\theta > 0$ for $g \to \infty$, just as that of free fermions
in $d = 1$. At the strong-coupling fixed point of noninteracting strings, 
the pair field $\Psi(t)$ is therefore a relevant perturbation inducing the
crossover to the ``fermionic'' behavior for $g \to \infty$. 

This argument can be made quantitative~\cite{KinzelbachLassig.fermions}.
We start from an expansion
\EQ
\Delta \overline {\cal C} (g,R)  =
- \beta^{-1} R^{(1 - \omega)/\zeta}  \sum_{N = 1}^\infty
\frac{ (-\beta g)^N }{N!}
\int {\rm d} t_2 \dots {\rm d} t_N \;
\overline{ \left\langle  \Psi (0)  \Psi (t_2) \dots \Psi (t_N)
           \right\rangle^c }  
\Label{Cbarseries2}
\EEQ
of the Casimir amplitude (\Ref{Cbar}) 
about the strong-coupling fixed point of noninteracting strings ($g = 0$). 
Since the overlap probability of the two strings 
\EQ
\overline{ \langle \Psi (t) \rangle } \sim R^{-x / \zeta}
\Label{Psi}
\EEQ
remains finite as $R \to \infty$, the local field $\Psi (t)$ has
 dimension $x = 0$. The connected
correlation functions of $\Psi (t)$ vanish at zero temperature, as
do those of the contact field $\Phi (t)$ of Section 3.3. 
One obtains an operator product expansion of the form (\Ref{ope1}),
\EQ
\Psi (t) \Psi (t') = C_2 \beta^{-1} |t - t'|^{-\omega} \Psi (t) + \dots \;.
\EEQ
with $C_2 > 0$.
Its leading term is again a correction to scaling proportional
to the dangerous irrelevant variable $\beta^{-1}$. 
Renormalization 
of the perturbation series (\Ref{Cbarseries2}) then leads to the
flow equation (\Ref{beta}) with $y = 1 - \omega$. 

It follows that an attractive pair interaction always localizes the strings. 
The  pair density $\overline P(r, \xi)$ of the localized state 
has the form 
\EQ
\overline P(r, \xi) \sim \beta^{-1} r^{-d - \omega/\zeta} 
                        {\cal P} (r / \xi) 
\hspace{1cm} \mbox{for} \;\; r 
\,\raisebox{-.6ex}{$\stackrel{\displaystyle >}{\sim}\,$} \rc \;;
\Label{P2xi}
\EEQ
the scaling function ${\cal P}$ has a finite limit at short distances
and decays exponentially on scales 
$r \,\raisebox{-.6ex}{$\stackrel{\displaystyle >}{\sim}\,$} \xi$. 
The localization length $\xi$ has the singularity
\EQ
\xi \sim (-g)^{-\zeta/y} =
   (-g)^{(1 + \omega) / 2 (1 - \omega)} \HS (g < 0) \;.
\EEQ
The transversal scale $\xi^{(m)}$ defined by the $m$th moment
of the pair density,
\EQ
\xi^{(m)} \equiv 
   \left ( \int \D \r \, r^m \, \overline P(r, \xi) \right )^{1/m} 
\HS (m = 1,2,\dots) \;,
\label{xim}
\EEQ
scales as
\EQ
\xi^{(m)} \sim \beta^{-1/m} (-g)^{-(\zeta - \omega/m) / y}
\HS (g < 0) \;.
\Label{ximsing}
\EEQ
Recall that at an ordinary fixed point, all the scales $\xi^{(m)}$
have the same exponent as the correlation length $\xi$. The 
dangerous irrelevant variable $\beta^{-1}$ breaks this universality
and induces the {\em multiscaling} (\Ref{ximsing}). A similar phenomenon
occurs for thermal directed strings in dimensions 
$d > 4$~\cite{Lipowsky.lines}.

Repulsive forces lead to an asymptotic scaling $\overline P(r) \sim
r^\theta$ with $\theta = x^\star/\zeta - d$ given in terms of the
dimension  $x^\star = 2(1 - \omega) + O((1 - \omega)^2)$
of the pair field at the nontrivial fixed point. A numerical transfer
matrix study of a pair of strongly repulsive strings ($g \to \infty$)
confirms the long-ranged depletion of the pair 
density~\cite{Willbrand.diplom}. The exponent
$\theta \approx 2$ is underestimated by the one-loop calculation.

The above considerations are valid only as long as $d$ is below the upper
critical dimension $d_>$ of a single string. This can be seen as follows.
It is possible to show that the ground state path $\r_0 (t)$ is unique 
(up to microscopic degeneracies of the order of the lattice spacing)
in almost all realizations of the 
disorder~\cite{FisherHuse.paths,HwaFisher.paths}. This uniqueness is also 
manifest in the pair density (\Ref{P2}): for any fixed $r_0 > 0$,
the probability of finding the strings at a relative distance 
$r > r_0$ remains finite for $R \to \infty$, and that limit
value tends to zero for 
$\beta^{-1} \to 0$~\cite{KinzelbachLassig.fermions}, 
\EQ
\int_{ r  > r_0 } \D \r \, \overline P(r)
    \sim \beta^{-1} r_0^{-\omega / \zeta} \;.
\Label{Pint}
\EEQ
 However, for $d \to d_>$
(i.e., $ \omega \to 0$), the pair distribution (\Ref{P2})
shows a singular broadening: for $R \to \infty$ and fixed $\beta^{-1}$, 
the probability (\Ref{Pint}) approaches one. Consequently,
the overlap probability  (\Ref{Psi}) goes to zero, 
$ \overline{\langle \Psi (t) \rangle } \sim \omega$. This suggests
that the statistics of string configurations becomes more complicated
for $d \geq d_>$. The strings no longer cluster in the vicinity of
the minimal path as expressed by (\Ref{Pint}), but exploit multiple 
near-minimal paths at any finite temperature. Their overlap 
$ \overline{\langle \Psi (t) \rangle}$ is expected to vanish for 
$R \to \infty$.

\bigskip

\subsection{Upper critical dimension of a single string}

The theory of strings with pair interactions discussed in 
Section~3.5 has an important implication for the single-string
system~\cite{LassigKinzelbach.ucd}: the upper critical dimension 
of a single string is less or equal to four. 
  As $d_>$ is approached from below, the exponents $\zeta$ and
$\omega$ tend to the Gaussian values $\zeta = 1/2$ and $\omega = 0$
continuously. Hence, the upper critical dimension could
serve as the starting point for a controlled expansion. The name 
 ``upper critical dimension'' is, however, quite misleading since $d_>$ 
does not mark the borderline to simple
mean-field behavior as in the standard theory of critical phenomena.
On the contrary, the state of the system in high dimensions
may even be more complicated, having  
presumed glassy characteristics~\cite{MezardParisi,MooreAl.modecoupling}.

The existence of an upper critical dimension has been controversial.
Numerical work seems to indicate that a strong coupling
phase with nontrivial exponents $z<2$, $\chi >0$ persists in dimensions
$d = 4$ and
higher~\cite{AlaNissilaAl.num,AlaNissila.ucd,Kim.ucd}. 
However, the results for $d > 3$ are not 
very reliable since the available system sizes are limited and corrections 
to scaling are not taken into account~\cite{LassigKinzelbach.reply}. 

Various theoretical arguments
favor the existence of a finite upper critical dimension $d_>$
at or slightly below four. Most of these approaches rest on approximation
schemes (such as functional 
renormalization~\cite{HalpinHealey.fren,NattermannLeschhorn.fren} or 
mode-coupling theory~\cite{MooreAl.modecoupling}) whose
status is not very well understood. The same is true for a recent
approximate real-space renormalization~\cite{CastellanoAl.kpz} challenging
these results.

The argument given here~\cite{LassigKinzelbach.ucd} is of a different nature; 
it is not tied to any of these approximation schemes. An important ingredient 
is (\Ref{ximsing}), a set of {\em exact} relations in the strong-coupling regime.
These relations describe the bound state (\Ref{P2xi}) of attractively
coupled strings in terms of the only independent single-string exponent 
$\omega = 2 \zeta - 1$. We focus on the temperature dependence of the scales
$\xi^{(m)}$. For fixed $g$, $\xi^{(m)}$ is
monotonically increasing with temperature according to (\Ref{ximsing}).
This is not surprising
since it is temperature-driven fluctuations of the strings around the
path $\r_0(t)$ that generate the pair distribution (\Ref{P2xi}). At the
roughening transition (i.e., for $\beta^{-1} = \beta_c^{-1}$), the
singularity of $\xi^{(m)}$ changes,
\EQ
\xi^{(m)}  \sim \xi 
                 \sim (-g)^{- \zeta^\star /  y^\star} \HS \mbox{with} \;\;
y^\star = (d - 2)/2 \;,
\EEQ
as discussed in  Section 3.1.
With the natural assumption that $\xi^{(m)}$ remains a monotonic function
of $\beta^{-1}$ for all $\beta^{-1} \leq \beta_c^{-1}$ and fixed $g$,
one then obtains the inequalities \linebreak
$ y / (\zeta - \omega/m) \geq y^\star / \zeta^\star $. These imply an upper
bound on the free energy exponent:
\EQ
\omega \leq \frac{4 - d}{d} \;.
\EEQ 
The result $ d_> \leq 4$ then follows immediately.  

It is tempting to speculate about the nature of the strong-coupling
regime in high dimensions. Below $d_>$, the pair distribution of noninteracting 
strings at fixed temperature has the  finite limit (\Ref{P2}) for $R \to
\infty$, and this limit distribution collapses to $\delta(\r_1 - \r_2)$ for 
$\beta^{-1} \to 0$. For $d \geq d_>$, the pair distribution is expected 
to depend on both $\beta^{-1}$ and $R$ in an essential way. Its asymptotic 
behavior will then depend on the order in which the zero-temperature
limit and the thermodynamic limit $R \to \infty$ are taken. Similar
properties are familiar from glassy systems.

\bigskip

\subsection{Discussion}

Interacting strings in a random background turn out to have a rich
scaling behavior. In one transversal dimension alone, there are no
less than six universality classes characterized by 
of the exponent $\theta$ of the disorder-averaged string density 
and the renormalization group eigenvalue $y$ of the contact coupling. 
These universality classes describe 
\begin{itemize}

\item 
the (de-)localization transition of a string at a linear defect 
($\theta = 0, y = 0$),

\item 
the (de-)localization transition of a string at an attractive wall
($\theta = -1/2, y = 1/3$),

\item
the (de-)localization transition of a pair of strings with contact
attraction 
($ \theta = -3/2, y = 2/3$),

\item
the scaling of a string at a barrier 
($\theta = 1, y = 1/3$),

\item
the scaling of a string at a repulsive wall
($\theta = 1, y = -2/3$),

\item
the scaling of a pair of strings with contact repulsion
($\theta = 2, y = -4/3$).

\end{itemize}

Clearly, the list could be continued by including higher dimensions, 
long-ranged interactions or disorder correlations etc. Recall from
Section 2 that without quenched disorder, the six cases are described
by just two universality classes, namely 
Gaussian strings ($\theta = 0, y = 1/2$) and
free fermions ($\theta = 2, y = -1/2$).

This scenario is consistent with an operator product
expansion~(\ref{ope1})
of the contact field, which is the basis for perturbation theory 
about the strong-coupling fixed point. The existence of an operator
product expansion is of conceptual importance since the  
correlations in the strong-coupling regime are generated by global
minimization of the free energy and not by a local action. 
The operator product expansion explicitly contains the temperature $\beta^{-1}$
as dangerous irrelevant variable. This variable, which is proportional
to the surface tension $\nu$ of the associated KPZ surface,
will prove dangerous in the growth problem as well: it generates the 
dynamical anomaly discussed in the next Section.

\bigskip

\section{Directed growth}
\setcounter{equation}{0}

All of the methods and results on directed polymers in a random medium
discussed in the previous Section have their correspondences in
KPZ surface growth via the Hopf-Cole transformation 
(\Ref{HopfCole}). In particular, the KPZ equation has an upper critical 
dimension less or equal to four (see Section~3.6).

To discuss dynamical renormalization, we rewrite Eq.~(\Ref{KPZ}) 
as a path integral for the height field
$h(\r,t)$ and the response field $\tilde h(\r,t)$. It proves necessary 
to distinguish carefully between renormalized fields and couplings
defined in a nonperturbative way, and perturbatively renormalized 
quantities defined,
for example, by minimal subtraction. The perturbative renormalization of the
dynamic path integral is seen to be equivalent to the replica
perturbation theory of Section 3: it captures the roughening transition
but does not produce a fixed point corresponding to the strong-coupling
regime~\cite{Lassig.kpz}. This failure of perturbation theory is explained by the fact
that the nonperturbatively renormalized coupling constant $u$ and the
corresponding height field $h$ (defined by conditions on correlation functions
at a renormalization point) have a singular dependence on their 
perturbatively subtracted  counterparts
$u_P$ and $h_P$.

The strong-coupling regime thus calls for nonperturbative methods.
In the second part of this Section, we analyze constraints on the
effective growth dynamics imposed by the consistency of correlation
functions. The consistency relations take again the form of an operator 
product expansion. 
The response field $\tilde h$ is found to obey an operator product
expansion similar to those of the previous 
Section~\cite{KinzelbachLassig.depinning}. The resulting structure of the 
response functions is tied to KPZ growth with additional deterministic
driving forces. In particular, a surface with a local variation 
$ -(g / \lambda) \delta (\r)$ in the deposition rate corresponds 
to a directed string interacting with a linear defect (see Section~3.3).
It turns out that (for $\lambda > 0$) a sufficiently enhanced deposition 
($g / \lambda < g_c / \lambda$) induces an
increased growth rate even in the thermodynamic limit $R \to \infty$,
while for $g / \lambda >  g_c / \lambda$ the growth rate is asymptotically 
independent of $g$~\cite{WolfTang.inh,TangLyuksyutov.depinning}. 
These two regimes are separated by a nonequilibrium phase
transition, the analogue of the string (de-)localization
transition.

Finally, we turn to the correlations of the height 
field~\cite{Lassig.anomaly}. These are 
assumed to obey an operator product expansion as well.
Consistently with the available numerical data, we further assume these
correlations do not show multiscaling, unlike the correlations 
in a turbulent fluid. This property constrains severely the 
possible form of the operator product expansion. A discrete set 
of values of the roughness exponent emerges:
\EQ
\chi =  \frac{2}{k_0 + 2} 
\HS \mbox{with $k_0 = 1,2, \dots$} \;.
\label{quant}
\EEQ
Only for these values, the
KPZ equation admits solutions with a {\em dynamical 
anomaly} in the strong-coupling regime. It is argued on phenomenological
grounds that the scaling of growing surfaces should be governed by
such solutions with an odd value of $k_0$. Hence the exact values 
of the growth exponents for two- and three-dimensional surfaces are derived.

\bigskip
 
\subsection{Field theory of the Kardar-Parisi-Zhang model}

The path integral formulation~\cite{dynpath} of the stochastic evolution 
(\Ref{KPZ}), (\Ref{etaeta}) is obtained in a straightforward way. We rewrite the 
Gaussian distribution of the driving force $\eta(\r,t)$,
\EQ
\int {\cal D} \eta \exp
     \left ( - \int \D \r \D t \,
     \frac{1}{2 \sigma^2} \, \eta^2 \right ) =
\int {\cal D} \eta {\cal D} \tilde h \exp 
     \left ( - \int \D \r \D t 
     \left( -\frac{ \sigma^2}{2} \, \tilde h^2 +  \tilde h \eta \right )
     \right )  \;,
\EEQ
introducing the purely imaginary ``ghost'' field $\tilde h(\r,t)$. Eliminating 
$\eta$ by using the equation of motion, we obtain the (\^Ito-dicretized)
path integral
\EQ
Z = \int {\cal D} h {\cal D} \tilde h 
     \exp \left [ - \int \D \r \D t \left (
    - \frac{ \sigma^2}{2} \tilde h^2 + \tilde h \left (
    \frac{\partial}{\partial t} h - \frac{\nu}{2} \nabla^2 h 
    - \frac{\lambda}{2} (\nabla h)^2  \right ) \right )
          \right ] \;,
\Label{Zdyn}
\EEQ
the generating functional of the dynamical correlations. Ghost field
insertions produce the response functions
\EQ
\left \langle \prod_{j = 1}^{\tilde N} \tilde h (\r_j, t_{j}) 
              \prod_{j = \tilde N + 1}^{\tilde N + N} h (\r_j, t_{j}) 
\right \rangle =   
\prod_{j=1}^{\tilde N} \frac{\delta}{ \delta \rho (\r_j, t_{j})}
\left \langle \prod_{j = \tilde N + 1}^{\tilde N + N} h(\r_j, t_{j}) 
\right \rangle \hs,
\EEQ
where $\rho(\r,t)$ is an additional source field in the equation of
motion,
\EQ
\partial_t h = \nu \nabla^2 h + \frac{\lambda}{2} (\nabla h)^2 
               + \eta + \rho \;.
\Label{KPZrho}
\EEQ 

There is a third way to express the KPZ dynamics.
Writing the equal-time height correlations in the form
\EQ
\langle h(\r_1,t) \dots h(\r_n,t) \rangle \equiv 
\langle h(\r_1) \dots h(\r_n) \rangle_t
 = \int \! {\cal D}h  \, h(\r_ 1) \dots h(\r_ n) \,
  P_t \;,
\Label{hn}
\EEQ
the time dependence has been shifted from the field variables $h(\r,t)$
to the configuration weight $P_t [h]$. The latter obeys the functional 
Fokker-Planck equation
\EQ
\partial_t P_t = \left. \int  d \r \left (
               \sigma^2 \frac{\delta^2}{\delta h(\r)^2} 
               - \frac{\delta}{\delta h(\r)} J(\r) 
               \right ) P_t \right. \;,
\Label{Pt}
\EEQ
where 
\EQ
J(\r) \equiv \nu {\bf \nabla}^2 h(\r) + 
   \frac{\lambda}{2} ({\bf \nabla} h)^2 (\r)
\EEQ
is the deterministic part of the current. 

In the linear theory, the fields $h(\r,t)$ and $\tilde h(\r,t)$ have canonical
dimensions 
$-\chi_0 = (d - 2)/2$ and
$\chi_0 + d = (d + 2)/2$, respectively, and the dynamical exponent
is $z_0 = 2$ (the basic scale is now that of $\r$).

A crucial role in the sequel will be played by the infrared divergencies
of the height correlations~\cite{Lassig.anomaly}. 
Their physical meaning can be seen already in 
the linear theory, which has the  response propagator  
\EQ
\langle \tilde h (\r_1, t_1) h (\r_2, t_2) \rangle =  
\frac{ \theta (t_2 - t_1) }{ (2 \pi \nu (t_2 - t_1))^{d/2} }
\exp \left ( - \frac{ (\r_1 - \r_2)^2 }{ 2 \nu (t_2 - t_1) } \right )
\Label{G0}
\EEQ
($\theta(t)$ denoting the step function) and the 
height-height correlation function         
\EQ
 \langle h (\r_1,t_1) h(\r_2,t_2) \rangle =                                               
    \sigma^2 \int \D \r \D t \,
    \langle \tilde h(\r,t) h(\r_1,t_1) \rangle 
    \langle \tilde h(\r,t) h(\r_2,t_2) \rangle  \;.
\Label{hh0}
\EEQ
However, the last expression is divergent in infinite space 
for $d < 2$ (i.e., $\chi_0 > 0$).
Therefore, it depends strongly on the initial or boundary conditions. For
stationary growth in a system of size $R$, the dominant part in the
limit $R \to \infty$ is
\EQ
 \langle h (\r_1,t_1) h(\r_2, t_2) \rangle_R \sim
  \langle h^2 \rangle_R \sim R^{2 \chi} \;,
\label{h20}
\EEQ
where $\langle h^2 \rangle \equiv \langle h^2 (\r_1,t_1) \rangle
                               =  \langle h^2 (\r_2,t_2) \rangle$
with periodic boundary conditions. The amplitude (\ref{h20})  
measures the {\em global} roughness of the surface, i.e., the size 
of its mountains and valleys. The regularized height 
correlation, however, can be written in terms of height differences
measuring the {\em local} roughness,
\EQ 
 \langle h (\r_1,t_1) h(\r_2,t_2) \rangle_R -
 \langle h^2 \rangle_R =
 -\frac{1}{2} \langle (h(\r_1,t_1) - h(\r_2,t_2))^2 \rangle_R \;.
\EEQ
The finite limit for $R \to \infty$,
\EQ
\langle (h(\r_1,t_1) - h(\r_2,t_2))^2 \rangle = 
|\r_1 - \r_2|^{2 \chi_0} 
 {\cal G}_0 \left( \frac{\nu |t_1 - t_2|}{(\r_1 - \r_2)^2} \right) \;,
\EEQ
shows (affine) scale invariance. In the interacting theory, the
infrared regularization is more complicated since also the higher
connected height correlations develop $R$-dependent singularities.
But the local statistics of the surface, measured by height differences,
is still expected to remain finite for $R \to \infty$, i.e., to 
decouple from the divergent amplitudes. 

The interacting theory has an ultraviolet singularity as well. Taking
the expectation value of the equation of motion, we obtain the growth
rate, which has the form
\EQ
\langle \partial_t h \rangle_R = 
  \frac{\lambda}{2} \langle (\nabla h)^2 \rangle_R \sim
  a^{2 \chi - 2} + O(R^{2 \chi - 2}) 
\Label{uvsing}
\EEQ
for stationary growth. The dependence on the short-distance cutoff $a$
can be removed by the subtraction
\EQ
(\nabla h)^2 \to (\nabla h)^2 - \langle (\nabla h)^2 \rangle_\infty \;,
\Label{normalord}
\EEQ
which affects the one-point function $\langle h(\r,t) \rangle$ but
leaves the higher connected correlation functions invariant. 
This normal-ordering of the interaction field will always be implied 
in the sequel. The nature of the cutoff dependent term in (\Ref{uvsing})
is clear from the mapping onto directed strings. Via
Eq.~(\Ref{HopfCole}), the subtracted growth rate is directly related
to the Casimir term (\Ref{Cbar}),
\EQ
- \lambda \langle \partial_t h \rangle_R = \overline f (R) \;,
\EEQ
while the $a$-dependent term contributes to the nonuniversal part of the
free energy.

\bigskip

\subsection{Dynamical perturbation theory}

Perturbation theory for the KPZ dynamics can be set 
up in different ways: 
\newline
(a) The standard formalism is a diagrammatic
expansion of the path integral (\ref{Zdyn})
with interaction vertex $ \tilde h (\nabla h)^2$ about
the Gaussian limit $\lambda = 0$. The propagators of the
Gaussian theory are the response function 
$\langle \tilde h (\r_1, t_1) h (\r_2, t_2) \rangle$
and the correlation function
$\langle h (\r_1,t_1) h(\r_2,t_2) \rangle$;
see Eqs.~(\ref{G0}) and (\ref{hh0}).
Renormalization is usually based on the expansion of  
Fourier transformed two-point  functions 
$\langle \tilde h(- {\bf k}, - \omega) h({\bf k}, \omega) \rangle$ and
$ \langle h(- {\bf k}, - \omega) h({\bf k}, \omega) \rangle$. 
This has been carried out to
first~\cite{TangNattermannForrest,NattermannTang} and 
second order~\cite{SunPlischke.kpz,FreyTaeuber}; see also the criticism of 
ref.~\cite{Wiese.kpz1}. In the Appendix, we discuss instead the
expansion for the real-space response function 
$\langle \tilde h(\r_1,t_1) h(\r_2, t_2) \rangle$ and
the universal part of the growth rate $\langle \partial_t h \rangle_R$
in order to exhibit the analogy to the replica perturbation
theory of Section~3.1. Beyond leading order, however, calculations 
become cumbersome in this formalism.
\newline
(b) By the Hopf-Cole transformation (\ref{HopfCole}) and
the corresponding transformation of the response
field~\cite{KinzelbachLassig.depinning},
\EQ
h = \frac{2 \nu}{\lambda} \log \phi \;,
\hspace{1cm}
\tilde h = \frac{\lambda}{2 \nu} \bar \phi \phi \;,
\label{HopfCole2}
\EEQ
the path integral (\ref{Zdyn}) can be brought to the form~\cite{Wiese.kpz2}
\EQ
Z = \int {\cal D} \phi {\cal D} \bar \phi 
                 \exp \left [ - \frac{1}{2 \nu} \int \D \r \D t
                   \left ( 
                   \bar \phi \left ( \partial_{t} 
                                     - \nu \nabla^2 
                              \right ) \phi -
                    \frac{\sigma^2 \lambda^2}{2 \nu} \bar \phi^2 \phi^2
                   \right ) 
           \right ] \;,
\label{Zdyn2}
\EEQ
which equals the replica action (\ref{F2quant}) introduced 
previously in ref.~\cite{Lassig.kpz}
(with $\beta = 1 / 2 \nu$ and $\lambda = 1$). We emphasize, however,
that (\ref{Zdyn2}) is not a faithful representation of the growth
dynamics unless it is defined with boundary 
conditions enforcing the limit $p \to 0$; see the
discussion in the Appendix. The theory is now Gaussian for 
finite $\lambda$ but $\sigma^2 = 0$. The propagator 
$\langle \bar \phi (\r_1,t_1) \phi(\r_2, t_2)  \rangle$ is again
given by (\ref{G0}); the interaction vertex $\bar \phi^2 \phi^2$
arises from the noise term $\tilde h^2$. The resulting
diagrammatic expansion is  identical to the replica
perturbation theory of Section~3 and
can hence be renormalized exactly to all orders, as shown
in the Appendix.

In both cases, the expansion parameter is again the 
dimensionless coupling constant $u_0 = g_0 R^{2 y_0}$ with 
\EQ
g_0 = \frac{\sigma_0^2 \lambda_0^2}{(2 \nu_0)^3} \, 
\EEQ
where the fields and couplings of the bare theory  are
labeled by the subscript~0.
The  poles in $y_0$ can be absorbed into
loopwise subtracted fields $h_P, \tilde h_P$ and the coupling
constant $u_P = \Z_P u_0$ given by (\Ref{Z0}). There is no renormalization
of $t_0 = t_P$, which would be required in the strong coupling regime according
to (\Ref{r2freg}).

The flow equation is again (\Ref{beta0}) with the fixed point $u_P^\star =
y_0/C_0$ representing the roughening transition for $2 < d < 4$. 
At the transition, the Casimir amplitude of the growth rate takes
a finite value regular in $y_0$,
\EQ
R^2 \, \frac{\lambda_c}{(2 \nu)^2} \langle \partial_{t_P} h_P \rangle_R 
 = \frac{1}{4} \frac{y_0}{C_0} + O(y_0^2) \;,
\EEQ
which is directly related to the Casimir amplitude (\Ref{Cbarstar}) 
of a string in a random medium. The scaling exponents at the transition,
\EQ
\chi^\star = 0 \;, \HS z^\star = 2 \;,
\Label{chistar}
\EEQ
are independent of $d$, in agreement with the finite-order calculations of 
refs.~\cite{TangNattermannForrest,NattermannTang,FreyTaeuber}. These exponents
have been predicted previously by a scaling
argument~\cite{DotyKosterlitz.roughening}.

\subsection{Renormalization beyond perturbation theory}

Of course, renormalization is not tied to a loopwise expansion and can 
be defined in a nonperturbative way~\cite{Lassig.kpz}. The correlation
functions show crossover scaling from the linear to the strong-coupling
regime ($d \leq 2$) and from the critical to the strong-coupling regime~($d >2)$.
The crossover has the characteristic scales $\rc$ and $\tc$ given by 
(\Ref{rctc}) with $\beta^{-1} = 2 \nu$.

For $d \leq 2$, the response and correlation functions have the scaling form 
\EQ
\left \langle \prod_{j = 1}^{\tilde N} \tilde h_0 (\r_j, t_{0j}) 
          \prod_{j = \tilde N + 1}^{\tilde N + N} h_0 (\r_j, t_{0j}) \right 
      \rangle =  
R^{-N \chi_0 + \tilde N (\chi_0 + d)} 
{\cal G}_{N \tilde N} \left ( \frac{\r_j - \r_k}{R}, 
                       \frac{t_{0j} - t_{0k}}{R^{z_0}}, 
                       u_0 
               \right ) \;.
\Label{cross}
\EEQ

Eq.~(\Ref{cross}) can be written as the bare Callan-Symanzik equation 
\EA 
\lefteqn{\BS \BS 
          \left ( R \left. \frac{\partial}{\partial R} \right |_{\lambda_0}
                  + \sum_j \r_j \frac{\partial}{\partial \r_j} 
                  + z_0 \sum_j t_{0j} \frac{\partial}{\partial t_{0j}} 
                  + \dot{u}_0 \, \frac{\partial}{\partial u_0} 
                  - N \chi_0 + \tilde N (\chi_0 + d) 
          \right )  }          \nonumber
\\ & & 
\times \left \langle \prod_{j = 1}^{\tilde N} \tilde h_0 (\r_j, t_{0j}) 
          \prod_{j = \tilde N + 1}^{\tilde N + N} h_0 (\r_j, t_{0j}) \right 
      \rangle =  0   
\Label{bareCS}
\EEA 
with
\EQ
\dot{u}_0 \equiv R \partial_R \, u_0 = 2 y_0 u_0 \hs.
\EEQ 
For the infrared-regularized correlators, the explicit dependence on $R$
vanishes in the thermodynamic limit. 

In the strong-coupling regime, these correlation 
functions develop anomalous scaling and hence a singular dependence on 
the bare coupling constant $u_0$. Renormalization consists in absorbing
these singularities into new variables
\EA 
h         & = &  \Z_h  h_0  \hs,  \;\;        
\tilde h    =   \Z_h^{-1}  \tilde h_0  \hs,     \nonumber   \\
t         & = &  \Z_t  t_0  \hs,                \label{R}   \\
u         & = &   \Z u_0 = \Z_h^{-2} \Z_t^{-2} u_0  \hs. \nonumber                   
\EEA
The $\Z$-factors can all be written as functions of $u$. 

Under this change of variables, Eq. (\Ref{bareCS}) transforms into the
renormalized Callan-Symanzik equation
\EA 
\lefteqn{\BS \BS \BS 
          \left ( R \left. \frac{\partial}{\partial R} \right |_{\lambda_0}
                  + \sum_j \r_j \frac{\partial}{\partial \r_j} 
                  + z (u) \sum_j t_{j} \frac{\partial}{\partial t_{j}} 
                  + \dot{u} \frac{\partial}{\partial u} 
                  - N \chi (u) + \tilde N (\chi (u) + d) 
          \right )  }          \nonumber
\\ & & 
\times \left \langle \prod_{j = 1}^{\tilde N} \tilde h (\r_j, t_{j}) 
          \prod_{j = \tilde N + 1}^{\tilde N + N} h (\r_j, t_{j}) \right 
      \rangle =  0   
\Label{renCS}
\EEA 
with
\EA
\dot{u} \equiv R \partial_R \, u  & = &
\frac{2 y_0 u}{1 - u (\D / \D u) \log \Z} \hs,   \Label{beta.pert} \\
z(u)                           & = & 
z_0 - \dot u \frac{\D}{\D u} \log \Z_t \hs,        \Label{z}    \\
\chi(u)                           & = & 
\chi_0 - \dot u \frac{\D}{\D u} \log \Z_h \hs.     \Label{zeta}
\EEA
Notice that only two of the $\Z$-factors  are independent. 
The ghost field renormalization is 
tied to that of $h$ since the response function 
$\langle \tilde h(\r_1,t_1) h(\r_2,t_2) \rangle$ always has dimension
$d$ by its definition. Furthermore, Galilei invariance implies 
the relation $\Z = \Z_h^{-2} \Z_t^{-2}$ obtained by inserting the 
exponent relation $\chi + z = 2$ into (\Ref{beta.pert}), (\Ref{z}), and
(\Ref{zeta}). A different but equivalent Callan-Symanzik equation is 
derived in ref.~\cite{FreyTaeuber}. 

The renormalized variables (\Ref{R}) can be defined in a nonperturbative
way by two independent normalization conditions. These are  imposed, e.g., 
on the stationary two-point functions in an infinite system,
\EA
R^{2 \chi_0} \langle (h   (0,t  ) - h   (R, t  ))^2 \rangle (u)
& = & 
R^{2 \chi_0} \langle (h_0 (0,t_0) - h_0 (R, t_0))^2 \rangle (0),  
\Label{CR}
\\
R^d \langle \tilde h   (\r,t  )  h  (\r, t   + R^2) \rangle (u)
& = & 
R^d \langle \tilde h_0 (\r,t_0)  h_0(\r, t_0 + R^2) \rangle (0) \hs.  
\Label{GR}
\EEA
$R$ is now an arbitrary normalization scale. An alternative to (\Ref{CR}) is  
the normalization condition 
\EQ
R^{2 - \chi_0} \langle \partial_{t} h \rangle_R (u) = - \frac{1}{2} \, u
\Label{hR}
\EEQ
on the universal finite-size correction to the stationary growth velocity in a system of 
size $R$. Comparing (\Ref{CR}) and (\Ref{GR}) with the asymptotic scaling of the bare
functions obtained from (\Ref{cross}), 
\EA
\langle (h_0 (\r_1,t_0) - h_0 (\r_2, t_0))^2 \rangle 
& \sim &  
\rc^{-2 (\chi - \chi_0)} |\r_1 - \r_2|^{2 \chi} \;,
\\
\langle \tilde h_0 (\r,t_{01})  h_0(\r, t_{02}) \rangle 
& \sim & 
{\tc_0}^{\; d/z - d_0/z_0} (t_{02} - t_{01})^{- d/z} 
\hs,  
\EEA
we infer the asymptotic behavior of the $\Z$-factors:
\EQ
\Z_h \sim ( \rc / R)^{\chi - \chi_0} \;, \HS
\Z_t \sim ( \rc / R)^{z - z_0} \;.
\label{Zas}
\EEQ
The renormalized height correlation and response functions have 
a finite continuum limit $\rc \to 0$. Of course, this does not imply
that all observables are finite in this limit since composite fields
may generate additional singularities. However, as discussed in Section~4.5,
the structure of composite fields of the KPZ
theory is expected to be quite simple. All negative-dimensional 
fields (for example, the monomials $h^k(\r)$) have scaling dimensions
given by the linear spectrum (\ref{xkN}). Consequently, the transformations
(\ref{R}) do renormalize the correlations of these fields.

The renormalized height field $h(\r,t)$ obeys again an equation of
motion of the form (\Ref{KPZ}) but the coefficients become singular 
in the limit $\rc \to 0$. By substituting the renormalized variables
(\ref{R}) into (\Ref{KPZ}), (\Ref{Zdyn}) or (\Ref{Pt}), we obtain
\EA
\nu       & = & \Z_t^{-1} \nu_0 
                \simeq \nu^* \times (\rc / R)^\chi \;,
\nonumber \\
\sigma^2  & = & \Z_h^2 \Z_t^{-1} \sigma_0^2
                \simeq \sigma^{* 2} \times (\rc / R)^{d - 2 + 3 \chi} \;,
\Label{coeffas} \\
u         & = & u_0 {\cal Z}_t^{-2} {\cal Z}_h^{-2} 
                \simeq u^* 
\nonumber
\EEA
with finite limit values $\nu^*$, $\sigma^{* 2}$, and $g^*$. 
Galilei invariance is reflected in the asymptotic scale invariance of
the dimensionless coupling, $u \simeq u^*$. This is the nonperturbative
strong-coupling fixed point. The other
coefficients become irrelevant as $\rc / R \to 0$. 
In particular, we recover the scaling (\ref{betareg}) of the renormalized
temperature $\beta^{-1} = 2 \nu$. It is easy
to verify that the strong-coupling asymptotics (\Ref{Zas}) and
({\Ref{coeffas}) remains unchanged for $d > 2$ although the scaling
functions (\ref{cross}) now contain singularities marking the roughening
transition. 

We emphasize again that the existence of this finite continuum limit
depends only on the existence of a well-defined scaling regime at large
distances. We have not assumed perturbative renormalizability, i.e., that the
$\Z$-factors can be computed as power series in $u_0$. It is
instructive, however, to compare the renormalized theory with the perturbatively
finite theory of Section~4.2. Recall that in a usual $\ep$-expansion,
imposing normalization conditions like (\Ref{CR}), (\Ref{GR}) is
equivalent to requiring finiteness order by order in perturbation theory. 
The respective coupling constants, $u$ and $u_P$, are related by  a
diffeomorphism that remains regular in the limit $\ep \to 0$ and is defined on a
domain of interaction space that contains the fixed points $u_P = u = 0$  and
$u^* (u_P^*)$. This equivalence is lost in the crossover to the
strong-coupling fixed  point. To see this, we integrate the flow equation 
(\Ref{beta0}) for $u_P$ in two dimensions with
the initial condition $ u_P (R_0) = u_1$. The solution 
\EQ 
u_P (R) = \frac{u_1}{1 - u_1 \log (R/R_0)}  \hs,  
\EEQ
diverges at a finite value 
$R  = R_1 \equiv R_0 \exp (1 / u_1)$. It follows
immediately that $u_P (R)$ is not a good parametrization of the crossover. 
The pole of $u_P(R)$ is only a ``coordinate  singularity''
\cite{Lassig.cth} of perturbative subtraction schemes; any renormalized coupling
$u (R) $ is expected to remain regular at $R = R_1$. Hence the function 
$ u_P (u) $ 
has a pole at the value $ u (R_1)$ between the 
fixed points $u = 0$ and $u^*$. Accordingly, the loopwise subtracted
correlation functions at the normalization point become singular as
well: perturbative subtraction fails to maintain a smooth crossover of the 
large-distance regime.

\bigskip

\subsection{Response functions in the  strong coupling regime}

The correspondence between directed strings in a random medium and 
growing surfaces extends to strings with additional interactions. 
We restrict ourselves here to the case of a string and a linear defect 
discussed in Section~3.3 to infer the properties of the renormalized
ghost field correlations at the strong-coupling fixed point. 

The action (\Ref{Sdefect}) translates into the growth equation
\EQ
\partial_t h = \nu \nabla^2 h + \frac{\lambda}{2} (\nabla h)^2 
               + \eta - \frac{g}{\lambda} \, \delta(\r) \;,
\Label{KPZdefect}
\EEQ
see refs.~\cite{WolfTang.inh,TangLyuksyutov.depinning}.
The extra term describes a spatial inhomogeneity in the average rate
of mass deposition onto the surface. 
The Casimir amplitude (\Ref{Cbarg}) of the free energy 
is that of the growth rate,
\EQ
\Delta \overline C(g,R) = - \lambda R^{-\chi + z} 
   \left ( \langle \partial_t h \rangle_R (g) 
          -\langle \partial_t h \rangle_R (0) \right ) \;.
\label{Cbarg.2}
\EEQ

For $g /\lambda < 0$ ($d \leq 1$) and 
$g/\lambda  < g_c/\lambda $ ($d > 1$), translational 
invariance is strongly broken. The string  bound state corresponds
to a surface state with a macroscopic stationary mountain
\EQ
H(r) \equiv \langle h (0,t) \rangle - \langle h(\r,t) \rangle 
\EEQ
generating an enhanced growth rate even in the limit $R \to \infty$. 

The inhomogeneity in Eq.~(\ref{KPZdefect}) contributes a term 
$- (g /\lambda) \int \D t \, \tilde h (0,t) $ 
to the dynamic action in (\Ref{Zdyn}). Hence,
the perturbation theory (\Ref{Cbarseries}) 
for the Casimir amplitude (\Ref{Cbarg}) 
can be reproduced in the dynamical formalism~\cite{KinzelbachLassig.depinning},
\EQ
\Delta \overline C(g,R) 
   = \lambda \lim_{L \to \infty}  \partial_L \,
 \langle \exp [ - (g/\lambda) \mbox{$ \int $} {\rm d}t \, \tilde  h (0,t) ] \,
         h(\r,L) \rangle \;.
\EEQ
The two series are related term by term
via the mapping (\ref{HopfCole2})~\cite{KinzelbachLassig.depinning}, 
\EQ
(\beta \lambda)^{m-1} \;
\overline{ \langle \Phi (t_1) \cdots \Phi (t_N ) \rangle^c } = 
\lim_{L \to \infty} 
  \langle \tilde{h} (0, t_1) \dots \tilde{h} (0, t_N) \, h(\r,L)
  \rangle  \;.
\EEQ
Eq.~(\Ref{ope1rt}) translates into an operator product 
expansion for the ghost fields,
\EQ
\tilde h(\r,t) \tilde h (\r',t') = 
  \lambda C |t - t'|^{-(\chi + d)/z} \, 
  {\cal H} \! \left ( \frac{ |t - t'|}{|\r - \r'|^z} \right ) 
  \tilde h(\r,t) \;,
\EEQ 
which produces an equal-time singularity
$\tilde h(\r,t) \tilde h (\r',t) \sim  
     |\r - \r'|^{-(\chi + d)} \, \tilde h(\r,t)$.    
It is quite plausible intuitively that the response to a pair of nearby 
sources reduces to the response to a single source times a divergent factor.  
This form of the operator product expansion follows also from the 
mode-coupling approach of ref.~\cite{HwaNattermann.depinning}.

\bigskip

\subsection{Height correlations in the strong coupling regime}

In the sequel, we discuss the renormalized and connected 
equal-time height correlations 
\EQ
\langle h(\r_1,t) \dots h(\r_n,t) \rangle \equiv 
\langle h(\r_1) \dots h(\r_n) \rangle_t
\label{hn2}
\EEQ
in the strong-coupling regime, following ref.~\cite{Lassig.anomaly}.
An infinite surface growing from a flat initial 
state $h(\r, 0) = 0$ develops height correlations depending on the
differences $\r_{ij} \equiv \r_i - \r_j$ and the  
correlation length $\xi_t \sim t^{1/z}$ increasing with time.
(In a finite system of size $R$, the correlation length will eventually
saturate at a value $\xi \sim R$. We restrict ourselves here to unsaturated
growth, i.e., to values $R \gg \xi_t$.)  
In the scaling regime 
\EQ
\rc \ll |\r_ {ij}| \ll \xi_t \;\;\; (i,j = 1, \dots, n) \;,
\label{scregime}
\EEQ
the height correlations will generically 
become singular as some of the points approach each other. 
For $d < d_>$, these singularities should be power laws. They are assumed 
to follow from an operator product expansion 
\EQ
 h(\r_ 1) \dots h(\r_ k) =  
            \sum_{\cal O} |\r_ {12}|^{-k x_h + x_{\cal O}} \,
            \, C_k^{\cal O} \!\!\left ( \frac{\r_ {13}}{|\r_ {12}|}, \dots, 
            \frac{\r_ {1k}}{|\r_ {12}|} \right ) \, 
            {\cal O} (\r_ 1) \;,
\Label{hope}
\EEQ
an identity expressing any $n$-point function as a sum
of $(n - k + 1)$-point functions in the limit 
$|\r_ {ij}| \ll |\r_ {il}| \ll \xi_t$ 
($i,j = 1, \dots, k$ and 
$l =  k+1, \dots, n$). 
The sum on the r.h.s. runs over all local scaling fields ${\cal O}(\r)$. Each term
contains a dimensionless scaling function $C_k^{\cal O}$ (a simple number
for $k = 2$) and a power of $|\r_ {12}|$ given by the scaling dimensions
$x_{\cal O}$ and $x_h = - \chi$ (such that the overall dimension equals 
that of the l.h.s.). The field ${\cal O}_k$ with the smallest dimension,
$x_k$, determines in
particular the asymptotic behavior of the $k$-point functions as 
$\xi_t \to \infty$,
\EQ
\langle h(\r_ 1) \dots h(\r_ k) \rangle_t 
        \sim \langle {\cal O}_k \rangle_t
        \sim \xi_t^{-x_k} \;.
\Label{hk2}
\EEQ
The amplitudes 
$\langle {\cal O}_k \rangle_t = \langle {\cal O}_k (\r) \rangle_t$ 
diverge with $\xi_t$, i.e., $x_k < 0$.
They measure the global roughness, which increases as the 
surface develops higher mountains and deeper valleys.
Local properties of the surface should,
however, behave quite differently. For example, the gradient correlation
functions are assumed to have a finite limit 
\EQ
\lim_{\xi_t \to \infty} 
 \langle {\bf \nabla} h(\r_ 1) \dots 
         {\bf \nabla} h(\r_ n) \rangle_t
 \equiv \langle {\bf \nabla} h(\r_ 1) \dots 
                {\bf \nabla} h(\r_ n) \rangle \;.
\Label{vk}
\EEQ
By writing 
$h(\r_ i) - h(\r_ i') = 
 \int_{\r_ i}^{\r_ i'}  {\bf ds} \cdot {\bf \nabla} h ({\bf s})$, the same
property follows for the height difference correlation functions
\EQ
\langle (h(\r_1) - h(\r_1')) \dots (h(\r_n) - h(\r_n')) \rangle_t \;.
\Label{hdiff} 
\EEQ
This implies a feature familiar from
simulations: one cannot recognize the value of $\xi_t$ from snapshots 
of the surface in a region much smaller than $\xi_t$. 

The stationarity condition (\ref{vk}) restricts severely the form of
the height correlations~(\ref{hn2}). This can be seen as follows.
The operator product expansion (\Ref{hope}) induces an expansion for the
gradient field ${\bf v} \equiv {\bf \nabla} h$ of the form
\EQ
{\bf v}(\r_ 1) \dots {\bf v}(\r_ k) =
      \sum_{\cal O} |\r_ {12}|^{- k x_{\bf v} + x_{\cal O}} \,
      \, \tilde C^{\cal O}_k 
      \!\!\left ( \frac{\r_ {13}}{|\r_ {12}|}, \dots, 
      \frac{\r_ {1k}}{|\r_ {12}|} \right ) 
      {\cal O}(\r_ 1)
\Label{vope}
\EEQ
with new scaling functions $\tilde C_k^{\cal O}$ and the dimension
$x_{{\bf v}} = - \chi + 1$. (Both sides of
(\Ref{vope}) are tensors of rank $k$ whose indices are suppressed.)
The fields ${\cal O}$ on the r.h.s.~govern the time-dependent 
amplitudes 
$\langle {\bf v}({\bf r}_ 1) \dots {\bf v}({\bf r}_ k) \rangle_t 
  \sim \langle {\cal O} \rangle_t 
  \sim \xi_t^{-x_{\cal O}}$ in analogy to (\ref{hk2}). 
Hence, the stationarity condition (\ref{vk}) allows in 
(\ref{vope}) only fields ${\cal O}$ with a {\em non-negative} scaling 
dimension $x_{\cal O}$, such as ${\bf 1}$ (the identity field), 
$({\bf \nabla} h)^2({\bf r})$, etc.
This in turn restricts the possible terms in~(\ref{hope}): 
\newline
(a) {\em Singular} terms involving fields ${\cal O}({\bf r})$ with 
$x_{\cal O} \geq 0$. In particular, we call the coefficient function 
of~${\bf 1}$ the {\em contraction} of the fields 
$h(\r_1), \dots, h(\r_k)$.
\newline
(b) {\em Regular} terms, where the coefficient 
$|{\bf r}_ {12}|^{-k x_h + x_{\cal O}} \, C_k^{\cal O}$ 
is a tensor of rank $N$ in the differences ${\bf r}_ {1i}$ ($i = 2,
\dots, k)$. Such terms do not violate (\ref{vk}) since
they have a vanishing coefficient $\tilde C_k^{\cal O}$ in (\ref{vope})
for $N < k$. They can readily be associated with composite fields of 
dimensions
\begin{equation}
x_{k,N} = - k \chi + N \;.
\label{xkN}
\end{equation}
The leading ($N = 0$) term involves the (normal-ordered) field 
${\cal O}_k ({\bf r}) = h^k  ({\bf r})$ and governs
the asymptotic singularity (\ref{hk2}) with $x_k = x_{k,0} = -k \chi$. 
The higher terms correspond to fields with $k$ factors $h({\bf r})$ and 
$N$ powers of ${\bf \nabla}$. 
\newline
The leading terms of (\ref{hope}) are shown in Fig.~8. 
\begin{figure}[t]
\vspace*{-2cm}
\epsfig{file=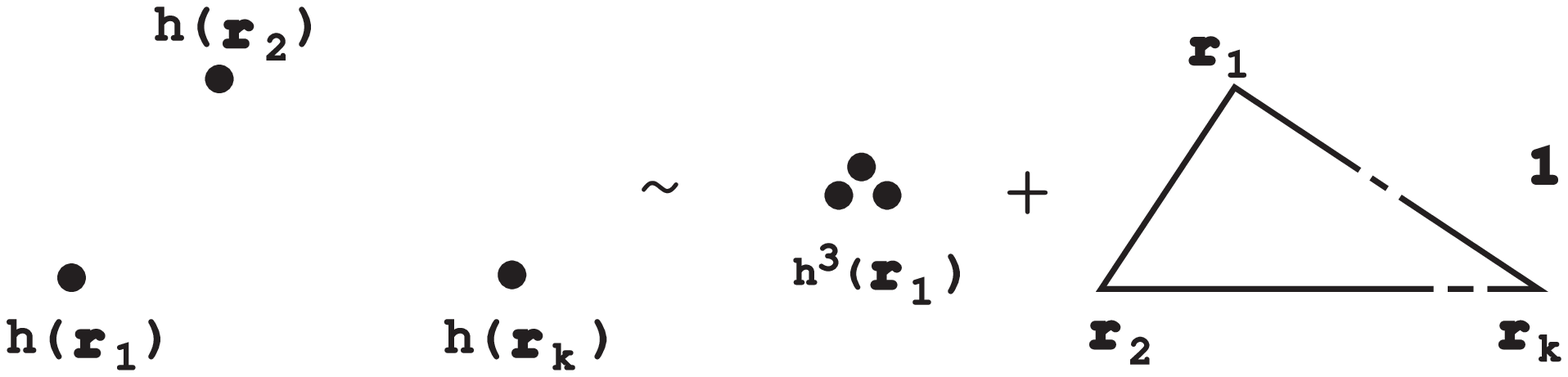,height=8cm}  

\vspace*{-2.3cm}
\baselineskip=12pt
{\small Fig. 8: Operator product expansion for a $k$-tuple of KPZ 
 height fields. The lines indicate the contraction of the fields, i.e., 
 the coupling to the identity $\bf 1$. Partial contractions would not
 contribute to connected correlation functions. Subleading singular
 and regular terms are omitted.}
\end{figure} 
It is useful to introduce the (normal-ordered) vertex fields 
\EQ
Z_q (\r) \equiv \exp [q h(\r)] \;,
\EEQ
which are the generating functions
of the fields $h^k (\r)$. Eq.~(\Ref{hope}) then implies
the operator product expansion
\EQ
Z_{q_1} (\r_ 1) Z_{q_2} (\r_ 2) =
     {\rm exp} \! \left ( \sum_{k,l} C_{k,l}^{\bf 1} w_1^k w_2^l \right ) 
     Z_{q_1 + q_2} (\r_ 1) 
     +  O ( C_{k,l}^{{\cal O} \neq 1}) \;,
\Label{Zope}
\EEQ
where  $C_{k,l}^{\cal O} \equiv 
      C_{k+l}^{\cal O} (0,\dots, 0, 
      \r_ {12}/|\r_ {12}|, \dots, \r_ {12}/|\r_ {12}|)$
with the first $k$ arguments equal to 0 and 
$w_i \equiv q_i |\r_ {12}|^\chi$~\cite{long}.
This is a generalization of Wick's theorem. In an
expansion of the exponential, each term 
$C_{k,l}^{\bf 1} w_1^k w_2^l$ represents
a contraction of $k+l$ fields $h$. Subleading singular terms 
(with positive-dimensional fields ${\cal O}$) and regular terms 
(with fields containing height gradients) are omitted. Eq.~(\Ref{Zope})
determines the asymptotic behavior of the vertex $n$-point functions,
\EQ 
\langle Z_{q_1} (\r_ 1) \dots Z_{q_n} (\r_ n) \rangle_t
\sim \exp \left ( \xi_t^\chi \sum_{i = 1}^n q_i \right ) \;. 
\EEQ
If $\sum_i q_i = 0$,
they  have a finite limit 
$\langle Z_{q_1} (\r_ 1) \dots 
         Z_{- q_1 \dots - q_{n-1}} (\r_ n) \rangle$. Since
these are precisely the vertex correlators that generate the height
difference correlation functions (\Ref{hdiff}) and since the vertex operator product 
expansion is analytic in the $q_i$, this leads back to the stationarity
condition (\Ref{vk}).

The operator product expansions (\ref{hope}) and (\Ref{Zope}) with the 
linear dimensions (\Ref{xkN}) are at the heart of the field theory for KPZ
systems. They describe the following structure of height correlations:
\newline
(a) The single-point amplitudes 
\EQ 
\langle h^k \rangle_t \sim \xi_t^{k \chi}
\label{mom1}
\EEQ
are the moments of a probability distribution,
\EQ
\langle h^k \rangle_t = \int \D h \, h^k P_{1,t} (h) \;.
\EEQ
The time dependence of the distribution takes the form 
\EQ
P_{1,t} (h) = \xi_t^{-\chi} \, {\cal P}_1 (h \xi_t^{-\chi} ) 
\EEQ
expressing global scaling of unsaturated growth. 
Recall that by normal-ordering Eq.~(\ref{KPZ}),
we have eliminated the nonuniversal part of the average height,
$\langle h \rangle_t \sim t$.
\newline
(b)  Since (\Ref{hope}) does not have
singular terms with negative-dimensional operators,
the local correlation functions (\ref{vk}) and (\ref{hdiff})
decouple from the global amplitudes (\ref{mom1}). In particular,
powers of the height difference $h_{12} \equiv h(\r_ 1) - h(\r_ 2)$
reach stationary expectation values for $\xi_t \to \infty$ determined
by the contraction term in (\ref{hope}),
\EQ
\langle h_{12}^k \rangle 
      \sim |\r_ {12}|^{ k \chi } \;.
\Label{hk}
\EEQ  
These can be written as moments of
a stationary probability distribution,
\EQ
\langle h_{12}^k \rangle = 
 \int \D h_{12} \, h_{12}^k \, P_2 (h_{12}, |\r_{12}|) \;,
\EEQ
which has the scaling form
\EQ
P_2 (h_{12}, r) = r^{-\chi} \, {\cal P}_2 (h_{12} r^{-\chi}) \;.
\EEQ
Eq.~(\Ref{hk}) 
can be verified exactly for $d = 1$ and seems to be consistent with
the presently available numerical data also in higher dimensions.

It is instructive to compare this structure with models of 
turbulence. Burgers' equation (\Ref{Burgers}) with force correlations 
\EQ
\overline{\eta(\r,t) \eta(\r',t')} 
     = \varepsilon r_0^2 \delta (t - t') \,
       \Delta \! \left ( \frac{|\r - \r'|}{r_0} \right )
\label{etaeta.2}
\EEQ
over large distances $r_0$ develops {\em multiscaling}: for example,
the longitudinal velocity difference moments 
\EQ
\langle [v_{\|} (\r_ 1) - v_{\|} (\r_ 2)]^k \rangle \sim
   |\r_ {12}|^{-k x_{{\bf v}} + \tilde x_k} \, r_0^{-\tilde x_k}
\Label{multisc}
\EEQ
have a $k$-dependent singular dependence on $|\r_ {12}|$ and $r_0$
for 
\EQ
r_0 / {\cal R} \ll |\r_ {12}| \ll r_0 \;,
\EEQ
where ${\cal R} \equiv r_0^{4/3} \varepsilon^{1/3} / \nu$ denotes the
Reynolds number~\cite{ChekhlovYakhot,BouchaudAl.dp}. This so-called 
inertial scaling regime should be compared to the KPZ strong-coupling
regime $\rc \ll |\r_ {12}| \ll \xi_t$. According to (\ref{multisc}),
local and global scaling properties are no longer decoupled.
Similar multiscaling is present in Navier-Stokes turbulence. 
Kolmogorov's famous argument predicts the exact scaling dimension of 
the velocity field, $x_{{\bf v}} = -1/3$, from dimensional analysis~\cite{K41}. This 
determines the scaling of the third moment in (\Ref{multisc})
since $\tilde x_3 = 0$. The higher exponents $\tilde x_4, \tilde x_5, \dots < 0$ cannot 
be obtained from dimensional analysis. Assuming the existence of
an operator product expansion (\Ref{vope}), the term (\Ref{multisc}) is
generated by the lowest-dimensional field $\tilde {\cal O}_k$ 
with a singular coefficient. A similar operator product expansion 
is discussed in ref.~\cite{Eyink}. Multiscaling thus implies
the existence of a (presumably infinite) number of composite fields with 
anomalous negative dimensions. For the velocity vertex fields
${\rm exp} [q v(r)]$ of Burgers turbulence in one dimension, 
Polyakov has conjectured an operator
product expansion similar to (\Ref{Zope})~\cite{Polyakov}.  
The distinguishing feature of 
KPZ surfaces is the absence of multiscaling expressed by (\ref{vk}).

\bigskip

\subsection{Dynamical anomaly and quantized scaling}

The operator product expansion (\Ref{Zope}) and the dispersion relation (\Ref{xkN})
have to be consistent with the underlying dynamics (\Ref{Pt}).  
However, as explained in ref.~\cite{Polyakov} for Burgers turbulence, the
equation of motion for the renormalized correlation functions
is quite subtle due to anomalies dictated by the operator product expansion. 
To exhibit the anomalies for the height correlations~\cite{Lassig.anomaly}, we  
introduce the smeared vertex fields
\EQ
Z_{q}^a (\r) \equiv 
       \exp \left ( q  \int d \r' \, \delta_a (\r - \r') h(\r')  \right)
\EEQ
(where $\delta_a (\r)$ is a normalized function with support in the 
sphere $|\r| < a$) and the abbreviations
$Z_i^a \equiv Z_{q_i}^a (\r_ i),  
 Z_i   \equiv Z_{q_i}   (\r_ i)$.

Using (\Ref{hn}) and (\Ref{Pt}), it
is straightforward to derive the equation of motion 
\EQ
\partial_t \, \langle Z_1^a \dots Z_n^a \rangle_t =
   \sum_{i = 1}^n q_i 
   \langle Z_1^a \dots {\cal J} \! Z_i^a \dots Z_n^a \rangle_t \;,
\Label{Znta}
\EEQ
where 
\EQ 
{\cal J} \! Z_i^a \equiv [q_i \sigma^2 \delta_a (0) + J(\r_ i)] Z_i^a \;.
\EEQ
The singularity structure of the current is determined by
(\Ref{hope}) and (\Ref{coeffas}):
\EQ
{\cal J}\! Z_i^a = g^* \hat Z_i + a^{2 \chi - 2} 
                   \left ( \sum_{k = 1}^{\infty} c_k a^{k \chi} q_i^k 
                   \right ) Z_i + O(a^\chi) 
\Label{JZ}
\EEQ
for $a, \rc \to 0$ with $a / \rc$ kept constant. 
The field
$\hat Z_q (\r) \equiv (\nabla h)^2 Z_{q} (\r)$ denotes the finite
part of the operator product 
$(\nabla h)^2 (\r) Z_q^a (\r)$ for $a \to 0$, and
$\hat Z_i \equiv \hat Z_{q_i} (\r_ i)$. 
The finite dissipation term $(\nabla^2 h) Z_{q_i} (\r_ i)$ 
becomes irrelevant in this limit since $\nu \sim a^\chi$. 
The singular part of (\Ref{JZ}) is a power series
in $q_i$ with asymptotically constant coefficients 
$c_1 = \sigma^{* 2} a^d \delta_a (0) 
       + \nu^* c_{1,1} + g^* c_{2,1}$ and 
$c_k = \nu^* c_{1,k} + g^* c_{2,k}$ for $k = 2,3,\dots$.
The terms of order $a^{(2 + k) \chi - 2}$ originate from contractions
$\nabla^2 h (\r_ i) h (\r_ 1') \dots h (\r_ k') \sim {\bf 1}$ and 
$(\nabla h)^2 (\r_ i) h (\r_ 1') \dots h (\r_ k') \sim {\bf 1}$. (Their
respective coefficients $c_{1,k}$ and $c_{2,k}$ are complicated integrals
involving the scaling functions in (\Ref{hope}), the regularizing functions
$\delta_a (\r_ i - \r_ j')$, and the ratio $a / \rc$.) 
Of course, divergent terms have to cancel
so that Eq.~(\Ref{Znta}) has a finite continuum limit
\EQ
\partial_t \, \langle Z_1 \dots Z_n \rangle_t =
   \sum_{i = 1}^n q_i 
   \langle Z_1 \dots {\cal J} \! Z_i \dots Z_n \rangle_t 
\Label{Znt}
\EEQ
with 
\EQ
{\cal J} \! Z_i = \lim_{a \to 0} {\cal J} \! Z_i^a \;.
\EEQ 
For generic values of $\chi$,
this implies $ {\cal J} \! Z_i = g^* \hat Z_i$. However, if $\chi$ satisfies
the condition (\Ref{quant}) for some integer $k_0$, the dissipation
current contributes an anomaly:
\EQ
{\cal J} \! Z_i = g^* \hat Z_i + \nu^* c_{1,k_0} q_i^{k_0} Z_i \;.
\Label{anomaly}
\EEQ

Eqs.~(\Ref{Znt}) and (\Ref{anomaly}) govern in particular the stationary
state of the surface. For $d = 1$, the stationary height distribution
$P[h]$ is known, 
\EQ
P \sim   {\rm exp} \left( - \frac{\sigma^2}{\nu}
         \int \D r \, (\nabla h)^2 \right ) \;.
\EEQ
It equals that of the linear theory, thus restoring the up-down symmetry
$ h(\r) - \langle h \rangle_t \to 
- h(\r) + \langle h \rangle_t$ broken by
the nonlinear term in (\Ref{KPZ}). The exponent $\chi = 1/2$ satisfies
(\Ref{quant}) with $k_0 = 2$ but the up-down symmetry forces the anomaly
to vanish ($c_{1,2} = 0$). In higher dimensions, this symmetry is
expected to remain broken in the stationary regime. The surface has
rounded hilltops and steep valleys, just like the upper side of a cumulus 
cloud (argued in ref.~\cite{Pelletier} to be a KPZ surface). 
Hence, the local slope of the surface is correlated with the relative 
height, resulting in nonzero odd moments 
$\langle (\nabla h)^2 (\r_ 1) [h(\r_ 1) - h(\r_ 2)]^k \rangle$.
However, this is consistent with Eqs.~(\Ref{Znt}) and (\Ref{anomaly})
only for odd values of $k_0$, where   
\EQ
\langle \hat Z_q (\r_ 1) Z_{-q} (\r_ 2) \rangle -
\langle \hat Z_{-q} (\r_ 1) Z_q (\r_ 2) \rangle 
    =  - (\nu^* / g^*) \, c_{1,k_0} q^{k_0} 
       \langle Z_q (\r_ 1) Z_{-q} (\r_ 2) \rangle 
\EEQ
and hence for odd values of $k \geq k_0$
\EQ
\langle (\nabla h)^2 (\r_ 1) [h(\r_ 1) - h(\r_ 2)]^k \rangle
    = - (\nu^* / g^*) \, c_{1,k_0} 
      \langle [h(\r_ 1) - h(\r_ 2)]^{k - k_0} \rangle \;.
\Label{direct}
\EEQ
The directedness of the stationary growth pattern thus requires 
a nonzero anomaly $c_{1,k_0}$ with an odd integer $k_0$. 
The roughness exponent is then determined by Eq.~(\Ref{quant}).
The values $k_0 = 3$ for $d = 2$ and $k_0 = 5$ for $d = 3$ 
give the exponents (\ref{chi2}) and (\ref{chi3}) quoted  below.
These appear to be consistent with the numerical 
results $\chi \approx 0.39$ and 
$\chi \approx 0.31$~\cite{TangAl.num,AlaNissilaAl.num}, 
respectively, and with the experimental value $\chi = 0.43 \pm 0.05$
for $d = 2$~\cite{PaniagoAl.exp}.

\bigskip

\subsection{Discussion}

The scaling of strongly driven surfaces has been determined by
requiring consistency of the effective large-distance field theory
subject to a few phenomenological constraints. 
These are the existence of a local operator product expansion
(\ref{hope}) and of a stationary state (\ref{vk}) that is directed
(i.e., has no up-down symmetry). The stationarity condition 
has an important consequence: the decoupling of local and global
scaling properties. The latter can be expressed by the time-dependent
height probability distribution at a single point, the former by
the stationary distribution of the height difference between nearby points.

The Galilei invariance of 
the dynamic equation conspires with these constraints to 
allow only discrete values of the roughness exponent in two and three
dimensions: $\chi = 2/ (k_0 + 2)$ with an odd integer $k_0$. Comparison
with numerical estimates then gives the exact values 
\EQ
\chi = 2/5, \hspace{1cm} z = 8/5 \hspace{1cm} \mbox{for $d = 2$}
\Label{chi2}
\EEQ
and
\EQ
\chi = 2/7, \hspace{1cm} z = 12/7 \hspace{1cm} \mbox{for $d = 3$} \;.
\Label{chi3}
\EEQ

The underlying solutions of the KPZ equation 
are distinguished by a dynamical anomaly in the strong-coupling regime:
the dissipation term contributes a finite part to the effective
equation of motion (\ref{Znt}) despite being formally irrelevant.
The anomaly manifests itself in identities like (\ref{direct}) 
between stationary correlation functions, which can be tested numerically.
 
There are at least three directions of future research where these
concepts and methods can be of use. As mentioned in the introduction,
the field of stochastic growth is rather diverse, and the theoretical
cousins of the KPZ equation could be analyzed in 
a similar way, aiming at a better understanding of these non-equilibrium universality
classes. In contrast to the KPZ
equation, some of these systems do show multiscaling~\cite{Krug.turbif},
which makes their field theoretic description certainly more complex.

Further applications to the theory of disordered systems are equally
important. The present theory should be extendable from a directed
string to higher-dimensional manifolds in a random medium.

Finally, this theory highlights both the theoretical similarities
and differences of surface growth and fluid turbulence. 
While it is quite surprising that the scaling exponents of an interacting
dynamical field theory can be predicted exactly in two and three
dimensions, there is one other such case in a related system:  the
exact Kolmogorov scaling of the third velocity difference moment in
Navier-Stokes turbulence. The higher velocity difference moments show
multiscaling (see Section 4.5). For the simpler case of Burgers turbulence,
dynamical anomalies are intrinsically connected to the multiscaling  
exponents~\cite{Lassig.Burgers}. This link is expected to be important
in a wider context of turbulence. 

Perhaps some of these fascinating scaling phenomena far from equilibrium
are not as inaccessible as they have appeared so far.

\bigskip

\section*{Appendix}
\renewcommand{\theequation}{A.\arabic{equation}}
\setcounter{equation}{0}
\addcontentsline{toc}{section}{Appendix}

This Appendix contains some details about the renormalization of the 
replicated string system (Section 3.1) and the dynamic functional (Section 4.2).
 
We start from the replica partition function
\EQ
Z = \int {\cal D} \phi {\cal D} \bar \phi 
      \exp \left [ - \int_0^{L_0} \D t_0 \int_0^R \D \r 
                   \left ( 
                   \bar \phi (\partial_{t_0} - \nabla^2 ) \phi +
                   g_0 \bar \phi^2 \phi^2
                   \right ) 
           \right ] \hs,
\Label{F2quant0}           
\EEQ 
with the effective coupling constant (\Ref{g0}),
obtained from (\ref{F2quant}) by the change of variables 
$t_0 \to \beta_0^{-1} t_0$.
The field $\phi$ is real-valued with the constraint $\phi > 0$; the field 
$\bar \phi$ is purely imaginary~\cite{Wiese.kpz2}. The normal-ordered
interaction $ g_0 \bar \phi^2 \phi^2 $ conserves the number of strings.
To evaluate (\ref{F2quant0}), we take periodic boundary conditions
for the components of $\r$ and define 
\EQ
Z_p = \langle p,\r | Z | p \rangle
\label{if}
\EEQ
with the initial state
$ | p \rangle \equiv (1/p!) 
  [\int \D \r \bar \phi (\r)]^p | 0 \rangle $
at $t_0 = 0$ and the final state
$ \langle p,\r | \equiv \langle 0 | \phi^p (\r) $
at $t_0 = L_0$. Correlation functions $\langle \dots \rangle_p$ are defined 
in a similar way. With an appropriate normalization of the propagator,
these boundary conditions normalize
$ Z_p (g_0,L_0 = 0,R) = Z_p (g_0 = 0,L_0,R) = 1$ for any value of $p$.

The Casimir amplitude is defined in analogy to (\Ref{C0}) and (\Ref{C0g}).
Its expansion in terms of the dimensionless coupling 
$u_0 \equiv g_0 R^{2y_0}$, 
\EQ
\Delta {\cal C}_p (u_0) = - R^{-2} \sum_{N=1}^\infty 
  \frac{(-g_0)^N}{ N!} \int \D t_{02} \dots \D t_{0N} 
  \langle \Phi_2 (0) \Phi_2 (t_{02}) \dots \Phi_2 (t_{0N}) \rangle_p^c \;,
\Label{Cpseries}
\EEQ
is a sum involving connected pair field correlations in the $p$-string
sector of the unperturbed theory ($u_0 = 0$). The integrals in
Eq.~(\Ref{Cpseries}) are infrared-regularized by the system width $R$; their
ultraviolet singularities are determined by the short-distance
structure of the pair field correlations and have to be absorbed into
the coupling constant renormalization.  Hence consider the asymptotic
scaling of the $N$-point  function $ \langle \Phi_2 (t_{01})) \dots
\Phi_2 (t_{0N}) \rangle_p^c $ as the points $t_{01}, \dots,
t_{0N}$ approach each other. More precisely, we define
$t_0$ and $\tau_{jk}$ by 
$t_{0j} - t_{0k} = t_0 \, \tau_{jk}$ and let
\EQ
 t_0 / R^2 \to 0 
\EEQ
with $\tau_{jk}$ and the ``center of mass''  
$t_0' = N^{-1} \sum_{j = 1}^N t_{0j}$ remaining fixed. 
The asymptotic scaling is given by the $p$-independent 
operator product expansion
\EQ
\Phi_2 (t_{01}) \dots \Phi_2 (t_{0N}) =  
\sum_{m = 2}^{N + 1} t_0^{- (N - m + 1) d / 2} 
   \left [ C_N^m (\tau_1, \dots, \tau_{N-2}) \, \Phi_m (t_0') + \dots 
   \right ] \hs,
\Label{OPA}
\EEQ
where $C_N^m$ are scaling functions of the $N - 2$ linearly independent 
distance ratios $\tau_{jk}$, the dots denote subleading terms down by 
positive integer powers of $t_0 / R^2$, and we have defined the 
normal-ordered $m$-string contact fields 
\EQ
\Phi_m (t_0) \equiv \int \D \r \, \bar \phi^m (\r,t_0) \phi^m (\r,t_0) \hs.
\EEQ
For $m = N = 2$, (\ref{OPA}) reduces to (\ref{ope0}); i.e., $C_2^2 = C_0$.
The operator product expansion dictates the leading singularities
of the integrals in (\ref{Cpseries}),
\EA
\lefteqn{\BS \BS
\int \D t_0 \, t_0^{N-2} \prod_{l = 1}^{N - 2} \D \tau_l \;
\langle \Phi_2 (t_{01}) \dots \Phi_2 (t_{0N}) \rangle_p = } \nonumber
\\ & & 
\sum_{m = 2}^{N + 1} 
\int J_N^m \, t_0^{m - 3 + y_0 (N - m + 1)} \D t_0 \; 
\langle \Phi_m (t'_0) \rangle_p + \dots
\Label{IN}
\EEA
with
\EQ
J_N^m = \int \prod_{l = 1}^{N - 2} \D \tau_l \, C_N^m (\tau_1, \dots, \tau_{N-2}) \hs.
\Label{J}
\EEQ 

The diagrammatic representation of the perturbation series (\ref{Cpseries})
is discussed in detail in ref.~\cite{BundschuhLassig.kpz}. At any integer
value of $p$, the series is somewhat simplified since 
\EQ
\langle \Phi_m \rangle_p = 0 \hspace{1cm} \mbox{for $m > p$.} 
\label{Phim}
\EEQ
For $p$ noninteger, all values
of $m$ contribute, e.g., to (\ref{IN}). Hence, the series becomes complicated 
in the random limit $p \to 0$. Its pole structure at $y_0 = 0$, however,
remains simple. Consider  the term
in (\ref{OPA}) with $N = m = 2$,  corresponding to the diagram of 
Fig.~9(a). The loop has the value 
$ R^{2 y_0} c(y_0)/y_0 $ with 
$ c(y_0) = C_0(y_0) + O(y_0) $. The pole  originates from the universal
short-distance singularity in (\ref{IN}), while the finite part depends also on
the infrared regularization. Hence, we obtain to one-loop order
\EQ
\Delta {\cal C}_p (u_0) = - R^d \langle \Phi_2 \rangle_p \;  
                     u_0 \left (1 - \frac{c}{y_0} u_0 \right )
+ 0 (y_0^0 u_0^2, u_0^3) 
\Label{Cpsing}
\EEQ
with 
$ R^d \langle \Phi_2 \rangle_p = p (p - 1) / 4 $. The pole 
can be absorbed into the definition $u_P = \Z_P u_0$ with the
$\Z$-factor (\ref{Z01}), which leads to the beta function (\ref{beta01})
and the field renormalization $\Phi_P = \tilde \Z_P \Phi$ with (\ref{Z01tilde}). 

\begin{figure}[t]
\psfig{file=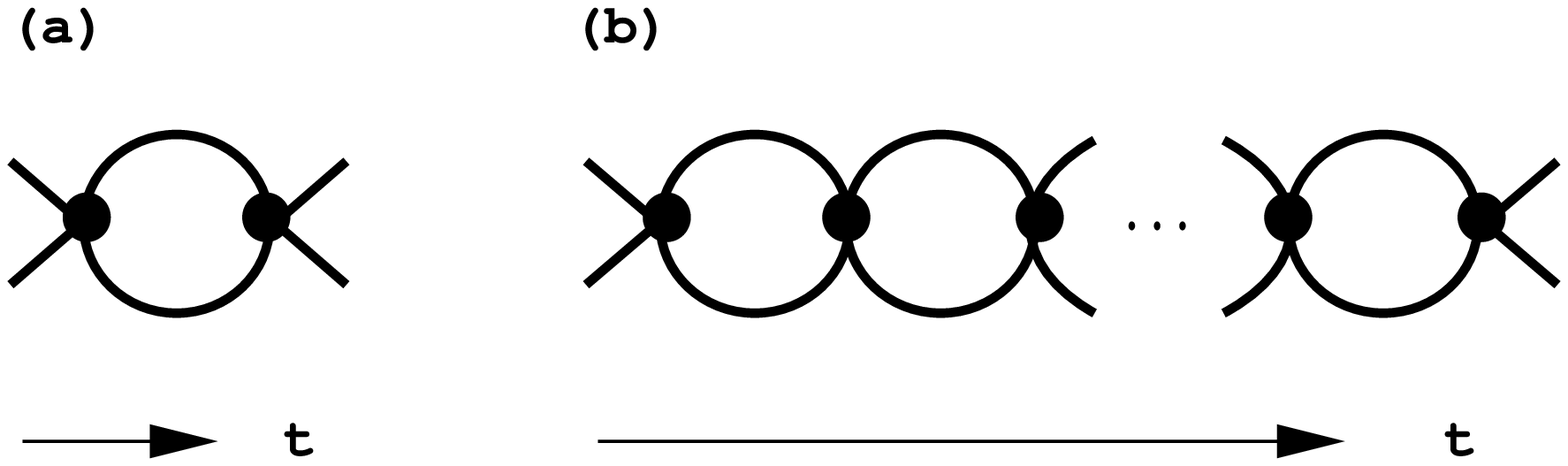,width=10cm}

\vspace{12pt}
\baselineskip=12pt
{\small Fig. 9: Singular diagrams contributing to the finite-size 
free energy $\Delta C_p(u_0)$ in the 
expansion (\Ref{Cpseries}).  
The lines denote unperturbed single-line propagators 
$\langle  \bar \phi (\r_1, t_{01}) \phi (\r_2, t_{02}) \rangle $, 
the dots pair contact vertices $\Phi_2 (t_0)$.
(a) One-loop diagram containing a primitive pole in $y_0$. 
(b) $N$-loop diagram containing a pole in $y_0$ of order $N - 1$.}
\end{figure}

It is not difficult to discuss higher orders.
The ultraviolet singularities of the $N$-th order integral (\Ref{IN}) are contained
in the coefficients $J_N^m$ or arise from the integration over $t_0$. In
the first case, they are due to a {\em proper subdiagram} and hence
already absorbed into the renormalized coupling constant at lower order.
Only the divergences from the integration over $t_0$, with $J_N^m$
denoting the regular part of (\Ref{J}), may contribute to the {\em
primitive singularity} at order $N$. Inspection of (\Ref{IN}) then
shows that a pole at $y_0 = 0$ only appears for $m = 2$. However, there
is only one diagram per order of this kind, which is shown in Fig.~9(b). 
Since this diagram factorizes into loops of the kind of Fig.~9(a), it
contributes a pole in $y_0$ of order $N - 1$. Therefore the pole at
order $N = 2$ is the only primitive singularity in the 
series (\Ref{Cpseries}) for the free energy; analogous arguments apply to the
expansions of the contact field correlation functions. It follows that 
the renormalization can be carried out to all orders; see
also refs.~\cite{Duplantier.oneloop,RajBhatt.oneloop}.  The loops of
Fig.~9 form a geometric series. Summing this series defines the 
$\Z$-factors~\cite{LassigLipowsky.Altenberg}
\EQ
\Z_P = 1 - \frac{c}{y_0} u_P \;, \HS
\tilde \Z_P = \frac{\D u_0}{\D u_P} 
            = \left (1 - \frac{c}{y_0} u_P \right )^{-2} 
\Label{Z02}
\EEQ
and the flow equation $\dot u_P = y_0 u_P - c u_P^2$.
The Casimir amplitude $\Delta {\cal C}_p$ becomes 
a regular function of $u_P$,
\EQ
\Delta {\cal C}_p (u_P) = \frac{p(p - 1)}{4} u_P + O(u_P^2) \;.
\label{Cpreg}
\EEQ

The coupling $u_P$ defined by (\ref{Z02}) is not unique. Any family of
diffeomorphisms $u_P \to u_P' (u_P, y_0)$ with fixed point 
$u_P'(0, y_0) = 0$ will preserve the regular functional dependence
of observables such as (\ref{Cpreg}). In particular, the linear 
transformation $u_P \to (c/C_0) u_P$ leads to the flow equation
(\ref{beta0}) and the $\Z$-factors 
\EQ
\Z_P = \frac{c}{C_0} \left ( 1 - \frac{C_0}{y_0} u_P \right ) \;,
\HS 
\tilde \Z_P = \frac{C_0^2}{c^2} \left (1 - \frac{C_0}{y_0} u_P \right )^{-2} \;.
\Label{Z0}
\EEQ
We may also substitute  $C_0(y_0)$
by $C_0(y_0 \!=\! 0)$ in (\ref{Z0}) and (\ref{beta0}) so that  
the dependence of the flow equation on $y_0$ is contained only in
the linear term.  This scheme
has been called ``minimal subtraction'' in ref.~\cite{Lassig.kpz}. However,
the $\Z$-factors (\ref{Z0}) still have a regular part as $y_0 \to 0$
generated by the function $c(y_0)$. Eliminating this part requires a 
non-linear transformation of $u_P$ and introduces cubic and higher order 
terms into the flow equation~\cite{Wiese.kpz1}. Of course,
these terms are spurious; i.e., they do not generate higher order 
terms in expressions like (\ref{x0star}) for local observables. (In the 
present case of a single coupling constant $u_P$, the value of the $C_0$
drops out of these expressions as well. In more general cases with
several coupling constants, ratios of operator product
coefficients do play a role; see the discussion in ref.~\cite{Cardy.book}.)

The replica trick is unproblematic within perturbation theory, since it
reduces to convenient bookkeeping of the averaging over disorder.
Eqs.~(\Ref{Z0}), (\Ref{beta0}) are  independent of $p$, and 
in Eq.~(\Ref{Cpreg}), the dependence on $p$ reduces to the combinatoric
prefactor. Hence, the random limit 
$\Delta \overline{\cal C} = \lim_{p \to 0} {\cal C}_p / p$ is trivial 
and leads to Eqs.~(\Ref{Cbarreg}) and (\Ref{Cbarstar}).

The exponents (\ref{expstar}) can also be obtained in a different way. They
follow from the fact that the two-string interaction $\bar \phi^2 \phi^2$
does not renormalize the ``mass term'' $\bar \phi \phi$ at any order.
Consider the (normal-ordered) density field 
$\Phi (t_0) \equiv \bar \phi \phi (\r \! = \! 0, t_0)$,
which has the two-point function 
\EQ 
\langle \Phi (t_0) \Phi (t'_0) \rangle_p^c (u_0,R) =  
\sum_{N = 0}^\infty \frac{(-g_0)^N}{N!}  
  \int \D t_{01} \dots \D t_{0N} 
  \langle \Phi (t_0) \Phi (t_0') \Phi_2 (t_{01}) \dots \Phi_2 (t_{0N}) 
  \rangle_p^c \hs.
\Label{Rpert}
\EEQ
Its short-distance asymptotics is related to the return probability of 
strings to the origin $ \r = 0 $. In the linear theory, 
\EQ
\langle \Phi (t_0) \Phi (t'_0) \rangle_p^c 
\sim |t_0 - t_0'|^{- d  \zeta_0} 
\Label{return}
\EEQ
for $ |t_0 - t_0'| / R^2 \ll 1$. Any perturbative correction to this exponent 
arises from the renormalization of the fields $ \Phi (t_0)$ and $ \Phi (t_0') $.
The renormalization of $\Phi (t_0)$ is due to a short-distance
coupling of the form
\EQ
\Phi (t_0) \Phi_2 (t_{01}) \dots \Phi_2 (t_{0N}) =
  t_0^{- N d / 2} C_N^1 (\tau_1, \dots, \tau_{N - 1}) \, \Phi (t_0) + \dots
\hs
\EEQ
for $t_0 / R^2 \to 0$ (with 
$t_{0j} - t = t_0 \tau_{j}$, $t_{0j} - t_{0k} = t_0 \tau_{jk} $ 
for $ j,k  = 1, \dots, N $, and  $\tau_1, \dots, \tau_{N-1}$ 
denoting a basis of the fixed ratios $\tau_{j}, \tau_{jk}$),
and there is a corresponding expression for $\Phi (t_0')$.   
However, it is obvious that the product on the l.h.s.~couples only to
contact fields of at least two lines, and therefore 
$C_N^1 = 0$ at all orders $N$. The singularity (\ref{return}) remains
unchanged, $\zeta^* = \zeta_0 = 1/2$. The strong-coupling 
form (\ref{ope1}) of the singularity cannot be reached by perturbation theory.

Now we turn to the dynamic path integral. 
After the change of variables 
$t_0        \to  \nu_0 t_0$,
$h_0        \to  (\nu_0 / \sigma_0^2)^{1/2} h_0$, and
$\tilde h_0 \to  (\sigma_0^2 / \nu_0)^{1/2} \tilde h_0$, 
(\Ref{Zdyn}) takes the normal form
\EQ
Z = \int {\cal D} h_0 {\cal D} \tilde h_0 
     \exp \left [ - \int \D \r \D t \left (
    - \frac{1}{2} \tilde h_0^2 + \tilde h_0 \left (
    \partial_{t_0} h_0 - \frac{1}{2} \nabla^2 h_0 
    - \frac{\lambda_0}{2} (\nabla h_0)^2 \right ) \right )
          \right ]
\Label{Zdyn0}
\EEQ
with $\lambda_0^2 = -g_0$. As discussed in ref.~\cite{Wiese.kpz2},
the transformation (\ref{HopfCole2}) leads back to (\ref{F2quant0}),
rendering dynamical perturbation theory identical to replica perturbation
theory. The path integral (\ref{F2quant0}) has to be supplemented 
with boundary conditions corresponding to the random  limit $p \to 0$. 
In this limit, the boundary
condition (\ref{if}) describes the one-point-function of 
an initially flat surface, 
$\langle h(\r, t_0 \! = \! 0) \rangle = 0$. 
At any integer value of $p$, the dynamical
correlations are artificially truncated,
$ \langle \tilde h(\r_1, t) \dots \tilde h (\r_m, t) h (\r, L) \rangle_p = 0$
for all $m > p$ by (\ref{Phim}). 

In the context of growth dynamics, the representation (\ref{Zdyn0}) 
has an advantage over (\ref{F2quant0}): it contains explicitly the
most interesting observable, the height field $h$. The price to pay
is that perturbation theory becomes more complicated. The differences
to replica perturbation theory are two-fold: 
(i) There is the cubic vertex $\tilde h (\nabla h)^2$. 
(ii) The vertex $\tilde h^2$, the image of the replica interaction 
$\bar \phi^2 \phi^2$, no longer has incoming lines. These vertices
can therefore be integrated out, giving the standard 
diagrammatics with the two kinds of propagators (\ref{G0}) and 
(\ref{hh0}). 

We now discuss the renormalization of (\ref{Zdyn0}) and show its 
equivalence to replica renormalization up to one-loop order.
The real-space response function has the diagrammatic expansion shown 
in Fig.~10(a). To order $u_0^2$, the expansion reads 
\EQ
R^d \langle \tilde h_0 (\r,t_0)  h_0(\r, t_0 + R^2) \rangle (u_0)  = 
R^d \langle \tilde h_0 (\r,t_0)  h_0(\r, t_0 + R^2) \rangle (0)
+ \, O (y_0^0 u_0, u_0^2) \hs.
\EEQ
The one-loop diagram does not have a pole at $d = 2$ since its short-distance 
singularity cancels with a geometric factor $2 - d$ 
\cite{SunPlischke.kpz,FreyTaeuber,Wiese.kpz1}. Constructing the strong-coupling fixed 
point in $d = 1$ requires taking into account the resulting finite 
renormalization of the response function~\cite{FreyTaeuber}. 
For the critical fixed point above $d = 2$, however, it can be ignored.  
\begin{figure}[t]
\psfig{file=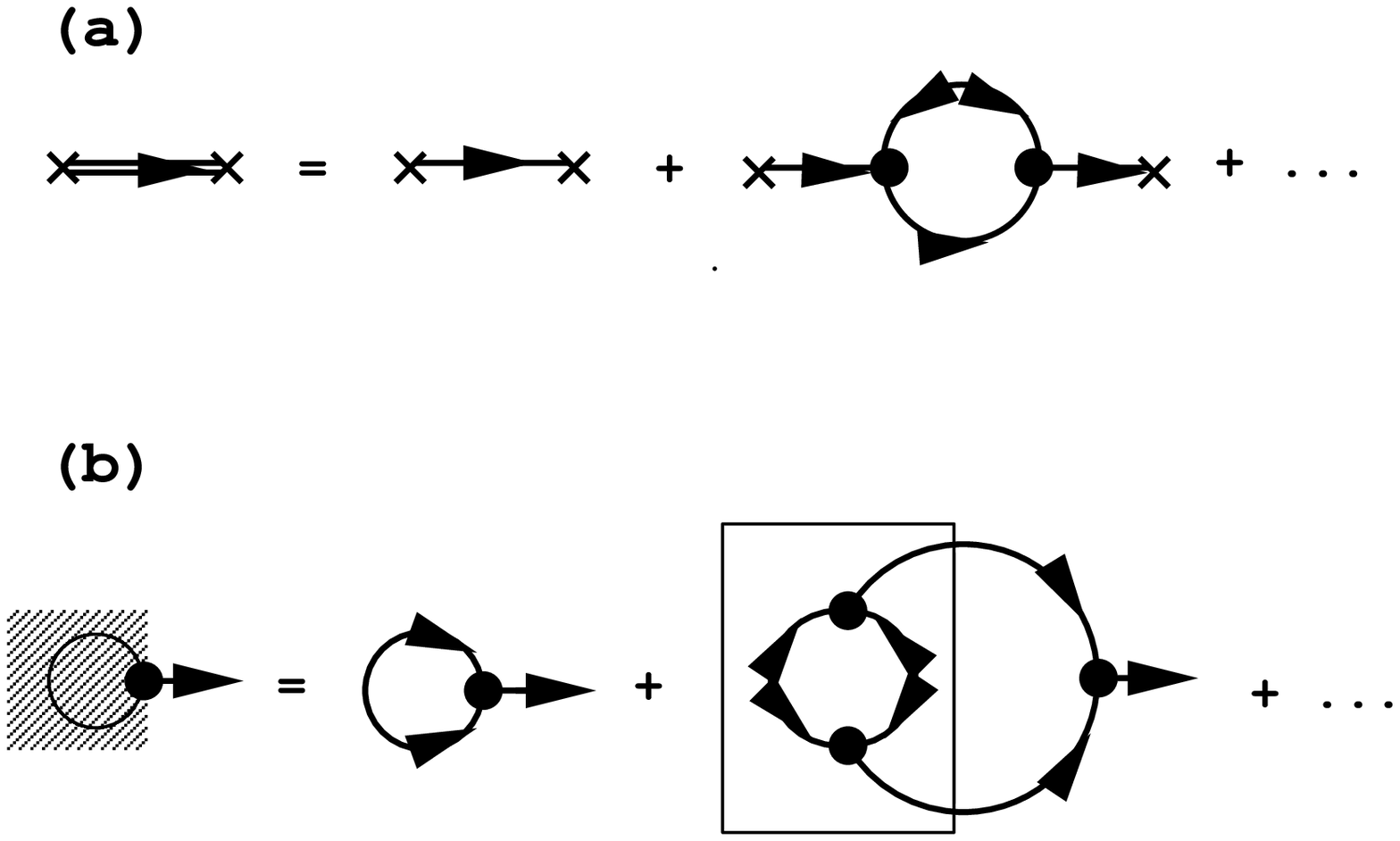,width=8cm}  

\vspace{12pt}
\baselineskip=12pt
{\small Fig. 10: Diagrammatic expansions generated by the dynamic functional
(\Ref{Zdyn}). 
The lines with one and two arrows denote the unperturbed  response function
(\Ref{G0}) and the unperturbed correlation function (\Ref{hh0}), respectively.
Dots represent the vertices $\tilde h (\nabla h)^2$;  
each incoming line to a vertex has to be differentiated 
with respect to $\r$. 
(a)~Response function 
$\langle  \tilde h (\r_1,t_{01}) h(\r_2,t_{02})  \rangle$. The one-loop diagram is
regular at $y_0 = 0$.
(b)~Stationary growth rate $\langle \partial_{t_0} h_0 \rangle_R$.
The boxed subdiagram contains a simple pole at $y_0 = 0$. A further
diagram with a loop as in (a) is regular at $y_0 = 0$ and has been
omitted.}
\end{figure}

The expansion for the growth velocity is shown in Fig.~10(b). The tadpole
diagram at order $\lambda_0$ has the form
\EQ 
\lambda_0 [ I(R) - \lim_{R \to \infty} I(R) ]   
= - \frac{\lambda_0}{2} \, R^{-d}
\label{IR}
\EEQ
with 
\EQ
I(R) = \int \D \r' \D t_0' \, 
 [\nabla_\r \langle \tilde h_0 (\r',t_0') h(\r,t) \rangle_R (u_0 \!=\! 0) ]^2
\;;
\Label{h1}
\EEQ
the second term in (\Ref{IR}) is generated by the normal-ordering 
(\Ref{normalord}) and cancels the ultraviolet divergence of $I(R)$. 
At order $\lambda_0^3$, the boxed subdiagram contributes a pole
originating from the integration region where the two
vertices approach each other. Hence, we have to this order
\EQ
R^{2 - \chi_0} \langle \partial_{t_0} h_0 \rangle_R ( u_0) =         
     - \frac{1}{2} \, u_0^{1/2} 
     \left ( 1 -  \frac{C_0}{y_0} u_0 + O(y_0^0 u_0, u_0^2) \right ) \;. 
\EEQ
The pole can be absorbed into the definition of the 
variables 
$h_P = \Z_{Ph} h_0$, 
$\tilde h_P = \Z_{Ph}^{-1} \tilde h_0$, 
$t_P = \Z_{Pt} t_0$, 
$u_P = \Z_P u_0$, with
\EA
\Z_{Ph} (u_P) & = & 1 + \frac{C_0}{2 y_0} u_P + O(u_P^2) \hs,
\Label{ZMh} \\ 
\Z_{Pt} (u_P) & = & 1 + O(u_P^2) \hs,
\Label{ZMt}
\\
\Z_P (u_P) & = & 1 - \frac{C_0}{y_0} u_P + O(u_P^2) \hs.
\Label{ZM}
\EEA
This reparametrization  respects (\Ref{R}) and renders both the response function 
and the growth rate regular as $y_0 \to 0$,
\EA
R^d \langle \tilde h_P (\r,t_P)  h_P(\r, t_P + R^2) \rangle (u_P)  & = & 
R^d \langle \tilde h_0 (\r,t_0)  h_0(\r, t_0 + R^2) \rangle (0) 
\nonumber \\
& &  +\, O (y_0^0 u_P, u_P^2) \;,
\EEA
\EQ
R^{2 - \chi_0} \langle \partial_{t_P} h_P \rangle_R (u_P) = 
- \frac{1}{2} \, u_P^{1/2} (1 + O (y_0^0 u_P, u_P^2)) \hs.
\EEQ
It leads to the beta function (\Ref{beta01}) and to
the exponents
\EQ 
\chi^\star = 2 - z^\star  = 
\chi_0 - \dot{u}_P \left. \frac{\D}{\D u_P} \log \Z_{Ph} (u_P) 
                   \right |_{u_P^\star} = O(y_0^2) \hs.
\EEQ

\bigskip

\section*{Acknowledgments}

This article is an extended version of my habilitation thesis (University of
Potsdam, 1997). I wish to express my warmest thanks to Reinhard Lipowsky  for
support and encouragement. Sections 3.2, 3.3, 3.5, 3.6, and 4.4 result from a
very enjoyable collaboration with Harald Kinzelbach.
Many people have generously shared their insights with me; I am  grateful in
particular to Ralf Bundschuh, John L. Cardy, Christin Hiergeist, Terence Hwa, 
Harald Kallabis,  Harald Kinzelbach, Thomas Nattermann, Reinhard Lipowsky,
Lei-Han Tang, Karen Willbrand, and Dietrich Wolf.




\begin{thebibliography}{999}
\addcontentsline{toc}{section}{Bibliography}

{\small
\baselineskip=13pt
\itemsep=-3pt





\bibitem{Cardy.book}
J.L. Cardy, Scaling and Renormalization in Statistical Physics, 
Cambridge University Press (1996). 


\bibitem{ItzyksonDrouffe.book}
C. Itzykson and J.-M. Drouffe, Statistical Field Theory, vol.~2,
Cambridge University Press (1989).


\bibitem{KrugSpohn.review}
J. Krug and H. Spohn, {\em in} Solids Far From Equilibrium, ed. 
C. Godr\`eche, Cambridge University Press (1990). 


\bibitem{HalpinHealeyZhang.review}
T. Halpin-Healy and Y.-C. Zhang, Physics Reports {254}, 215 (1995).


\bibitem{HerrastiAl.gold}
P. Herrasti et al., Phys. Rev. A 45, 7440 (1992).


\bibitem{KahandraAl.electrochem}
G. Kahanda et al., Phys. Rev. Lett. 68, 3741 (1992).


\bibitem{BakAl.ff}
P. Bak, K. Chen, and C. Tang, Phys. Lett. A 147, 297 (1990).


\bibitem{DrosselSchwabl.ff}
B. Drossel and F. Schwabl, Phys. Rev. Lett. 69, 1629 (1992).


\bibitem{KPZ}
M. Kardar, G. Parisi, and Y.C. Zhang, Phys. Rev. Lett. 56, 889 (1986).


\bibitem{WolfVillain.cons}
D. Wolf and J. Villain, Europhys. Lett. 13, 389 (1990).


\bibitem{Krug.review}
J. Krug, {\em in} Scale Invariance, Interfaces, and Non-Equilibrium
Dynamics, ed. A. McKane et al., Plenum, New York (1995).


\bibitem{ImbrieSpencer}
J.Z. Imbrie and T. Spencer, J. Stat. Phys. {52}, 609 (1988).


\bibitem{CookDerrida}
J. Cook and B. Derrida, Europhys. Lett. {10}, 195 (1989);
J. Phys. A {23}, 1523 (1990).


\bibitem{EvansDerrida}
M.R. Evans and B. Derrida, J. Stat. Phys. {69}, 427 (1992). 


\bibitem{TangNattermannForrest}
L.-H. Tang, T. Nattermann, and B.M. Forrest, Phys. Rev. Lett. 65,
2422 (1990).


\bibitem{NattermannTang}
T. Nattermann and L.-H. Tang, Phys. Rev. A 45, 7156 (1992).


\bibitem{FreyTaeuber}
E. Frey and U.C. T\"auber, Phys. Rev. E 50, 1024 (1994).


\bibitem{Lassig.kpz}
M. L\"assig, Nucl. Phys. B448, 559 (1995).


\bibitem{FNS}
D. Forster, D.R. Nelson, and M. Stephen, Phys. Rev. {A 16}, 732
(1977).


\bibitem{GwaSpohn.Bethe}
L.-H. Gwa and H. Spohn, Phys. Rev. Lett. 68, 725 (1992).


\bibitem{KimKosterlitz.num}
J.M. Kim and J.M. Kosterlitz, Phys. Rev. Lett.  62, 2289 (1989).


\bibitem{ForrestTang.num} 
B.M. Forrest and L.-H. Tang, Phys. Rev. Lett.  64, 1405 (1990).


\bibitem{KimAl.num}
J.M. Kim, J.M. Kosterlitz, and T. Ala-Nissila, J. Phys. A 24,
5569 (1991).


\bibitem{KimMooreBray.num}
J.M. Kim, M.A. Moore, and A.J. Bray, Phys. Rev. A 44, 2345 (1991).


\bibitem{TangAl.num}
L.-H. Tang, B.M. Forrest, and D.E. Wolf, Phys. Rev. A {45}, 7162
(1992).


\bibitem{AlaNissilaAl.num}
T. Ala-Nissila et al., J. Stat. Phys {72}, 207 (1993).


\bibitem{AlaNissila.ucd}
T. Ala-Nissila, Phys. Rev. Lett. 80, 887 (1998).


\bibitem{Kim.ucd}
J.M. Kim, Phys. Rev. Lett. 80, 888 (1998).


\bibitem{HalpinHealey.fren}
T. Halpin-Healy, Phys. Rev. Lett. 62, 445 (1989);
Phys. Rev. A {42}, 711 (1990).


\bibitem{NattermannLeschhorn.fren}
T. Nattermann and H. Leschhorn, Europhys. Lett. {14}, 603 (1991).



\bibitem{MooreAl.modecoupling}
M.A. Moore et al.,  Phys. Rev. Lett. {74}, 4257 (1995),
and references therein.


\bibitem{MaunukselaAl.paper}
J. Maunuksela et al., preprint (1998).


\bibitem{PaniagoAl.exp}
R. Paniago et al., Phys. Rev. B 56, 13442 (1997). 


\bibitem{MeakinAl.scal}
P. Meakin et al., Phys.Rev. A 34, 3390 (1986).


\bibitem{BouchaudAl.dp}
J.P. Bouchaud, M. M\'ezard, and G. Parisi, Phys. Rev. E {52}, 3656
(1995). 


\bibitem{HuseHenley.paths}
D.A. Huse and C.L. Henley, Phys. Rev. Lett. 54, 2708 (1985).


\bibitem{Kardar.dp}
M. Kardar, Phys. Rev. Lett. 55, 2235 (1985);
Nucl. Phys. B 290, 582 (1987).


\bibitem{FisherHuse.paths}
D.S. Fisher and D.A. Huse, Phys. Rev. B 43, 10728 (1991).


\bibitem{FreyAl.modecoupling}
E. Frey, U.C. T\"auber, and T. Hwa, Phys. Rev. E 53, 4424 (1996).


\bibitem{ForgacsAl.DG}
G. Forgacs, R. Lipowsky, and T.M. Nieuwenhuizen, 
{\em in} Phase transitions and
Critical Phenomena, Vol. 14,  ed. C. Domb and J.L. Lebowitz,  
Academic Press, London (1991).


\bibitem{LassigLipowsky.depinning}
M. L\"assig and R. Lipowsky, Phys. Rev. Lett. 70, 1131 (1993).


\bibitem{Lassig.fermions}
M. L\"assig, Phys. Rev. Lett.  73, 561 (1994).


\bibitem{LassigLipowsky.Altenberg}
M. L\"assig and R. Lipowsky, {\em in} Fundamental Problems of
Statistical Mechanics VIII, Elsevier, Amsterdam (1994). 


\bibitem{BundschuhLassig.kpz}
R. Bundschuh and M. L\"assig, Phys. Rev. E 54, 304 (1996). 


\bibitem{KinzelbachLassig.depinning}
H. Kinzelbach and M. L\"assig, J. Phys. A 28, 6535 (1995). 


\bibitem{KinzelbachLassig.fermions}
H. Kinzelbach and M. L\"assig, Phys. Rev. Lett. 75, 2208 (1995).


\bibitem{LassigKinzelbach.ucd}
M. L\"assig and H. Kinzelbach, Phys. Rev. Lett. 78, 903 (1997).


\bibitem{LassigKinzelbach.reply}
M. L\"assig and H. Kinzelbach, Phys. Rev. Lett. 80, 889 (1998).


\bibitem{Lassig.anomaly}
M. L\"assig, Phys. Rev. Lett. 80, 2366 (1998).










 \bibitem{Jayapakrash.TSK}
C. Jayaprakash, C. Rottmann, and W.F. Saam, Phys. Rev. B 30, 6549 (1984).


\bibitem{NelsonAl.fluxlines}
D.R. Nelson, Phys. Rev. Lett. 60, 1973 (1988); \\
D.R. Nelson and H.S. Seung, Phys. Rev. B 39, 9153 (1989).


\bibitem{CuleHwa.tribology}
D. Cule and T. Hwa, Phys. Rev. Lett. 77, 278 (1996).


\bibitem{HwaLassig.dna}
T. Hwa and M. L\"assig, Phys. Rev. Lett. 76, 2591 (1996).


\bibitem{DrasdoAl.global}
D. Drasdo, T. Hwa, and M. L\"assig, preprint physics/9802023.


\bibitem{HwaLassig.local1}
T. Hwa and M. L\"assig, {\em in} Proceedings of the second annual 
conference on computational molecular biology (RECOMB 98),
ACM Press, New York (1998).


\bibitem{DrasdoAl.local2}
D. Drasdo, T. Hwa, and M. L\"assig, {\em in} Proceedings of the sixth 
international conference on intelligent systems for molecular biology 
(ISMB 98), AAAI Press, Menlo Park, California (1998).


\bibitem{Olsen}
R. Olsen, T. Hwa, and M. L\"assig, {\em in} Proceedings of the 
Pacific Symposium on Biocomputing '99 (PSB 99), to appear. 



\bibitem{Lipowsky.lines}
R. Lipowsky,  Europhys. Lett. 15, 703 (1991).


\bibitem{Duplantier.oneloop}
B. Duplantier, Phys. Rev. Lett. 62, 2337 (1989).


\bibitem{RajBhatt.oneloop}
J.J. Rajasekaran and S.M. Bhattacharjee, J. Phys. A 24, L371 (1991).


\bibitem{fermions}
P.G. de Gennes, J. Chem. Phys. 48, 2257 (1968); \\
V.L. Pokrovski and A.L. Talapov, Phys. Rev. Lett. 42, 65 (1979).


\bibitem{RedfieldZangwill}
A.C. Redfield and A. Zangwill, Phys. Rev. B 46, 4289 (1992). 


\bibitem{SongMochrie}
S. Song and S.G.J. Mochrie, Phys. Rev. Lett. 73, 995, (1994);
Phys. Rev. B 51, 10068 (1995). 


\bibitem{Lassig.vicinal}
M. L\"assig, Phys. Rev. Lett. 77, 526 (1996).


\bibitem{vanDijkenAl.SudohAl}
S. van Dijken, H.J.W. Zandvliet, and B. Poelsema, 
Phys. Rev. B 55, 7864 (1997);
Phys. Rev. B 56, 1638 (1997); \newline
K. Sudoh et al., Phys. Rev. Lett. 80, 5152 (1998).










\bibitem{McKaneMoore.rep}
A.J. McKane and M.A. Moore, Phys. Rev. Lett. 60, 527 (1988).


\bibitem{Zhang.rep}
Y.-C. Zhang, J. Stat. Phys. 57, 1123 (1989).


\bibitem{FeigelmanAl.fluxlines}
M.V. Feigel'man, V.B. Geshkenbein, A.I. Larkin, and V.M. Vinokur,
Phys. Rev. Lett. 63, 2303 (1989).


\bibitem{NelsonLedoussal.fluxlines}
D. Nelson, P. Le Doussal, Phys. Rev. B 42, 10113 (1990).


\bibitem{Hwa.fluxlines}
T. Hwa, Phys. Rev. Lett. 69, 1552 (1992).


\bibitem{CivaleAl.defects}
L. Civale et al., Phys Rev. Lett. 67, 648 (1991).


\bibitem{BudhaniAl.defects}
R.C. Budhani et al., Phys. Rev. Lett. 69, 3816 (1992).


\bibitem{TangLyuksyutov.depinning}
L.-H. Tang and I.F. Lyuksyutov, Phys. Rev. Lett. 71, 2745 (1993).


\bibitem{BalentsKardar.depinning}
L. Balents and M. Kardar, Europhys. Lett. 23, 503 (1993);
Phys. Rev. B 49, 13030 (1994).


\bibitem{KolomeiskyStraley.depinning}
E.B. Kolomeisky and J.P. Straley, Phys. Rev. B 51, 8030 (1995).


\bibitem{HwaNattermann.depinning} 
T. Hwa  and T. Nattermann, Phys. Rev. B 51,  455 (1994). 


\bibitem{Mezard.pair}
M. M\'ezard, J. Phys. (France) 51, 1831 (1990).


\bibitem{NattermannAl.pair}
T. Nattermann, M. Feigel'man, I. Lyuksyutov, Z. Phys. B84, 353 (1991).


\bibitem{Tang.pair}
L.-H. Tang, J. Stat. Phys. 77, 581 (1994).


\bibitem{Mukherji.pair}
S. Mukherji, Phys. Rev.  E 50, R2407 (1994).


\bibitem{HwaFisher.paths}
T. Hwa and D. Fisher, Phys. Rev. B 49, 3136 (1994).


\bibitem{Lassig.barrier}
M. L\"assig, unpublished.


\bibitem{HwaLassig.unpubl}
T. Hwa and M. L\"assig, unpublished.


\bibitem{KrugTang.dp}
J. Krug and L.-H. Tang, Phys. Rev.  E 50, 104 (1994).


\bibitem{Willbrand.diplom}
K. Willbrand, diploma thesis, Technical University Aachen (1996).


\bibitem{MezardParisi}
M. M\'ezard and G. Parisi, J. Phys. I (France) 1, 809 (1991).


\bibitem{CastellanoAl.kpz}
C. Castellano, M. Marsili, and L. Pietronero, 
Phys. Rev. Lett. 80 (1998), 3527. 











\bibitem{WolfTang.inh}
D.E. Wolf and L.-H. Tang, Phys. Rev. Lett. 65, 1591 (1990).


\bibitem{dynpath}
P.C. Martin, E.D. Siggia, H.A. Rose, Phys. Rev. A8, 423 (1973); \\
R. Bausch, H.K. Janssen, H. Wagner, Z. Phys. B24, 113 (1976).


\bibitem{SunPlischke.kpz}
T. Sun and M. Plischke, Phys. Rev. E 49, 5046 (1994).


\bibitem{Wiese.kpz1}
K.J. Wiese, Phys. Rev. E 56, 5013 (1997).


\bibitem{Wiese.kpz2}
K.J. Wiese, J. Stat. Phys., to appear (1998).


\bibitem{DotyKosterlitz.roughening}
C.A. Doty and M. Kosterlitz, Phys. Rev. Lett. 69, 1979 (1992).


\bibitem{Lassig.cth}
M. L\"assig, Nucl. Phys. B 334, 652 (1990).


\bibitem{long}
M. L\"assig, to be published.


\bibitem{ChekhlovYakhot}
A. Chekhlov and V. Yakhot, Phys. Rev. E {51}, R2739 (1995).


\bibitem{K41}
A. Kolmogorov, Dokl. Akad. Nauk. SSR {32}, 16 (1941), reprinted in
Proc. R. Soc. Lond. A 434, 15 (1991). 


\bibitem{Eyink}
G.L.~Eyink, Phys. Lett. A {172}, 355 (1993).


\bibitem{Polyakov}
A.M. Polyakov, Phys. Rev. E {52}, 6183 (1995).


\bibitem{Pelletier}
J.D. Pelletier, Phys. Rev. Lett. 78, 2672 (1997).


\bibitem{Krug.turbif}
J. Krug, Phys. Rev. Lett. {72}, 2907 (1994).


\bibitem{Lassig.Burgers}
M. L\"assig, preprint cond-mat/9811223.


}

\end{thebibliography}
\end{document}